\documentclass[aop]{imsart}
\usepackage{graphicx}                      

\usepackage{amsfonts}
\usepackage[margin=1.0 in]{geometry}

\RequirePackage{amsthm,amsmath}  

\usepackage{amssymb}

\startlocaldefs
\numberwithin{equation}{section}
\theoremstyle{plain}
\newtheorem{thm}{Theorem}[section]
\newtheorem{lemma}{Lemma}[section]
\newtheorem{proposition}{Proposition}[section]  

\newtheorem{cor}[thm]{Corollary}
\endlocaldefs

\long\def\/*#1*/{}  

\usepackage{placeins} 
\usepackage{bm} 
\usepackage{enumitem} 
\usepackage{caption}

\newtheoremstyle{boldhead}
  {\topsep}
  {\topsep}
  {\itshape}
  {}
  {\bfseries}
  {.}
  {.5em}
  {\thmname{#1}\thmnumber{ #2}\thmnote{ (#3)}}

\theoremstyle{boldhead}
\newtheorem{boldthm}[thm]{Theorem}

\begin{document}

\begin{frontmatter}
\title{Concentration Inequalities for Branching Random Walks with Applications to Phase Transitions in CSP\lowercase{s}}
  
 \begin{center}
  
 \begin{tabular}{cc}
 \begin{tabular}{l}
         Changqing Liu  \\ ClinTFL Ltd. \\  \textnormal{c.liu@ClinTFL.com}  
 \end{tabular}
 &
 \begin{tabular}{l}
         Bo Zhang  \\ Dept. of Comp. Sci.,  Tsinghua Univ. \\  {dcszb@tsinghua.edu.cn}
 \end{tabular}   
 \end{tabular}
 
 \end{center}

\begin{abstract}
\/**************************************************************************************
A new framework called PDEM (to vs Wormald's DEM) is proposed to study phase transition in CSPs.

The classic user-friendly Azuma-Hoeffding inequality and Chernoff bounds are widely used, but their applicability is restricted by two persistent limitations: 
(a) requirement of independence or martingale increments in the random process and 
(b) the assumption of bounded increments. Wormald's differential equation method (DEM) framework eliminates the independence and martingale conditions, but provides weaker concentration (sacrificing the sub-Gaussian tail bound) and still requires bounded increments. Motivated by phase transition problems in CSPs, we develop a more general concentration inequality that retains the classical sub-Gaussian tail bounds employing a mild global linear-growth condition $|S_n| < Cn$, relaxing the bounded-increment assumption to finite exponential moments and requiring neither independence nor the martingale property.

We further extend the concentration inequality to the setting of branching random walks (BRW), obtaining the first concentration inequality for BRW. 
  
Applying the BRW concentration inequality to $K$-SAT and $q$-COL, we derive partial differential equations (PDEs) for their backbones and obtain numerous new results, including  
(a) for $(2+p)$-SAT, a resolution of the long-standing open question of where the phase transition changes from second order (2-SAT like) to first order (3-SAT like);  
(b) the prefactor in the 2-SAT critical window; and
(c) for $K$-SAT, rigorous results for $\alpha_d$, which give new lower bounds on the phase transition for 3-SAT (4.0029 vs. 3.51) and 4-SAT (8.360 vs. 7.91). 
****************************************************************************************/  

\/**************************************************************************************
A new framework is developed for studying phase transitions in CSPs.

Classical concentration inequalities such as Azuma-Hoeffding and Chernoff bounds typically require independence or martingale increments and bounded increments, while Wormald's (DEM) weaker concentration bound eliminates the former conditions but still requires the boundedness condition. Motivated by phase transition problems in CSPs, we prove a more general concentration inequality that retains the classical sub-Gaussian tail bounds employing a mild global linear-growth condition $|S_n| < Cn$, assuming only finite exponential moments and neither independence nor the martingale property.

We further extend the concentration inequality to the setting of branching random walks (BRW), obtaining the first concentration inequality for BRW. 

As applications, we derive partial differential equations (PDEs) for the $K$-SAT and $q$-COL backbones, yielding numerous new results, including  
\textbf{(a)} a resolution of the long-standing $(2+p)$-SAT question of where the transition changes from second to first order;  
\textbf{(b)} rigorous results for $\alpha_d$ in $K$-SAT, which give new lower bounds on the phase transition for 3-SAT (4.0029 vs. 3.51) and 4-SAT (8.360 vs. 7.91); and  
\textbf{(c)} the prefactor in the 2-SAT critical window. 
****************************************************************************************/
 
A new framework is developed for studying phase transitions in random CSPs, from which a range of new results follows.

Motivated by phase transition problems in CSPs, a more general concentration inequality is proved that retains classical sub-Gaussian tails under a mild global linear-growth condition $|S_n| < Cn$, relaxing the bounded-increment assumption to finite exponential moments and requiring neither independence nor the martingale property.

We further extend the concentration inequality to branching random walks (BRW), obtaining the first concentration inequality for BRW. 

As applications, we derive partial differential equations (PDEs) for the $K$-SAT and $q$-COL backbones, yielding new results, including  
\textbf{(a)} a resolution of the long-standing open question of where $(2+p)$-SAT transition changes from second to first order;  
\textbf{(b)} rigorous results for $\alpha_d$ in $K$-SAT, which give new lower bounds on the phase transition for 3-SAT (4.0029 vs. 3.51) and 4-SAT (8.360 vs. 7.91); and  
\textbf{(c)} the prefactor of the 2-SAT critical window. 
 
\end{abstract}   






  

\end{frontmatter}

\section{\textbf Introduction}

The Boolean satisfiability problem (SAT) is a cornerstone of computational complexity theory, since Stephen Cook proved its NP-completeness in 1971. A canonical family of satisfiability problems is $K$-satisfiability ($K$-SAT), which is among the most studied. An instance of $K$-SAT is defined as a $K$-CNF formula comprising a collection of $M$ clauses (constraints) over $N$ Boolean variables, $\{x_1, x_2,\ldots, x_N\}$. Each clause $C_j$ is a logical disjunction (OR) of $K$ literals, where each literal is either a variable $x_i$ or its negation $\neg x_i$; so there are totally $2N$ literals. A CNF formula is satisfiable if there exists an assignment  $\sigma \in \{0,1\}^N$ that makes every clause true.

The Boolean satisfiability problem asks whether a Boolean formula is ``satisfiable''---that is, whether there exists an assignment to its input that makes the formula yield $1$ (true). It is well known empirically that the computational hardness of constraint satisfaction problems (CSPs) occurs around a critically constrained regime where the SAT/UNSAT transition takes place abruptly. \cite{Monasson_Nature}\cite{Kirkpatrick&Selman1994}. 
Relatedly, a category of theoretical questions concerns understanding the typical behavior of instances of CSP such as
where and how the phase transition takes place and how the structures of the solutions evolve as the constraint density varies.
This critical behaviour bears a strong resemblance to phase transitions in disordered physical systems \cite{MonassonZecchina96}\cite{paper alpha_d}. Given that a close analogy exists between $K$-SAT and the spin glass models of statistical physics, methods from the latter have been used and have yielded fruitful insights into the structure of the solution space of $K$-SAT, in particular the phase transition phenomenon, which may explain the computational hardness around the critical points \cite{Mezard Parisi Zecchina2002}. Nevertheless, from the strict mathematical perspective, these proposed phase diagrams should be considered as conjectures \cite{Mezard Parisi Zecchina2002}. In this paper, we develop a rigorous mathematical framework that yields partial differential equations (PDEs) from which the phase diagrams of the underlying CSPs are derived.

In many situations, a problem $P(\alpha)$ studied in one dimension may in fact be the restriction of a whole object 
$P(\, \cdot \, , \alpha)$ to a lower dimensional subset in its argument space---say $P(0,\alpha)$;  the problem is intrinsically defined in a higher-dimensional space, while its readily observable form is one-dimensional.
Lifting the original problem to its higher-dimensional generalization $P(\, \cdot\, , \alpha)$, it often reveals underlying structure and, in turn, gives results that would be difficult, or even impossible, to obtain within the restricted space. Moreover, placing a problem within a broader framework often renders it more tractable. 

The satisfiability probability $\Pr(\alpha)$ can be thought of as the restriction of a complete object defined in 
a larger domain; namely, the apparent one-variable problem is therefore a slice of a higher-dimensional structure. 
In light of this observation, we study $\Pr(i/N, \alpha)$, where $i$ is the number of preassigned variables $\big($later we will write $x \equiv i/N$ and thus $\Pr(x, \alpha) \big)$. This is equivalent to a $K$-SAT instance formed by adding $i$ unit clauses---$i$ (distinct) literals (i.e. Boolean variable or its negation)---to $M$ clauses. 

\begin{figure}[ht] 
    \begin{minipage}{0.4\textwidth} 
    
    \centering
       \hspace{-3em} 
        \includegraphics[height=15em, width=23em, trim=24 30 20 20, clip]{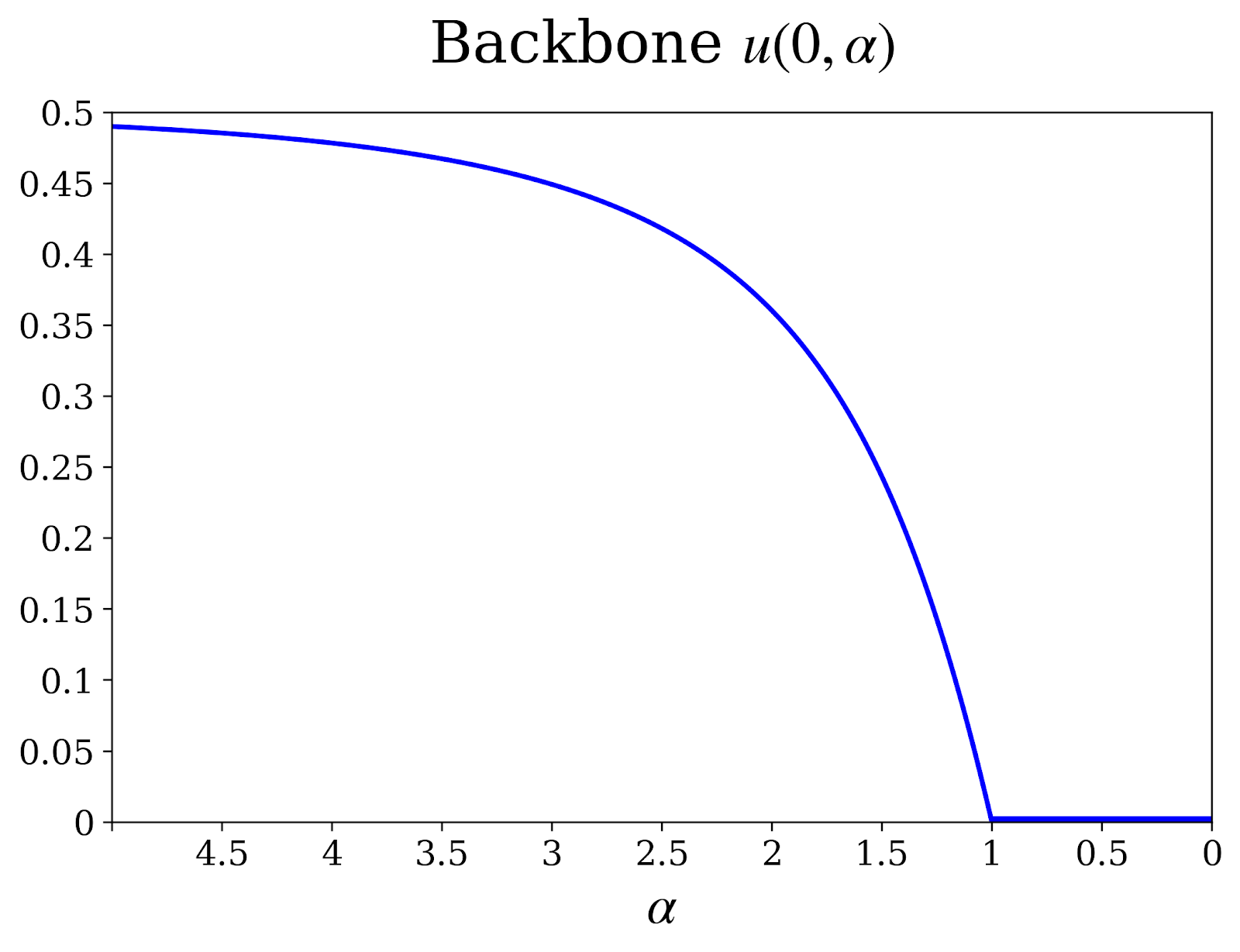}        
        \label{fig:2SAT_u0}

    \end{minipage}     
    \hfil
    \begin{minipage}{0.4\textwidth}   

       \hspace{-1em}           
        \includegraphics[height=15em, width=23em, trim=0 0 5 25, clip]{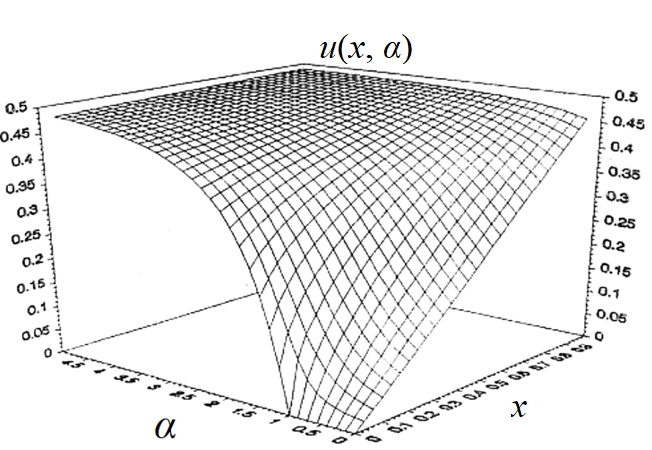}
        \label{fig:2SAT_ux}        
        
    \end{minipage}
\end{figure}
\vspace{-1em}

As an illustration, consider 2-SAT, whose SAT-UNSAT threshold occurs at $\alpha=1$. The state-of-the-art result is the $\Theta(N^{-1/3})$ scaling of the critical window \cite{scaling window of the 2-SAT 2001}, whereas from $u(0,\alpha)$ the prefactor can be identified explicitly as a function of the satisfiability probability\,\footnote
{Specifically, 
$\displaystyle \frac{3}{\sqrt[3]{4}}\sqrt[3]{ \ln \frac{\Pr(1)}{\Pr(y)}} \cdot N^{-1/3}
\quad (\text{see Subsection \ref{subsection 2-SAT} for details%
            }
      )
$.
}; for example, the $50\%$ threshold occurs approximately at a distance 
$
          1.67N^{-1/3} 
$
from the critical point. 
This precise prefactor seems only to be accessible through the full function $u(x,\alpha)$ over its entire two-dimensional domain; indeed, the recursion in Subsection \ref{Derivation of K-SAT} shows that an exact determination $\Pr(\alpha)$ requires the two-dimensional $\Pr(i, \, \cdot\, )$.

\section {Motivation for Extending Concentration Inequalities}\label{Motivation} 
The arguments are formulated in the context of the $K$-satisfiability problem, though they apply naturally
to many other CSPs. We consider the ensemble of randomly generated $K$-SAT formulae, in which each clause is randomly selected from all the $ (2N)^K$ $K$-clauses. For the study of satisfiability probability, the difference between this setting and the standard random $K$-SAT (where complementary literals in a clause are forbidden) is negligible. An instance of $K$-SAT can be constructed by selecting clauses sequentially, each step having $(2N)^K$ branches of choices. Therefore, it can be viewed as a path in a construction tree, where each node has a branching factor of $(2N)^K$. 

As aforementioned, we study generalized $K$-SAT, namely the satisfiability of a CNF formula $F$ in a partial assignment space. Specifically, define
\[
        \mathcal{S}_i = \{x \in \{0, 1\}^N: x_1 = \cdots = x_i = 1 \}
\]
i.e, the $(N-i)\text{-dimensional}$ subcube of $\{0,1\}^N$ defined by fixing the first $i$ variables.\footnote{Because of symmetry, only the number $i$ matters.} Let $P(i,m)$ represent the satisfiability probability of $F_m$ (a formulae of $m$ constraints) in $\mathcal{S}_i$. Then the satisfiability probability $\Pr(F_m)$ in the classical random $K$-SAT is the special case of $P(0,m)$. 
 
Let $u(x,z)$ ($z = \frac m N$, i.e. $\alpha$) be the fraction of the \emph{backbones} of $F_m$ (a variable is called backbone if in all solutions to the CSP the variables is assigned the same value; also called frozen variable).   If $u$ is concentrated around its mean $\bar u$ with high probability, we may treat it as effectively deterministic (i.e., neglect the fluctuations) and obtain the following system of recurrences for $P(i,m)$:
\[
   \frac{P(i,m)}{P(i,m-1)} = 1 - u^K, \qquad   \frac { P(i+1, m) }{ P(i, m) } = \frac{1 - u - x/2}{1 - x}
\]
In the large $N$ limit, the recursion converges to a first-order PDE (for convenience, here and elsewhere, ``PDE'' may refer either to the equation or to its solution)\footnote{
For a rigorous derivation, see Subsection \ref{Derivation of K-SAT}, {\textbf{\textup{Derivation of}  $\bm{K}$\textup{-SAT PDE}}}.
}\,: 
\[
            z = \frac{2(1-u^K)}{\rule{0pt}{10pt} Ku^{K-1}} \ln  \frac {1 - u - \frac{x}{2}}  {1-2u} , \quad x = \frac i N \text{, } z =\frac m N       
\]    
\FloatBarrier 
\begin{figure}[ht] 
  \begin{minipage}{0.49 \textwidth} 

    \centering
    \includegraphics[height=14em, width=20em, trim=0 0 10 0, clip]{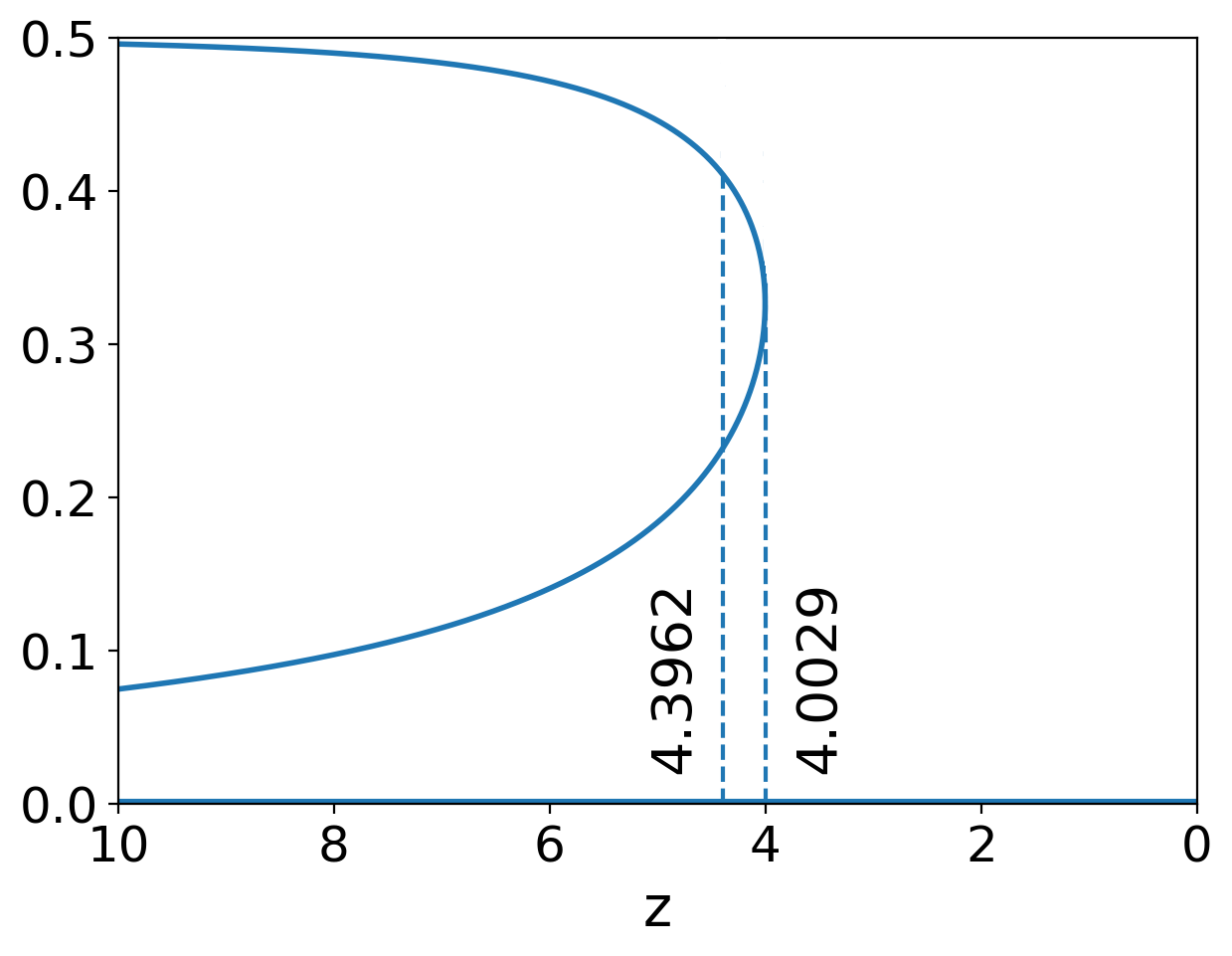} 
    \caption{$u(0,z)$, 3-SAT}
    \label{fig: u(0,z), 3-SAT}  

  \end{minipage}
  \hfill 
  \begin{minipage}{0.49\textwidth}     

    \centering       
    \includegraphics[height=13.5em, width=18em, trim=10 4 35 14, clip]{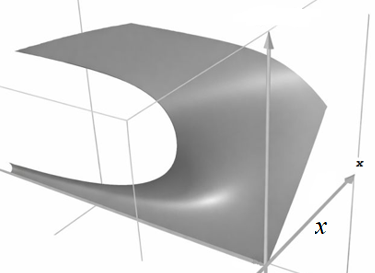}
    \caption{$u(x,z)$, 3-SAT}
    \label{fig: 3-SAT}  
    
  \end{minipage}
\end{figure}

Within this framework, we obtained the $K$-SAT PDE in the late 1990s \cite{My dissertation} and applied it to obtain some analytical results on $K$-SAT phase transitions, including 2-SAT and 3-SAT. Yet without rigorous proof of backbone concentration, all the results, while interesting, remain conjectural. To resolve this issue and ground the framework on a rigorous foundation, we develop a concentration inequality for the random sum $S_n = X_1 + X_2 + \cdots X_n$, where the underlying stochastic process has the following features: 
\begin{enumerate}
\item[(a)] The bounded-increment condition fails, since the backbone increments $X_i$ are not uniformly bounded (nor is the number of occurrences of a variable even); nevertheless $X_i$ have uniformly bounded exponential moment. 
\item[(b)] The variables $X_i$ are neither independent nor martingale differences. 
\item[(c)] The ensemble of all random $K$-SAT instances is generated sequentially by a branching random walk (BRW), with each instance corresponding to a spine of the BRW.
\end{enumerate}
As for (a), although Bernstein-type inequalities do not require boundedness of $X_i$, they are often less convenient to verify. 
Instead, we assume finite exponential moments for $X_i$, an assumption comparable in strength to Bernstein-type moment conditions but easier to verify, for example, in SAT, random graphs, and many other applications. For (b), as far we know, except the Wormald's framework which also rely on boundedness assumptions, most known concentration results require independence, martingale difference, or the like. For (c), to our knowledge, \cite{C. Liu Chernoff Bound BRW}, by the first author of the present paper, is the only work in this direction, which develops a Chernoff-type bound for a special BRW whose branching factors are independent of both history and position. In CSPs such as SAT, however, the branching factor depends on the parent's position which in turn on the past evolution. Essentially, in BRW of CSPs, the proportion of parameters---such as the backbone fraction---are influenced not only by the past evolution but also by the subsequent evolution; backbone fraction at the current stage of the process may be amplified or attenuated by its descendant population that rely upon it, unlike random walk. In short, the process is future-dependent, which has received little attention.

\section {Contributions of this paper}
\subsection*{}
The main contributions of this paper are threefold.
\begin{enumerate}


\item \text{Extension of concentration inequalities.}  We develop a general concentration inequality that removes both the independence and martingale assumptions, as well as the boundedness of increments, while retaining sub-Gaussian tail bounds. 
 
\begin{enumerate}[label=${\bullet}$]
\item Compared with classical concentration inequalities, our approach eliminates both independence/martingale conditions and bounded increments.
 
\item Compared with classic concentration inequalities, our approach eliminates (a) the conditions of independence and martingale increment and (b) the boundedness of increments, without significantly sacrificing strength of the results---retaining sub-Gaussian bound for large deviation. 
\end{enumerate}
\/***************************************************************************************************************************** 
\begin{enumerate}[label=${\bullet}$]
\item Compared with classic concentration inequalities, our approach eliminates (a) the conditions of independence and martingale increment and (b) the boundedness of increments, without significantly sacrificing the strength of the results---retaining sub-Gaussian bound for large deviation. 

\item Compared with Wormald's DEM, our framework shares the same advantage in that neither independence nor martingale conditions are assumed, while differing in that (a) the boundedness condition is relaxed to a finite exponential moment, and (b) the sub-Gaussian large deviation bound is preserved. 
\end{enumerate}
*******************************************************************************************************************************/
\item  \text {Concentration inequalities for branching random walk (BRW).} 

The second contribution is Theorem~\ref{con_ineq_BRW}, a general concentration inequality for BRW. 
 
\vspace{1em} 
\item  \text {Satisfiability problem (SAT).}   
      We obtain (a) the prefactor function in the phase transition critical window of 2-SAT, (b) for $(2+p)$-SAT, an answer to the open question concerning the critical point separating first order and second order of phases transitions, and (c) for $K$-SAT, rigorous results for $\alpha_d$ and new lower bounds for the phase transition of 3-SAT and 4-SAT.
\end{enumerate}

\section*{Organization of the rest of the paper}
\begin{enumerate}[label=]
\item {Pertaining to Extension of concentration inequalities:}
   
  \begin{enumerate}[label=-] 
  \item Subsection~\ref{intro con ine}: reviews concentration inequalities.
  
  \item subsection~\ref{Notation and first contribution} introduces notation for ``independence'' and ``martingale increments'', and explains our contribution. 
  
  \item Lemma \ref{Lemma: Q_Liu&Watble} $\Rightarrow$ Theorem \ref{Azuma} 
  \\ \hspace*{10em} $\Rightarrow$ Corollary of \ref{coro Azuma}  (a more general Azuma inequality). 
   
  \item Lemma~\ref{Lemma: Q_Liu&Watble}, Lemma~\ref{Lip_f}, and Lemma~\ref{con_extension}  
                                              $\Rightarrow$  Corollary of \ref{coro con_extension} 
  \\ \hspace*{10em} $\Rightarrow$   Theorem~\ref{con_ineq}  (a general concentration inequality). 
  \end{enumerate}
  
\item Pertaining to concentration inequalities for branching random walk (BRW): 
      \begin{enumerate}[label=-] 
            \item subsection~\ref{Branching random walk (BRW)} reviews branching random walks, 
            \item subsection~\ref{Motivation and notation} introduces notation,             
            \item subsection~\ref{Concentration inequality for BRW}: Lemma \ref{con_inequality_BRW} $\Rightarrow$ Theorem \ref{con_ineq_BRW} (concentration inequality in BRW)
     \end{enumerate} 

\item  Pertaining to the applications: 
      \begin{enumerate}[label=-] 
            \item  Section \ref{Applications to CSP Thresholds} reviews related work on $K$-SAT phase transition problems, derives the $K$-SAT partial differential equation, and obtains new results. 
      \end{enumerate}
\end{enumerate}

\section {Concentration Inequalities}
\subsection{Introduction}\label{intro con ine}
\textbf{}Let $(S_i)_{i=1, 2, \ldots}$ be a real-valued random process (a one-dimensional random walk). $S_n$ can be formulated as $S_n=X_1 + X_2 + \ldots + X_n$; $X_i=S_i-S_{i-1}$ are called increments. Under certain conditions, in particular independence, $S_n$ converges to a Gaussian distribution within a $\sqrt{n}$ neighborhood of its mean---the central limit theorem (CLT). For $S_n$ beyond the neighborhood, i.e., $|S_n - \mathbb{E}(S_n)| \gg \sqrt{n}$, Chernoff inequality (\cite{Chernoff 1952} \cite{Hoeffding 1963}),  also referred to as Chernoff bound,
\begin{equation}\label{chernoff bound}
         \Pr(|S_n - \mathbb{E}(S_n)| \geq \lambda) \leq e^{- c\frac{\lambda^2}{n}}
\end{equation}
tells us how unlikely this occurrence is.
Chernoff inequality extends to the setting of bounded martingale difference (e.g., $E(X_n \,|\, S_{n-1}) = 0$), known as Azuma-Hoeffding inequality  \cite{Azuma}  (see \cite{Chung} and \cite{McDiarmid89} for surveys and references therein). From a practical perspective, the Azuma-Hoeffding inequality and Chernoff inequality give the same tight probability bound: square-exponential decay\footnote{Therefore, the constant `$c$' in the Chernoff bound is immaterial and the same symbol is used for simplicity.}, while the former has wider scope of applicability.
After the classical results of Hoeffding and Azuma(\cite{Hoeffding 1963}, \cite{Azuma}), many refinements and extensions followed (e.g. \cite{Talagrand 1995}, \cite{McDiarmid89}, \cite{Bentkus 2004}, \cite{bentkus 2008}, \cite{Fan & Grama & Q. Liu 2012}, \cite{Kontorovich 2014}). These concentration inequalities have proven very useful and have a wide variety of applications in computer science, combinatorics, information theory (see e.g., \cite{Dubashi&Panconesi 2009}, \cite{Alon2000},  \cite{Raginsky&Sason 2013}, \cite{Motwani&Raghavan 1995}).

For a long time, no substantive progress was made in eliminating the boundedness condition (the trivial idea of truncation gives weak results, though)\footnote {In fact, it was not until 1995 that the first major improvement in Hoeffding inequality appeared \cite{Talagrand 1995}. The reason might be twofold. One reason may be the lack of strong motivation. Another, nifty techniques employed in the pioneering work, and in the subsequent literatures as well, achieving finer results somehow obscured the fundamental mechanism underlying the concentration inequalities.}. In the early 2000s \cite{Lesigne&Volny} relaxed the boundedness to finite exponential moments; i.e. for a $\delta > 0$, $\mathbb{E}(e^{\delta X_i}) $ is uniformly bounded for all $i$. They obtained  
\[
     \Pr(\,|S_n - \mathbb{E}(S_n)| \geq \lambda|\,) \leq \exp\left(-c \Big( \frac{\lambda^2}{n} \Big)^{1/3}\right)
\]
which falls short of a sub-Gaussian bound. Later, motivated by study of the free energy of a directed polymer in a random environment,  \cite{Q_Liu&Watble} gave a Chernoff bound under the hypothesis that, for some $\delta > 0$, $\mathbb{E}(e^{\delta |X_i|} \mid \mathcal{F}_{i-1}) $ is uniformly bounded for all $i$ \big(where $\{\mathcal{F}_i\}$ is a filtration for the martingale difference sequence $\{X_i\}$\big). Other improvements to concentration inequalities in the relaxation of the boundedness condition can be found in \cite{bentkus 2008} \cite{Fan & Grama & Q. Liu 2012} \cite{Kontorovich 2014} \cite{Fan & Grama & Q. Liu 2015}. 

Wormald's differential equation method (DEM) \cite{Wormald Ann. App Prob} is the first general framework that eliminates the the need for independence and martingale assumptions. To our knowledge, no existing work relaxes both independence/martingale and boundedness conditions while achieving sub-Gaussian concentration. 
The first major contribution of this paper is to do so: informally, with finite exponential moments of the increments, the small per-step effect in a random process can be controlled by the Lipschitz continuity of the mean increment with respect to the history, not necessarily by ``almost independence''. Our approach and the differential equation method (DEM) of Wormald (\cite{Wormald Ann. App Prob}, \cite{Wormald in Warsaw}) share similarities in that both dispense with assumptions of independence or martingale differences in the underlying process, instead employing a martingale construction together with smoothness conditions on the expectation. Although in our framework the absence of the boundedness assumption leads to technically distinct arguments, our approach and Wormald's can be viewed as mutually explanatory, each shedding a different light on how concentration arises without independence or martingale-difference assumptions in the underlying process.

\subsection {Notation and first contribution}\label{Notation and first contribution}
Let $U_i$ denote the partial sum of the first $i$ terms of  $S_n = \sum_{j=1}^n X_j$, namely, $U_i = X_1 + X_2 + \cdots + X_i$. We use lower case for variables scaled  by $1/N$, where $N$ is a constant integer (in the context of $K$-SAT, $N$ represents the total number of variables). For instance, $u_i := \sum_{j=1}^i X_j/N$, $x_i := X_i/N$. For the sake of readability, in some places we use $Y_n$ for $S_n$. 

We let $\mathbb{E}_{i}(\,\cdot \,)$ abbreviate $\mathbb{E}(\, \cdot \mid X_1, ..., X_i \, )$. 
To express the stochastic process in calculus notation, we define
\begin{equation}\label{Py/Pt}
    \frac{\partial y} {\partial t} := 
    \frac{\mathbb{E}_i(y_{t}) - \mathbb{E}_i(y_{t-1})}{  1/N}
     =  \mathbb{E}_i(Y_{t}) - \mathbb{E}_i(Y_{t-1}) = E_i(X_{t}) 
     , \quad (\text{for $ t > i$})
\end{equation}
where $Y_t= X_1 + X_2 + \cdots + X_t $. In the martingale-difference case, by the law of iterated expectations, we have  $E_i(X_t) = 0$ for $ t > i$, and thus
\[ 
    \displaystyle \frac { \partial y } {\partial t} = 0
\]
In the case of independence, $\mathbb{E}_i(X_t) = \mathbb{E}(X_t)$, i.e. 
\[
   \displaystyle \frac { \partial y } {\partial t}= \text{constant (with respect of $u_i$)}
\] 
Thus in  both cases, $\partial y / \partial t$ is independent of $u_i$, and we may write
\begin{equation}\label{vanishing mixed derivative}
              \frac { \partial } {\partial u_i} \left( \frac { \partial y} {\partial t} \right)
              =
              \frac { \partial^2 y } {\partial u_i \, \partial t} =0
\end{equation} 
Here, for $t > i$, we define $\mathbb{E}(X_t \mid u_i) := \mathbb{E}\big[X_t \mid \sigma(u_i)\big]$ which is a function of $u_i$ 
\vspace{0.1\baselineskip} 
by Doob–Dynkin lemma.

While classical concentration inequalities require the condition (\ref{vanishing mixed derivative}),  one of the primary contributions of this work is the relaxation of this classical constraint to the condition:
\[  \left| \frac { \partial^2 y } {\partial u_i \, \partial t} \right| \le L, \quad i=1, 2, \ldots, n
\]
i.e., 
$
\displaystyle \frac { \partial^2 y } {\partial u_i \, \partial t} 
$ 
is required only to be uniformly bounded, thereby allowing the drift $\displaystyle\frac { \partial y } {\partial t}$ to vary with $u_i$. This extends concentration results to general dependent processes with variable drift. 
For instance, the generation of random 3-SAT instances can be modelled as a branching random walk where, at each time step, a set of random clauses is applied to the current formula, producing $O(N^3)$ progeny formulae from the current state; for details, see later in this paper. In this process, certain properties of the formula, such as the number of frozen variables, are history-dependent.

As a side benefit, this translation of classical probabilistic conditions from expectation-based to derivative-based formulations provides a calculus-based perspective on stochastic concentration phenomena.

%
%

\section{Extending the Hoeffding-Azuma inequality}\label{Extending the Hoeffding-Azuma inequality}
As a pedagogical warm-up, we prove a ``non-sharp'' version of Azuma inequality without assuming bounded increments. 
To begin with, we prove a simple inequality drawing on the method introduced in \cite{Q_Liu&Watble} and \cite{Watbled2012}. 
\begin{lemma}\label{Lemma: Q_Liu&Watble} 
If $\mathbb{E}(X)=0$, then for all $\delta > 0$ and $|\, t \,|/\delta \in [0,1]$
\[
     \mathbb{E} \left[ e^{tX} \right] 
                                  \le \exp \left (t^2/\delta^2 \cdot \mathbb{E} \big[ e^{\delta |X|} \, \big] \right)
\]
\end{lemma}
In particular, when $\delta = 1$,
\begin{equation}\label{E<t^exp} 
\mathbb{E}\left[ e^{tX} \right ] \leq \exp{ \left( t^2\mathbb{E} \big[ e^{|X|} \big]  \right) }   
\end{equation}
{\allowdisplaybreaks
\begin{proof}
   \begin{align*}
   \mathbb{E}(e^{tX}) &= 1 + \mathbb{E}\left[ tX \right]  
                    +  \mathbb{E} \left[ \frac{t^2 X^2}{2!} + \frac{t^3 X^3}{3!} + \cdots \right] 
                                                            \\[5pt]
      & =  1 +  \mathbb{E} \left[ \frac{(t/\delta)^2 \cdot (\delta X)^2}{2!} +
                        \frac{(t/\delta)^3 \cdot (\delta X)^3}{3!}  +  \cdots  \ \right]
                                                                                 \\[5pt]
      &\leq  1 +  (t/\delta)^2\mathbb{E} \left[ \frac{|\delta X|^2}{2!} + \frac{|\delta X|^3}{3!} + \cdots \ \right]
                                   \quad    \text {\big(employing $ |\, t \,|/\delta \leq 1 $\big)}
                                                                                 \\[5pt]
      &\leq  1 +  (t/\delta)^2\mathbb{E} \left [ e^{\delta |X|} \right]         
                                                                                 \\[5pt]      
      &\le \exp \left (t^2/\delta^2 \cdot \mathbb{E} \big[ e^{\delta |X|} \, \big] \right) 
   \end{align*}
The proof is concluded.
\end{proof}
}%

\begin{thm}\label{Azuma}
Let $S_n = X_1 + X_2 + \cdots + X_n$, and $\{X_i\}_{i=1}^n$ be a sequence of martingale differences {\normalfont(thus $\mathbb{E} S_n = 0$)} with $\mathbb{E}[ e^{|X_i|} ] \le K$ for all $i = 1, \dots, n$. 

Then, for every $\lambda > 0$, 
\begin{equation}\label{Pr < min}
       \Pr(|S_n - \mathbb{E} S_n| \ge \lambda) \le 2\exp 
                                  \left(
                                           - \min  \left \{\frac{\lambda^2}{4Kn}, \frac{\lambda}{2} \right \}
                                  \right)    
\end{equation}            
\end{thm}
\noindent This is equivalent to 
\[  
                \Pr(|S_n - \mathbb{E} S_n| \ge \lambda) \le
\begin{cases}
              2\exp  \left(  - \displaystyle \frac{1}{4K}\frac{\lambda^2}{n}  \right)      & \lambda \le 2nK    \\[18pt]                   
      \displaystyle    2\exp  \left(  - \displaystyle \frac{\lambda}{2}  \right)  & \lambda > 2nK   
       
\end{cases}
\] 
since
\[
              \lambda \le 2nK  \iff  \frac{\lambda^2}{4Kn} \le \frac{\lambda}{2}
\]                 
\/**********************************************************************************************************************
To prove the inequality, Azuma (\cite{Azuma}) employed the following inequalities, for $|X| \le c $,
\begin{equation}\label{cosh}
      e^{tX} \le \frac {e^{tc} - e^{-tc}}{2c}X + \frac {e^{tc} + e^{-tc}}{2} \quad \text{(due to convexity of $e^{tX}$)}
\end{equation}
and
\begin{equation}\label{et_e2t}
         \frac {e^{tc} + e^{-tc}}{2} \le e^{t^2c^2/2}  \quad \text{($\cosh(x) < e^{x^2/2}    $)}
\end{equation}
to get sub-gaussian condition 
\[
     E(e^{tX}) \le e^{t^2c^2/2}  ~~~~~~ \mbox {for}~ \mathbb{E}(X) = 0
\]
as adopted by textbooks. Sub-gaussian condition is a key ingredient in the proof of concentration inequalities, as this  ultimately leads to a bound of the form 
\[   
   \exp({\sim n t^2 - t\lambda}),\quad 
\]
for $\lambda \sim n^{(1+\varepsilon)/2}$ with $0 < \varepsilon < 1$, 
which 
can be made exponentially small---of order $\exp(-\Th eta(n^\varepsilon))$---by choosing sufficiently small $t$. Intuitively, provided $\lambda$ is not too small, the exponential term can be made exponentially small by choosing small $t$, as $t^2$ is of higher order than $t$ and thus $nt^2$ is vanishing faster than $t\lambda$ when $t$ is approaching $0$. 
\begin{proof}[\textnormal{\textsc{Proof} (\text{Extended Azuma inequality}})]
******************************************************************************************************************/
%

\begin{cor}[Extended Azuma inequality]\label{coro Azuma} Suppose, in addition, that $|S_n| < Cn$ for some constant $C > 0$ {\normalfont (as in $K$-SAT, where $S_n$ is the number of backbones; indeed, a formula with $n$ $K$-clauses contains at most $Kn$ backbones)}. Then, for every $\lambda > 0$,
\begin{equation}
       \Pr(|S_n - \mathbb{E} S_n| \ge \lambda) \le 2\exp 
                                  \left(
                                           -  c\frac{\lambda^2}{n}  
                                  \right)                                   
\end{equation}  
where 
\[
       c = \min\left\{\frac{1}{4K}, \frac{1}{2C} \right \}
\]     
\end{cor}
Henceforth, the condition $|S_n| < Cn$ is assumed throughout.  

\begin{proof} [Proof of Theorem \ref{Azuma}]   
We only consider ${ \Pr }(S_n - \mathbb{E} S_n  \ge \lambda)$, as the proof of the reverse ${ \Pr }(S_n - \mathbb{E} S_n  \le -\lambda)$ is exactly the same. First, let $\lambda \le 2nK$. By Markov's inequality, inductive application of Lemma \ref{Lemma: Q_Liu&Watble} and optimization of $t$, we have
{\allowdisplaybreaks
\begin{align*}
      { \Pr }(S_n - \mathbb{E} S_n  \ge \lambda)  &\le e^{-t\lambda} \cdot \mathbb{E} \left[ e^{t(S_n - \mathbb{E}S_n)} \right]  \quad  \text{\big(Note $\mathbb{E} S_n=0$\big)} 
      \\[6pt]       
       &\le
       e^{-t\lambda} \cdot \mathbb{E} \Big[
                                           e^{tS_{n-1}} \cdot \mathbb{E} (
                                           e^{tX_n} \mid X_1, \ldots, X_{n-1}
                                      )
                            \Big]    
       \\[6pt]
       &\le e^{-t\lambda} \cdot \mathbb{E}\Big[
                        e^{tS_{n-1}} \cdot \exp{ \Big( t^2\, \mathbb{E}  \, [\, e^{|X_n|} \mid X_1, \ldots, X_{n-1} \, 
                                                                          ]   \Big) }
                      \Big] \tag{By (\ref{E<t^exp})  of  Lemma \ref{Lemma: Q_Liu&Watble}} 
                                    \\[8pt] 
       &\le e^{-t\lambda} \cdot \mathbb{E}\Big[
                        e^{tS_{n-1}} \cdot e^{t^2 K}
                      \Big]   \tag{$\mathbb{E}[e^{|X_i|}] \le K$}    
      \\
      & \tag{Induction over $n$} \le   \exp (-t\lambda \, + \, nt^2 K) 
      \\
      & = \exp{ \left ( - \frac{\lambda ^ 2}{2n} \cdot \frac{1}{2K}  \right )  } \tag{Optimal $t = \frac{\lambda}{2nK}$}  
      \\  &\hspace{0em}  
           \text{
                  \big(cf. Azuma's  $ \exp { \left(\displaystyle - \frac{\lambda ^ 2}{2n} \cdot \frac{1}{c^2}\right)} $ 
                   \!\!, assuming  $|X_i| \le c $ 
                  \big)
                } 
\end{align*}
}%
Note that $\lambda \le 2nK$ ensures $t \le 1$, as required in Lemma \ref{Lemma: Q_Liu&Watble}. 

\vspace{.5\baselineskip}
For $\lambda > 2nK$ $\big($and hence $ -\lambda + nK < -\lambda/2$$\big)$, Markov's inequality together with $\mathbb{E}[e^{|X_i|}] \le K$  and the tower property yields
{\allowdisplaybreaks
\begin{align*}
    \Pr (S_n - \mathbb{E} S_n \ge \lambda ) & \le   e^{-\lambda} \mathbb{E} e^{S_n}     \le   e^{-\lambda}   K^n      
                                               =   \exp(-\lambda + n\ln K) 
     \\[5pt]
     &\le    \exp(-\lambda + n K)       
     \\[5pt]
     & \le \exp(-\lambda/2)  \qquad \left(\text{since $-\lambda + nK < -\lambda/2$} \right)   
\end{align*}
Combining this with the result for $\lambda \le 2nK$ completes the proof. 
}%
\end{proof} 
  
\begin{proof}[Proof of Corollary \ref{coro Azuma}]   
Bear in mind that $ \mathbb{E} S_n = 0 $. For $\lambda > Cn $, the proof is trivial, since $|S_n| < Cn$ and hence the left-hand side of the concentration inequality is zero. For $ 0 < \lambda \le Cn $, recalling
\[
       c = \min\left\{\frac{1}{4K}, \frac{1}{2C} \right \}
\] 
we have 
\[
         c\frac{\lambda^2}{n} \le \frac{\lambda^2}{4Kn}
\]
and
\[
   c\frac{\lambda^2}{n} \le \frac{\lambda^2}{2Cn} \le \frac{\lambda}{2} 
\] 
Therefore 
\[ 
        c\frac{\lambda^2}{n} \le \min \left \{   
                                                 \frac{\lambda^2}{4Kn}, \frac{\lambda}{2} 
                                      \right \}
\]
Hence 
\[
                            \exp\left(    - \min \left \{   
                                                 \frac{\lambda^2}{4Kn}, \frac{\lambda}{2} 
                                      \right \}
                                \right)      
                       \le   \exp \left(
                                      -c\frac{\lambda^2}{n} 
                                  \right)    
\]                                      
By Theorem (\ref{Azuma}), the desired follows. 
\end{proof}

The preceding exposition foregrounds the fact that the condition $X_i < \infty $ can be relaxed to $\mathbb{E} (e^{|X_i|}) < \infty $ without compromising the strength of the results in the sense that from a practical point of view, the difference between the constant factors in the exponent, i.e., $\displaystyle \exp{\left(- \frac{\lambda ^ 2}{2n}\cdot \frac1 {2K}\right)}$ versus $\displaystyle \exp{\left(- \frac{\lambda ^ 2}{2n}\cdot \frac 1 {c^2} \right)}$, is immaterial---both are square-exponential decay. In addition, it is demonstrated that the sub-Gaussian tail bounds can be derived directly from the finite exponential moment without relying on ``nifty tricks'' (such as Hoeffding's Lemma); particularly when bounded increments are considered (i.e., $|X_i| \le c$), we see that the sub-Gaussian bound is obtained straightforwardly this way, taking the form $\displaystyle \exp{\left(- \frac{\lambda ^ 2}{2n}\cdot \frac1 {2e^c}\right)}$.

Next, we identify weaker conditions that ensure concentration inequalities, beyond independence, martingale differences, and boundedness. We begin by defining Doob's (or McDiarmid's) martingale, 
\begin{align*}\label{Doob's} 
  \Delta_i &= \mathbb{E}_{i}(X_i + \cdots + X_n) - \mathbb{E}_{i-1}(X_i + \cdots + X_n)
\end{align*}
where $\mathbb{E}_i(\,\cdot\,) := \mathbb{E} (\, \cdot \mid X_1, ..., X_i) $,  and similarly $\mathbb{E}_{i-1}(\,\cdot\,) := \mathbb{E} (\, \cdot \mid X_1, ..., X_{i-1}) $.
It is easy to check
\[       S_n - \mathbb{E} S_n = \sum_{i=1}^n \Delta_i \,
\begingroup
\renewcommand{\thefootnote}{$\S$}
\footnote{  $ \mathbb{E}_n[S_n] - \mathbb{E}_0[S_n] = S_n - \mathbb{E}S_n $    
} 
     \addtocounter{footnote}{-1}     
\endgroup%
                    \, , \quad \mathbb{E} _{i-1}(\Delta_i) = 0
\] 
i.e. $\Delta_i$ is a martingale difference. 
Note 
\begin{align} \label{di}  
      \Delta_i &=  X_i- \mathbb{E}_{i-1}X_i+\sum_{\ell = i+1}^n\big( \mathbb{E}_i X_\ell - \mathbb{E}_{i-1}X_\ell \big) 
\end{align}

%
%
%

%
%
%
%
 
\begin{lemma}[L-Lipschitz]\label{Lip_f} If 
\begin{equation}\label{L|X|/N}
  \left| \,  
          \mathbb{E}_{i-1}(X_\ell \mid X_i) - \mathbb{E}_{i-1}(X_\ell \mid X_i')  
        \, 
  \right | 
              \le L \displaystyle \frac{|X_i - X_i'|}{N} 
              ,\quad \text{for $i < \ell $}  
\end{equation} 
then
\begin{equation}\label{E_LXi} 
                    |\mathbb{E}_i(X_\ell)-\mathbb{E}_{i-1}(X_\ell) | 
                    \le L \frac{|X_i|}{N} + L \frac{\mathbb{E}_{i-1}(\,|X_i|\,)}{N}
                    ,\quad \text{for $ i < \ell $}  
\end{equation}
 
\end{lemma}   

\begin{proof}[\textnormal{\textsc{Proof} (\textsc{Lemma}} \ref{Lip_f})]
   
Let $X_i'$ be a conditionally independent copy of $X_i$ given $(X_1,\ldots,X_{i-1})$; that is,
\[
      X_i'\sim \mathcal{L}(X_i\mid X_1, ..., X_{i-1})
\]
Define
\[
     \mathbb{E}_{X_i'}(\,\cdot\,) \equiv \mathbb{E}_{X_i'}(\, \cdot \mid X_1, ..., X_{i-1})      
\]  
where $\mathbb{E}_{X_i'}(\cdot)$ averages over $X_i'$.  
With this notation and the tower property of conditional expectation, we have, 
{\allowdisplaybreaks
\begin{align*}
      \bigl|\mathbb{E}_i \left( X_\ell \right) - \mathbb{E}_{i-1} \left( X_\ell \right)\bigr| 
      &= \Bigl |\, \mathbb{E}_{i-1}\left( X_\ell \mid X_i \right)  
                         -  
                   \mathbb{E}_{X_i'}\!\left[\,   
                                                 \mathbb{E}_{i-1}\!\left(\, X_\ell \mid X_i'\right)   
                                      \right]
         \Bigl | 
      \\[5pt] &
       = \Bigl |\, \mathbb{E}_{X_i'} \big[\,
                   \mathbb{E}_{i-1} \left( X_\ell \mid X_i \right)  -  \mathbb{E}_{i-1}\!\left(\, X_\ell \mid X_i'\right)  
                                     \big]
        \, \Bigl | 
      \\[5pt] &
       \le \mathbb{E}_{X_i'} \Big[\,
       \bigl |\,  
                 \mathbb{E}_{i-1} \left( X_\ell \mid X_i \right )  -  \mathbb{E}_{i-1}\!\left(\, X_\ell \mid X_i'\right)  
         \bigl |
                             \, \Big]
      \\[5pt] &                             
       \le \mathbb{E}_{X_i'}\! \left[\,   
                                       L \frac{|X_i - X_i'|}{N}  
                             \, \right] 
         \qquad (\text{by \eqref{L|X|/N}})
      \\[5pt] & 
       \le                     
       \frac L N   |X_i| +  \frac L N   \mathbb{E}_{X_i'} \left(\, |X_i'| \,\right) 
      \\[5pt] & 
      =  L \frac {|X_i|} N +  L \frac {\mathbb{E}_{i-1} \left(\, |X_i| \,\right)   } N      
\end{align*}      
}%
The claim is true.
\end{proof}


\begin{lemma}\label{con_extension}  
Let $A$ denote $1+L\frac{M}{N}$, where $M=O(N)$. If the Lipschitz condition is satisfied, i.e. 
\begin{equation}\label{Lipschitz}
  \left| \,  
          \mathbb{E}_{i-1}(X_\ell \mid X_i) - \mathbb{E}_{i-1}(X_\ell \mid X_i')  
        \, 
  \right | 
              \le L \displaystyle \frac{|X_i - X_i'|}{N} 
              ,\quad \text{for $i < \ell $}  
\end{equation} 
and 
\[
       \mathbb{E} _{i-1} \left[\, e^{\delta A|X_i|}  \, \right] \le K  \quad \textnormal{(Finite MGF)} 
\]       
for a constant $\delta > 0$ , $i=1, 2, \ldots , n$, 
then for $n \le M$\renewcommand{\thefootnote}{$\dagger$} \footnotemark 
\begingroup
                   \renewcommand{\thefootnote}{$\dagger$}
                  \footnotetext{Actually, the lemma holds for $n = \Theta(M)$; we assume $n \le M$ here to simplify the argument. 
                               }
                  \addtocounter{footnote}{-1}%
\endgroup 
\begin{equation}\label{Extension for RW}
       \Pr(|S_n - \mathbb{E} S_n| \ge \lambda) \le 2\exp 
               \left(
                     - \min  \left \{
                                       \frac{\delta^2}{4K^2} \frac{\lambda^2}{n}, \frac{\delta \lambda}{2} 
                            \right \}
              \right)  
\/************************                                      
              \renewcommand{\thefootnote}{$\S$} \footnotemark          
              \begingroup
                   \renewcommand{\thefootnote}{$\S$}
                  \footnotetext{Generally, $c(\delta)$ admits a maximum, denoted by $c(\delta^\star)$; heuristically, $c(\delta)$ is close to zero for both very small and very large $\delta$. Hence 
                     $c(\delta^\star) \ge c(1)  \ge  \displaystyle {1}/{(4{K_M}^2) }, 
                                                       $
                   where $ \displaystyle K_{M} := \max_{i}  \mathbb{E} \left[\, e^{A|X_i|}\mid u_{i-1} \, \right]$.
                               }
                  \addtocounter{footnote}{-1}%
              \endgroup  
**************************/              
\end{equation}
\end{lemma}

\begin{cor}\label{coro con_extension} Suppose, in addition, that $|S_n| < Cn$ for some constant $C > 0$ {\normalfont(as in $K$-SAT discussed in this paper, and many other CSPs, including all cases where Wormald's DEM applies)}. 
%
Then, for every $\lambda > 0$,  
\begin{equation}\label{con ineq sub Gaussian}
       \Pr(|S_n - \mathbb{E} S_n| \ge \lambda) \le 2\exp 
                                  \left(
                                           -  c\frac{\lambda^2}{n}  
                                  \right)                                   
\end{equation}  
where 
\[   
       c = \min \left\{\frac{\delta^2}{4K^2}, \frac{\delta}{4C}  \right \}       
\]     
\end{cor}
Throughout the remainder of the paper, by ``concentration inequality'' we mean the sub-Gaussian tail bound \eqref{con ineq sub Gaussian}.

\textbf{Remarks}. 
\begin{itemize}

\item The finite MGF assumption is equivalent to the Bernstein-type moment condition $\mathbb{E}|X|^k < k!\, K^k$, up to the choice of constants, but more user-friendly than the latter; in practice, Bernstein-type bounds are often derived from the existence of an exponential moment.

\item Condition (\ref{Lipschitz}) in the Lemma \ref{con_extension} is equivalent to the following Lipschitz continuity
\begin{equation}\label{Lipschitz continuity} 
       \left| \mathbb{E}(X_\ell \mid u_i) - \mathbb{E}(X_\ell \mid u_i') \right| \le L\left| u_i - u_i' \right| 
                         \quad \mbox{for $i < l$} 
\end{equation}
where 
\[ 				u_i  = \displaystyle \frac{1}{N}  (X_1 + \cdots + X_{i-1} + X_i), \quad
          u_i' = \displaystyle \frac{1}{N}  (X_1 + \cdots + X_{i-1} + X_i').
\]

Informally, the Lipschitz condition (\ref{Lipschitz}) states that the drift is Lipschitz continuous with respect to history, including the initial state $S_0$. We will later show this holds for smooth drift functions. 

\item
   Independence and martingale difference assumptions both imply the L-Lipschitz condition (\ref{Lipschitz}). That is, the L-Lipschitz condition is weaker than those assumptions.

\item
\noindent Poisson$(d)$, Binomial$(N, d/N)$, and the variable occurrence distribution in SAT/COL models, for example, all satisfy the finite exponential moment condition, as does the Gaussian distribution.

\item 
  For $n = O(N)$,  the L-Lipschitz condtion may imply
                               \[ \displaystyle
                                  \left | \sum_{\ell=i+1}^n \mathbb{E}_{i-1}(X_\ell \mid X_i) 
                                        - \sum_{\ell=i+1}^n \mathbb{E}_{i-1}(X_\ell \mid X_i')   
                                   \right| 
                                \le  L \displaystyle {|X_i - X_i'|} 
                               \]
             When the increment $X_i$ (and $X_i'$) are bounded, this reduces to McDiarmid's bounded difference condition \cite{McDiarmid89}. 
     
\item For sufficiently small $\omega$, $\Delta_i = \mathbb{E}_{i}(X_i + \cdots + X_\omega) - \mathbb{E}_{i-1}(X_i + \cdots + X_\omega)$ is bounded if each $X_i$ is bounded. In this case, by the Azuma–Hoeffding inequality, $ \sum_{i=1}^\omega \Delta_i$ satisfies a sub-Gaussian concentration bound. The whole process of $n$ steps can be segmented into a number of small pieces of size $\omega$\,; each is concentrated around its expectation, and thus the entire process exhibits concentration. A ``piecewise concentration'' argument of this kind is used in Wormald's DEM framework (\cite{Wormald Ann. App Prob},
\cite{Wormald in Warsaw}).

\end{itemize}

\begin{proof}[Proof of Lemma \ref{con_extension}]   
First, consider the small $\lambda$ regime, i.e. 
\[ 
          \lambda \le   2nK^2/\delta 
\] 
By definition (\ref{di}), we have
{\allowdisplaybreaks
\begin{align*}
              &\mathbb{E}_{i-1}  \left[ e^{  \delta  |\Delta_i|   }  \right]
               = \mathbb{E}_{i-1} \exp \left( \delta \, 
          \Big |  
                  X_i- \mathbb{E}_{i-1}X_i \,+ \sum_{\ell = i+1}^n\big( \mathbb{E}_i X_\ell - \mathbb{E}_{i-1}X_\ell \big) 
        \, \Big| 
                                         \right) 
                                       \\[6pt]
                      &\le \mathbb{E}_{i-1}
             \exp \left( 
                         \delta | X_i |  \,+\, \delta | \mathbb{E}_{i-1}(X_i) | 
                               \,+\,  \delta \sum_{\ell=i+1}^n \left|\, \mathbb{E}_i X_\ell -\mathbb{E}_{i-1} X_\ell  \,\right|
                  \right)        
                                       \\[6pt]
\tag{Lemma \ref{Lip_f}}                      & \le e^{\mathbb{E}_{i-1}(\delta |X_i|)} \, \mathbb{E}_{i-1} 
                            \exp   \left (
                                         \delta | X_i |  \,+\,
                                         \delta \sum_{\ell=i+1}^n L \frac{|X_i|}{N}  
                                                         \,+\,
                                         \delta \sum_{\ell=i+1}^n L \frac{ \mathbb{E}_{i-1} |X_i|}{N} 
                                     \right )    
                                       \\[6pt] 
                      & \le e^{\mathbb{E}_{i-1}(\delta |X_i|)}   \, \mathbb{E}_{i-1} 
                              \exp \left(
                                         \delta |X_i| \,+\,
                                         \delta |X_i| L\frac{M}{N} 
                                                      \,+\,
                                         \delta \, \mathbb{E}_{i-1}(\,|X_i|\,)\cdot L \frac{M}{N}
                                   \right) 
                                       \\[6pt]
                      & = \exp \left( \delta \mathbb{E}_{i-1} |X_i| \, + \, \delta L\frac{M}{N} \mathbb{E}_{i-1}|X_i| \right)
                          \cdot
                             \mathbb{E}_{i-1}  
                                                \exp \left( \delta |X_i| \Big( 1 + L\frac{M}{N} \Big) 
                                                    \right)   
                                       \\[6pt]
                      & = \exp \big(\delta A \mathbb{E}_{i-1} |X_i| \big) 
                             \cdot \mathbb{E}_{i-1} \left[  e^{\delta A |X_i| }  \right] \le K^2   
                                      \\  &\hspace{10em}  \text{(Jensen's inequality and the hypothesis)}  
\end{align*}
}%
Thus, for every $i$  
\begin{equation}\label{<= K^2}
                \mathbb{E}_{i-1}  \left[ e^{  \delta  |\Delta_i|   }  \right]       \le K^2   
\end{equation} 
Applying the exponential Markov inequality $\Pr[X \ge \lambda] \le e^{-t\lambda}\, \mathbb{E} (e^{tX})$ and noting
                 $  \mathbb{E}_{i-1}(\Delta_i) = 0   $, 
we have
{\allowdisplaybreaks
\begin{align*}          
      \Pr \big( S_n - \mathbb{E} S_n   \ge \lambda \,\big)   
    &\le  e^{-t\lambda}\,
    \mathbb{E}\!\left[ e^{\,t(S_n - \mathbb{E}S_n)} \right]  
    \\[0.5em] 
    &=  e^{-t\lambda}\,
    \mathbb{E}\!\left[
       e^{t(\Delta_1 + \cdots + \Delta_{n-1})}\;
       \mathbb{E}_{n-1}  \left( e^{t \Delta_n} \right)
    \right]                                                    \\[0.5em]  
    &\le
    e^{-t\lambda}\,
    \mathbb{E}\!\left[
       e^{t(\Delta_1 + \cdots + \Delta_{n-1})}\;
           \exp \Big( t^2/\delta^2 \cdot \mathbb{E}_{n-1} \big[\, e^{\delta |\Delta_n|} \, \big]\, \Big)  
                \right]
    \quad\text{(by Lemma \ref{Lemma: Q_Liu&Watble})}           \\[0.5em] 
    &\le
    e^{-t\lambda}\, e^{\,t^2 K^2/\delta^2}\;
    \mathbb{E}\!\left[
       e^{t(\Delta_1 + \cdots + \Delta_{n-2})}\;
       \mathbb{E}_{n-2}\!\left(e^{t \Delta_{n-1}} \right)
    \right]
    \quad \text{(by (\ref{<= K^2}))}
    \\
    &\quad \vdots
    \\
    &\le
    e^{-t\lambda}\, e^{\,n t^2 K^2/\delta^2}. 
\end{align*} 
}%
The optimal $t$ gives
\[ {\Pr} (S_n - \mathbb{E} S_n \ge \lambda)
            \le \exp \left( - \frac{\delta^2}{4K^2} \frac{\lambda^2}{n} \right) 
\]   
The value of the optimal $t$ is  
\[\displaystyle
    t = \frac{\lambda}{2nK^2/\delta^2}  
\] 
Thus
\[
      \frac{t}{\delta} = \frac{\lambda \delta}{2nK^2}
\]
Hence, in the small $\lambda$ regime, i.e., 
\[ 
            \lambda \le   \frac{2nK^2}{\delta }    
\]
the optimal $t$ satisfies the hypothesis $t/\delta \le 1$ of Lemma \ref{Lemma: Q_Liu&Watble}. 

For the other side of the concentration inequality, note that Lemma \ref{Lemma: Q_Liu&Watble} holds for negative $t$ as well. The proof is then identical to the above.  
\\

Next, in the large $\lambda$ regime, 
\[
    \lambda >  \frac{2nK^2}{\delta }  
                 \text{, \  which implies \ } -\delta \lambda + n {K^2}   < - \frac{\delta\lambda}{2}, 
\] 
Note that \eqref{<= K^2} 
\[
                \mathbb{E}_{i-1}  \left[ e^{  \delta  |\Delta_i|   }  \right]       \le K^2   
\]
holds independently of $\lambda$. 
The exponential Markov inequality and the law of total expectation give
{\allowdisplaybreaks
\begin{align*}          
      \Pr \big( S_n - \mathbb{E} S_n   \ge \lambda \,\big)   
    &\le  e^{-\delta \lambda}\,
    \mathbb{E}\!\left[ e^{\,\delta (S_n - \mathbb{E}S_n)} \right]  
     =  e^{-\delta \lambda}\,
    \mathbb{E}\!\left[
       e^{\delta (\Delta_1 + \cdots + \Delta_{n-1})}\;
       \mathbb{E}_{n-1}  \big( e^{\delta  \Delta_n} \big)  
    \right]   
    \\[0.5em]   
    & \le  e^{-\delta \lambda}\,
    \mathbb{E}\!\left[
                             e^{\delta (\Delta_1 + \cdots + \Delta_{n-1})}\;  K^2   
               \right] \tag{By \eqref{<= K^2}}   
    \\[0.5em]   
    & \le  e^{-\delta \lambda}\,           (K^2)^n   \tag{Induction on $n$} 
      =  \exp( -\delta \lambda\, + n\ln K^2 )         
    \\[0.5em]   
    &  <  \exp( -\delta \lambda\, + nK^2 )   
    \\[0.5em]   
    &  <  \exp \left( - \frac{\delta\lambda}{2}  \right)        
\end{align*} 
}%
The argument for the opposite side is identical. 

Combining the estimate of the small $\lambda$ case, we have 
\begin{equation} 
       \Pr(|S_n - \mathbb{E} S_n| \ge \lambda) \le 2\exp 
               \left(
                     - \min  \left \{
                                       \frac{\delta^2}{4K^2} \frac{\lambda^2}{n},\; \frac{\delta \lambda}{2} 
                            \right \}
              \right)    
\end{equation} 
completing the proof. 
\end{proof}

\begin{proof} [Proof of Corollary \ref{coro con_extension}]   

It is easy to check that the global bound $|S_n| < Cn$ implies  $|S_n - \mathbb{E} S_n| < 2Cn$. For $\lambda > 2Cn $, the proof is trivial, since the left-hand side of the concentration inequality is zero. We now consider the case $ 0 < \lambda  \le 2Cn $. Recall that 
\[   
       c = \min \left\{\frac{\delta^2}{4K^2},\, \frac{\delta}{4C}  \right \}       
\]     
Thus,
\[
         c \le \frac{\delta^2}{4K^2} \text{ \ and \ } c \le \frac{\delta}{4C}
\]
We have 
\[
         c\frac{\lambda^2}{n} \le \frac{\delta^2}{4K^2} \frac{\lambda^2}{n}
\]
and, using $ 0 < \lambda  \le 2Cn $, 
\[
   c\frac{\lambda^2}{n} \le  \frac{\delta}{4C} \frac{\lambda^2}{n} \le \frac{\delta \lambda}{2} 
\] 
Therefore 
\[ 
        c\frac{\lambda^2}{n} \le \min \left \{   
                                               \frac{\delta^2}{4K^2} \frac{\lambda^2}{n}, \frac{\delta \lambda}{2} 
                                      \right \}
\]
and hence
\[
               \exp\left(  - \min \left \{   
                                             \frac{\delta^2}{4K^2} \frac{\lambda^2}{n}, \frac{\delta \lambda}{2} 
                                \right \}
                 \right)      
                       \le   \exp \left(
                                      -c\frac{\lambda^2}{n} 
                                  \right)    
\]                                      
By Lemma   (\ref{con_extension}), the desired follows. 

\end{proof}

\textbf{Remarks}. 

\begin{itemize} 

\item  The Lipschitz condition (\ref{Lipschitz continuity}) may be interpreted as Lipschitz continuous
dependence of the trajectory on the history; e.g., small perturbations in the initial position do not cause drastic changes in the trajectory.

\item  In the case of independence, $\mathbb{E}_i[ X_\ell ] = \mathbb{E}[X_\ell]$, and hence $\mathbb{E}_i[X_\ell]-\mathbb{E}_{i-1}[X_\ell]$ = 0. In the case of martingale difference sequence, $\mathbb{E}_i[X_\ell]=0$ for $\ell>i$. In both cases,
 \[
      \sum_{\ell > i} \big( \mathbb{E}_i X_\ell - \mathbb{E}_{i-1}X_\ell \big) = 0
 \] 
and therefore $\Delta_i$ (see, (\ref{di})) reduces to $X_i - \mathbb{E}X_i$ (independence) and to $X_i$ (martingale difference), respectively. In these cases, the condition of bounded difference or the existence of exponential moment alone suffices to imply the concentration inequalities. 

\item  Recall the notation introduced in the Subsection \ref{Notation and first contribution}. The condition (\ref{Lipschitz continuity}) is 
\begin{equation} \label{cond: Lipschitz} 
       \left|    \frac { \partial } {\partial u_i} \left( \frac { \partial y } {\partial t} \right)  \right| \le L .  
\end{equation} 

\vspace{0.5em} 
\item Two special cases of (\ref{Lipschitz continuity}) are one of independence where $\displaystyle \frac { \partial y } {\partial t}$ is constant (w.r.t. $u_i$), and one of martingale difference where  $ \displaystyle \frac { \partial y } {\partial t} = 0$.

\end{itemize}
 
The following theorem provides readily verifiable conditions for concentration inequality in general settings where
$$ 
        \frac { \partial } {\partial u_i} \left( \frac { \partial y } {\partial t} \right) \ne 0 
$$


\vspace{1\baselineskip}
Let $\mu_n(u_n)$ denote $\mathbb{E} (X_{n+1} \mid u_n)$\!\!\!
\begingroup
\renewcommand{\thefootnote}{$\star$}
\footnote{
          $\mu_n(\cdot)$ corresponds to the trend function $f$ in Wormald's DEM \cite{Wormald Ann. App Prob}, in which it is assumed Lipschitz continuous.\vspace*{.5em} 
         } 
     \addtocounter{footnote}{-1}     
\endgroup%
\!and $\nu_n (u_n)$ denote  $\mathbb{E}(X_{n+1}^2 \mid u_n)$; they are functions of $u_n$ by Doob–Dynkin lemma.\footnote{
\vspace{0.0\baselineskip} 
Precisely, 
$
    \mathbb{E}\big( X_{n+1} \mid u_n \big) := \mathbb{E} \big( X_{n+1} \mid \sigma(u_n) \big)
$, 
$ 
    \mathbb{E}\big( X_{n+1}^2 \mid  u_n \big) := \mathbb{E}\big( X_{n+1}^2 \mid \sigma(u_n) \big).
$ 
}
$\mathbb{E} (X_{n+1} \mid u_n)$ is the directional derivative of $u_n$ along the trajectory---e.g.
\vspace{0.2\baselineskip} 
$\displaystyle\frac{ \partial u_n }{\partial z }$, 
$\displaystyle\frac{ \partial u_n }{\partial x }$   
\vspace{0.2\baselineskip} 
or along other directions (see Fig.~\ref{fig: 3-SAT} of $3$-SAT, for a heuristic illustration of $u_n$, which is concentrated around $\bar u$). 
\vspace{0.2\baselineskip}

Let $\mu_{n}'$ denote $\displaystyle \frac {d\mu_{n}(u_n)}{du_n}$
\vspace{0.2\baselineskip}
, $\mu_{n}''$ denote $\displaystyle \frac {d^2\mu_{n}(u_n)}{d^2u_n}$, $\nu_n'$ denote $\displaystyle \frac{d\nu_n(u_n)}{du_n}$, and so on. We shall use term ``smooth function" to refer that (a) the first several order derivatives (with respective to $u_n$, briefly $u$) exist and (b) they are uniformly bounded (in other words, the domain in question is bounded away from the singularities).

\begin{boldthm}[General concentration inequality]\label{con_ineq}  
If $\mathbb{E} _{i-1}\big(e^{\delta A|X_i|} \big)\le K $ and $S_n \le Cn$, then in the domain where $\mu_n$ and $\nu_n$ are smooth, the sub-Gaussian concentration inequality holds. 
\end{boldthm}
In terms of partial derivatives, the conditions in Theorem \ref{con_ineq} states, in addition to the exponential moment condition and a mild global linear-growth condition $|S_n| < Cn$, that 
\[
    \frac { \partial \mu(t,u)} {\partial u}, \quad  
    \frac { \partial^2 \mu(t,u)} {\partial u^2}, \quad 
    \frac { \partial \nu(t,u)} {\partial u}, \quad \cdots
\]     
are uniformly bounded on the domain (i.e., bounded away from the singularities). These conditions are practically easy to verify, because one can examine $\bar {u}$, the mean of $u$, which in particular satisfies a partial differential equation (PDE); e.g., Fig.~\ref{fig: 3-SAT} for 3-SAT. 
Note, the Lipschitz condition (\ref{Lipschitz continuity}) says that small changes in distant past generations won't lead to violently change in the current step.

\begin{proof}[\textnormal{\textsc{Proof} of Theorem \ref{con_ineq}}] 
By Lemma \ref{con_extension}, we only need to show that the Lipschitz condition (\ref{Lipschitz continuity}) holds, i.e.
\[
       \left| \mathbb{E}(X_n \mid u_j) - \mathbb{E}(X_n \mid u_j') \right| \le L_n \left| u_j - u_j' \right| 
                         \quad \mbox{for $j < n$   and $L_n = O(1)$}
\] 
given the smoothness of  $\mu_j$ and $\nu_j$\!\!%
\begingroup
    \renewcommand{\thefootnote}{\fnsymbol{footnote}}
    \setcounter{footnote}{1}
    \footnotemark
    \footnotetext{In most cases of interest, when $\mathbb{E} (X|u)$ is smooth, $\mathbb{E}(X^2|u)$ is smooth as well. Generally, we assume that, in domains bounded away from singularities, the law of the random variable $X$ depends smoothly on $u$, in the sense that for every smooth function $f$, $\mathbb{E}\!\left[f(X)\mid u\right]$ is smooth in $u$. 
    The regularity of $\mathbb{E}(X^2|u)$ is the price paid for dispensing with a boundedness assumption on $X$.}
\endgroup%
\!\!. In particular, when $j=0$, as at the start point of $K$-SAT formula generation, 
\[
    \Bigg|\, \mathbb{E}\left(X_n \mid u_0  =  \frac   i N \right) 
              - 
             \mathbb{E}\left(X_n \mid u_0' =  \frac{i'}{N}\right)     
    \Bigg| 
                               \le L  \frac{|i-i'|} N  
\]

We consider $\big| u_j - u_j' \big| = o(1)$ because the case where $\big| u_j - u_j' \big | = O(1) $ is trivial since $S_n \le CN$.  
We proceed by induction. The base case, i.e. 
\[ 
        \Big| \mathbb{E} (X_{j+1}|u_j) - \mathbb{E} (X_{j+1} | u_j') \Big| \le L_j \left|u_j - u_j' \right| 
        ,\quad L_{j+1} = O(1) 
\]
holds because of the smoothness of $\mu_{j}$, where the Lipschitz constant $L$ is controlled by the maximum gradient up to $j$.   

In the following, we investigate stochastic process of $\{u_n \}_{n > j}$ starting from a fixed $u_j$, which therefore treated as deterministic.  We write $\mathbb{E}_{u_j}[\, \cdot \,]$ for expectation under the conditional law given $u_j$ which is a function $u_j$. Let 
\[
      \Delta \mathbb{E}\left[ X_{j+1} \right] := \mathbb{E}_{u_j}[\, X_{j+1} \,] - \mathbb{E}_{u_j'}[\,  X_{j+1}  \,]       
\]
Taking the difference between the two trajectories gives
\[
            \Delta \bar u_{j+1}  = u_j - u_j'  +  \frac{1}{N} \Delta \mathbb{E}[\, X_{j+1} \,]
\]
Then the base estimate gives
\begin{align*}
        \left| \Delta \bar u_{j+1} \right| & \le \left| u_j - u_j' \right|  +  \frac{1}{N} L_j \left| u_j - u_j' \right|  
                               \\[3pt]
                               & = \left( 1 + \frac{L_j}{N}  \right) \left| u_j - u_j' \right | 
\end{align*}
More precisely, by the mean value theorem, there exists some $\xi_j$ between $u_j$ and $u_j'$ such that
\begin{align}\label{Delta bar u_j+1}
         \Delta \bar u_{j+1}  & =  \left( 1 + \frac{\mu'_j(\xi_j)}{N}  \right) (u_j - u_j') 
         \\[3pt]\nonumber
                              :&= L_{j+1}(u_j - u_j')
\end{align}

Bearing in mind that $\mathbb{E}(X_{n+1} \mid u_n)$ is a function of $u_n$, we consider a general case of function $f(u_n)$ which is smooth. Write $\Delta_n := u_n - \bar u_n$. We restrict $u_n$ to the neighbourhood of $\bar u_n$: 
 \[ 
      A_n = \Big \{|\Delta_n| \le  \frac{1}{N^{1/2-\varepsilon }} \Big \}, \quad 0<\varepsilon < \frac 1 2
\]   
where $\bar u$ is bounded away from the singularities and hence $\mu_n$ is also bounded away from the singularities, and $f(u_n)$ is smooth on this region. Thus, there is no concern about singularities; all relevant functions are smooth. By the inductive hypothesis, $u_n$ is concentrated around $\bar u_n$ with sub-Gaussian tail bound, implying
\[ 
        \Pr\left( |\Delta_n| > \frac{1}{N^{1/2-\varepsilon } } \right) \le  e^{-c N^\varepsilon}
\]        
which means the unrestricted and the restricted expectations differ only by an exponentially small amount $\exp\!\left(-cN^\varepsilon\right)$, and the discrepancy is negligible. This fact admits a singularity-free discussion. 

Taking expectation of the Taylor series expansion of $f_n(u_{n})$ around the mean of $u_n$ ($u_n$ itself depends on the initial $u_j$ through the evolution of the process)  gives  
\begin{equation} \label{Expanding f}  
    \mathbb{E} \left[ f(u_n) \right] = f (\bar u_n) ~ +
   \frac {1}{2}  f''(\bar u_n) \mathbb{E}\Delta_n^2
                                           + \frac{1}{3!} \mathbb{E}  \left[ f'''(\xi )\Delta_n^3 \right]
\end{equation}
Similarly, since $\mathbb{E}(X_{n+1} \mid u_n)$ is a function of $u_n$, denote by $\mu_n$, by the law of total expectation
\begin{align*}
   \mathbb{E}(X_{n+1})\, \renewcommand{\thefootnote}{$\S$}   \footnotemark  
    & = \mathbb{E} \left[ \mathbb{E}(X_{n+1} \mid u_n) \right]  
      = \,  \mathbb{E} \left[ \mu_n(u_n) \right] \\
    & = \mu_{n} (\bar u_n) ~ +
   \frac {1}{2}  \mu_{n}''(\bar u_n) \mathbb{E}\Delta_n^2 
                                           + \frac{1}{3!} \mathbb{E}  \left[ \mu_n'''(\xi )\Delta_n^3 \right]
\end{align*}
The sub-Gaussian tail bound 
yields
\begingroup
    \renewcommand{\thefootnote}{$\S$}
    \footnotetext{Given $u_j$, $\mathbb{E}(X_{n+1})$ is a function of $u_j$. We may return to this notation  $\mathbb{E} (X_{n+1} \mid u_j)$ later. 
                 }
\endgroup 
\[
   \mathbb{E}|\Delta_n|^p \le \frac {C_p} {N^{p/2}}, \;   \mbox {for  $p \ge 1$ } 
                    \text{
                          (see Appendix \ref{sub-Gaussian<=> moment ≤ C_p/N^{p/2}} for a proof) 
                         } 
\]
By the hypothesis of ``smoothness'', $\mu_{n}'''(\xi)$ exists. Hence
\[
     \Big|\, \mathbb{E} \left[ \mu_{n}'''(\xi )\Delta_n^3 \right] \Big|   \lesssim  \;    {N^{-3/2}}
\]
and therefore  
\begin{equation} \label{E(X_{n+1})}
  \mathbb{E} (X_{n+1}) = \mu_{n} (\bar u_n) ~ +
               \frac {1}{2} \mu_{n}''(\bar u_n) \mathrm{Var}_n   +  \sim \frac{1}{N^{3/2}}
\end{equation}
where $\mathrm{Var}_n$ denotes $\mathbb{E} (u_n - \bar u_n)^2$.
Similarly, for $\nu$ we have
\begin{align}\label{EX{n+1}^2)}
  \mathbb{E} (X_{n+1}^2)   =& \nu_{n} (\bar u_{n})   + \frac {1}{2}   \nu_{n}''(\bar u_{n}) \mathrm{Var}_n
               +   \sim \frac{1}{N^{3/2}}
\end{align}
To estimate $\mathrm{Var}_n$, consider
\begin{align*}
    \mathrm{Var}_{n+1} &= \mathbb{E} \Big(u+x_{j+1} \cdots  + x_{n} -  
                                                                 (u + \bar x_{j+1}+ \cdots + \bar x_{n})
         + (x_{n+1} - \bar x_{n+1}) \Big)^2
         \\
                       &:= \mathbb{E} \Big( \Delta_n 
                                + (x_{n+1} - \bar x_{n+1}) \Big)^2         
\end{align*}
which yields the following recurrence equation about $\mathrm{Var}_{n}$
\begin{equation}\label{Var_n}
    \mathrm{Var}_{n+1} =  \mathrm{Var}_n + 2\mathbb{E} [ (x_{n+1}-\bar x_{n+1})\Delta_n]
        + \frac {1}{N^2}\mathbb{E} \left(X_{n+1} - \overline X_{n+1} \right)^2
\end{equation}
Applying the law of total expectation with $\mathbb{E} [X_{n+1} \,|\, u_n] = \mu_n(u_n)$, and noting $\bar x_{n+1} = \mathbb{E}[x_{n+1}]$, the expectation in the cross term above is estimated:
\begin{align*}
   \mathbb{E} \left[ (x_{n+1}-\bar x_{n+1}) \Delta_n \right] 
   = &
   \frac{1}{N} \mathbb{E} \left[\, X_{n+1} \Delta_n\, \right] 
     - 
   \bar x_{n+1} \mathbb{E} \left[\, \Delta_n \, \right]
   \\
   = & \frac{1}{N} \mathbb{E} \Big[ \mathbb{E} [ X_{n+1} \,|\,  u_n] \, \Delta_n \Big]
   \\
   = & \frac{1}{N}  \mathbb{E} \Big[\mu_{n}(u_{n})\Delta_n \Big]
\end{align*}
By the mean value theorem, 
we have 
\[
   \mu_{n}(u_{n})   =   \mu_{n}(\bar u_n) + \mu_{n}'(\xi) \Delta_n
\] 
where $\xi$ lies between $u_n$ and $\bar u_n$. 
Let 
\[
      I_n = \left[\bar u - \frac{1}{N^{(1-\varepsilon)/2}}, \bar u + \frac{1}{N^{(1-\varepsilon)/2}} \right]
\]
\[
              m_\xi := \min_{t\in I_n} \mu'_n(t) ,  \quad M_\xi := \max_{t\in I_n} \mu'_n(t)
\]
It holds that
\[
     m_\xi  \mathbb{E}     \Delta_n^2   
              \le  \mathbb{E} \left[ \mu_{n}'(\xi) \Delta_n^2  \,\right] \le  
     M_\xi  \mathbb{E}   \Delta_n^2   
\]
Therefore, there exist $\xi_n^* \in I_n$ such that
\[
                \mathbb{E} \left[ \mu_{n}'(\xi) \Delta_n^2    \,\right]
              = \mu_{n}'(\xi_n^*) \mathbb{E} \Delta_n^2 
\]
Plugging in, 
\begin{align*}
         \mathbb{E} &\left[\,  \mu_{n}(u_{n}) \Delta_n   \, \right]    
          =   \mu_{n}'(\xi_n^*) \mathbb{E} \Delta_n^2 
\end{align*}
Hence,  
\begin{align*}
   2\mathbb{E} \left[ (x_{n+1}-\bar x_{n+1}) \Delta_n  \right]     
    = & \frac{2\mathrm{Var}_n}{N}   \mu_{n}'(\xi_n^*)     
\end{align*}

\vspace{0.5\baselineskip}
Now let's estimate $N^{-2} \mathbb{E} \left( X_{n+1} - \overline  X_{n+1} \right)^2 $, the third term in \eqref{Var_n}. 
Combining the identities (\ref{EX{n+1}^2)}) and (\ref{E(X_{n+1})}) and neglecting terms of order $O(N^{-3/2})$ in $\mathbb{E} \left( X_{n+1} - \overline  X_{n+1} \right)^2$, we have 
\begin{align*}
     \mathbb{E} \left( X_{n+1} - \overline  X_{n+1} \right)^2 = &  \mathbb{E} (X_{n+1}^2) -  \overline  X_{n+1}^2
    \\      
       = & \nu_{{n}} (\bar u_{n})   +
  \frac {1}{2}
   \nu_{{n}}''(\bar u_{n}) \mathrm{Var}_n -
                                 \left[ \mu_{{n}} (\bar u_{n}) + \frac {1}{2} \mu_{{n}}''(\bar u_{n}) \mathrm{Var}_n \right]^2
      \\
    = & \nu_{{n}} (\bar u_{n}) + \frac {1}{2} \nu_{{n}}''(\bar u_{n}) \mathrm{Var}_n
        -    \mu_{{n}}^2 (\bar u_{n}) - \mu_{{n}} (\bar u_{n}) \mu_{{n}}''(\bar u_{n}) \mathrm{Var}_n
   \\
    = & \nu_{{n}} (\bar u_{n}) - \mu_{{n}}^2 (\bar u_{n})
         + \left( \frac {1}{2} \nu_{{n}}''(\bar u_{n})
         - \mu_{n} (\bar u_{n})\mu_{n}''(\bar u_{n})
           \right) \mathrm{Var}_n
\end{align*}
In particular, at $j+1$ 
\[
     \mathbb{E}\left(X_{j+1} - \overline X_{j+1} \right)^2 = \mathbb{E}\left(X_{j+1}^2 \right) - \mu_{j}^2
                                      = \nu_j(u_j) - \mu_{j}^2(u_j)
                                      = \mathrm{Var}(X_{j+1})
\]
where $\mathrm{Var}(X_j)=0$ because $u_j$ (such as $u_0$) is treated as a deterministic variable.

For final variance recurrence, substituting $\mathbb{E} (X_{n+1} - \overline  X_{n+1} )^2$ and $\mathbb{E} [(x_{n+1}-\bar x_{n+1})(u_{n} - \bar u_{n})] $ back into (\ref{Var_n}), we have the recurrence relation for $n \ge j$,
\begin{align*}
    \mathrm{Var}_{n+1} &=  \mathrm{Var}_n  + \frac{2\mathrm{Var}_n  }{N} 
                                                      \mu_{n}'(\xi_n^*)    
    \\
    & 
       ~ +   
       \frac{1}{N^2}  \left[ \nu_{{n}} (\bar u_{n}) - \mu_{{n}}^2 (\bar u_{n})
         + \left( \frac {1}{2} \nu_{{n}}''(\bar u_{n})
         - \mu_{n} (\bar u_{n})\mu_{n}''(\bar u_{n})
           \right)\mathrm{Var}(u_{n})
         \right]
    \\
    = &  \mathrm{Var}_n + \frac {\mathrm{Var}_n}{N} \left( 2\mu_{n}'(\xi_n) 
       ~ +      \frac {1}{2N} \nu_{{n}}''(\bar u_{n})
              - \frac {1}{N} \mu_{n} (\bar u_{n})\mu_{n}''(\bar u_{n})              
             \right)            
   \\  & 
        + \frac{1}{N^2}  \left( 
                                  \nu_{n} (\bar u_{n}) - \mu_{n}^2 (\bar u_{n})
                          \right)
\end{align*}
and the base case is $\mathrm{Var}_j=0$. With the arguments $\bar u_{n}$ and $\xi_n^*$ omitted for brevity:
%
\begin{align*}
    \mathrm{Var}_{n+1}   = & \mathrm{Var}_{n} 
               \left(1 + \frac {2\mu_{n}'}{N} 
             \right)
      +  \frac{1}{N^2} (\nu_{n}  - \mu_{n}^2 ) 
\end{align*}
where higher order terms in $N^{-1}$ inside the parenthesis have been neglected. 
We define  
\[ 
  a_n =   1 + \frac {2\mu_{n}'}{N}  
       , \qquad   b_n = \frac{\nu_{n}  - \mu_{n}^2 }{N^2} 
\]
Then, the variance recurrence relation for $n \ge j$ takes the form 
\[
       \mathrm{Var}_{n+1}   = a_n   \mathrm{Var}_{n} + b_n
\] 
It follows  
\begin{align*}
       \mathrm{Var}_{n+1}
                = & \mathrm{Var}_{j} \prod_{k=j}^{n} a_k + \sum_{r=j}^{n} \left( b_r \prod_{k=r+1}^{n} a_k \right)
                \qquad (\text{where $\mathrm{Var} _j=0$})
            \\
                = & \frac {1}{N^2} \left[  \sum_{r=j}^{n}    \left( \nu_{r}  - \mu_{r}^2 \right)
                          \prod_{k=r+1}^{n}  \left(1 + \frac {2\mu_{k}'}{N} \right)  
                                   \right]                                     
\end{align*}

\vspace{0.2\baselineskip} 
From the above expression, it is easy to check that, on any domain where $\mu_r$, $\nu_r$ and $\mu'_r$ are uniformly bounded for
\vspace{0.15\baselineskip}
$r=1, 2, \ldots, n$, $\bm{\mathrm{Var}_n = O(N^{-1})}$. Heuristically, since each step contributes $O(N^{-2} )$ to the total variance, provided each variance exists,   $n=O (N )$ steps result in a cumulative variance of $O ( N^{-1}  )$. 

We now estimate the derivative of $\mathrm{Var}_n$ and show that $\bm { \mathrm{Var}_n' } $ \textbf{is also} $\bm {O(N^{-1}) }$, which is needed for the rest of the proof and will also be used later in deriving the $K$-SAT PDE. Employing the expression of $\mathrm{Var}_{n+1}$, hereafter denoted by $V_{n+1}$, we have
\[
       V_{n+1} - V_n = O(1) \frac {1} {N^2}
\]
More generally, over $\Delta n$ steps, 
\[
       V_{n+\Delta n } - V_n = O\left( \frac {\Delta n} {N^2} \right) 
\]
and thus
\[
           \frac{V_{n+\Delta n}  - V_n}{\Delta y} = \frac {O(1)} N, \qquad \Delta y = \frac{\Delta n}{N}
\]
and thus
\[
            \frac{dV}{dy} = O(N^{-1})
\]
Note that in the preceding argument, $n$ is a generic index for successive steps of the stochastic process, which therefore may be considered along either the longitude ($m$) direction, or the latitude ($i$) direction, or any diagonal path combining both directions. Therefore, in the $i$-direction (corresponding to an increase in the number of fixed variables, in $K$-SAT), and in the $m$-direction (corresponding to the addition $K$-clauses), respectively, we have
\begin{equation}\label{Var'}
    \frac  {\partial V_n} { \partial x } =  O(N^{-1}), \qquad \frac  {\partial V_n} { \partial z } =  O(N^{-1})
\end{equation}
where
\[ 
      \Delta x = \frac{\Delta i} N, \qquad   \Delta z = \frac{\Delta m} N
\] 
Let $\Lambda_j :=  \displaystyle \frac {d \bar u_j} {d y}$ (in the $K$-SAT setting, ${d \bar u_0}/{d x} = \frac 1 2$). 
By the inductive hypothesis, i.e., $\Delta \bar u_n = L_n (u_j - u_j') $, the chain rule calculation gives
\[    
      \frac {dV_n}{d\bar u_n} 
      = \frac{1}{\Lambda_j}\frac {1}{L_n}  \frac{\partial V_n }{\partial y}      
      = O\left(\frac{1}{N}\right)
\]

Consequently, by (\ref{Expanding f}) we have
\begin{align*}  
  \Delta \mathbb{E} [f(u_n) ]  &=  \Delta f(\bar u_n) ~   +  ~  \Delta  \frac {1}{2} f_n'' \mathrm{Var}_n 
                                                                                                 ~ + ~  \frac{C}{N^{3/2}}\,  
       \\                                                                                                  
        & =   f_n' \Delta \bar u_n
           +   \frac 1 2 f_n''' \mathrm{Var}_n \Delta \bar u_n  +
                    \frac {1}{2} f_n''  \mathrm{Var}_n' \Delta \bar u_n ~ + ~  O(\Delta \bar u_n^2)
                    ~ + ~  \frac{C}{N^{3/2}}\, 
\tag{Expanding}                    
        \\
               & =   f_n' \Delta \bar u_n  
                         + \frac {1}{2} f_n''' O \Big(\frac {1}{N} \Big) \Delta \bar u_n  +
                        \frac {1}{2}  f_n''O\Big(\frac {1}{N} \Big) \Delta \bar u_n   ~ +  ~ O(\Delta \bar u_n^2)
                        ~ + ~  \frac{C}{N^{3/2}}\, 
        \\[5pt]
        & =   \Delta \bar u_n \Big( f_n' +  O\left( N^{-1} \right)  + O(\Delta \bar u_n)  \Big)
                  ~ + ~  \frac{C}{N^{3/2}}\, 
\end{align*} 
where $f_n'$ is shorthand for $f_n'(\xi)$, with $\xi$ lying between $\bar u_n$ and $\bar u_n'$.

Similarly, from (\ref{E(X_{n+1})}) where $f(\cdot) = \mu(\cdot)$,   
\[
  \Delta \mathbb{E} (X_{n+1})   =   \Delta \mu_{n} ~   +  ~  \Delta  \frac {1}{2} \mu_{n}'' \mathrm{Var}_{n} 
                                                                                                 ~ + ~  \frac{C}{N^{3/2}}\, . 
\] 
yielding
\begin{equation}\label{Delta E X_n+1}
     \Delta \mathbb{E} (X_{n+1})  
     =   \Delta \bar u_{n} \Big(\mu_{n}' + O\left(N^{-1}\right) + O\left(\Delta \bar u_{n}\right) \Big) 
     ~ + ~  \frac{C}{N^{3/2}}\, 
\end{equation}
It follows from  
\[
  u_{n+1} = u_{n} + \frac{X_n}{N}, \qquad    \bar u_{n+1}= \bar u_{n} +  \frac{1}{N}  \mathbb{E}(X_{n+1})
\] 
that
\begin{align*}
 \Delta \bar u_{n+1}&= \Delta \bar u_{n} + \frac {1}{N}  \Delta \mathbb{E}(X_{n+1}) \\
                  &= \Delta \bar u_{n} + 
                          \frac {1}{N}  \Delta \bar u_{n} \Big(\mu_{n}' + O(N^{-1}) + O(\Delta \bar u_{n}) \Big) 
                           + O\left(N^{-5/2}\right)
              \\
                  &= \Delta \bar u_{n} \left(  1 + \frac {1}{N}   \mu_{n}' + \frac{o(1)}{N}   \right) 
                   +  O\left(N^{-5/2}\right)
\end{align*}   
By iteration on $n$ and neglecting the term of $o(1)N^{-1}$, it follows
\begin{align*}
           \Delta \bar u_{n+1} &= \Delta \bar u_{j+1} \prod_{k=j+1}^n \left(1 + \frac{\mu'_k}{N} \right)  
            \\[3pt] 
\tag{By \eqref{Delta bar u_j+1}}    &= (u_j - u_j') L_{j+1}  \prod_{k=j+1}^n \left(1 + \frac{\mu'_k}{N} \right)  
            \\[3pt]                                               
                                    :&= L_{n+1} (u_j - u_j')
\end{align*} 
where the additive term $O\big(N^{-3/2}\big)$ is omitted. 
Incorporating the induction base, we have
\[
           \Delta \bar u_{n+1} \le \prod_{k=j}^n \left(1 + \frac{\mu'_k}{N} \right) \cdot  \left| u_j - u_j' \right|  
\] 
In particular, when $\left| u_0 - u_0' \right| = O\left(N^{-1}\right) $
\[
           \Delta \bar u_{n+1} = O\left(N^{-1}\right)
\] 
This is consistent with the fact that the solution of a PDE is continuous with respect to its initial condition (see, e.g.,  Fig.~\ref{fig: 3-SAT} in section~\ref{Motivation}). 

Finally, writing
\[
            M_n :=   \sup_{u}|\mu_n'(u)|
\]
and noting 
\begin{align*}
  \Big| \mathbb{E} (X_{n+1}  \mid u_j) - \mathbb{E} (X_{n+1}  \mid u_j') \Big|
                &=   \Big|   
                              \Delta   \mathbb{E}  \left[ \mu_n(u_n)   \right]  
                    \Big|
               \\[4pt]     
                &= \big|  \Delta \bar u_n \left( \mu_n'(\xi) + o(1) \right)  \big |   
                                                 \quad (\text{see \eqref{Delta E X_n+1}})
                \\[4pt]
                &\le
                   \big(M_n + o(1) \big) \big| \Delta \bar u_n  \big | 
\end{align*}
we obtain,
\begin{align*}   
                \Big|\,   \mathbb{E} (X_{n+1}  \mid u_j) - \mathbb{E} (X_{n+1}  \mid u_j') 
               \,\Big| 
            & \le 
                \big(M_n + o(1) \big) L_{n} \left| u_j - u_j' \right|  
\end{align*}
with  $L_n = O\left(L_{j+1}\right)$ when $n = O(N)$.  
This concludes the proof.
\end{proof}

\section{\textbf Concentration inequalities for BRW}\label{Concentration inequalities for BRW}

\subsection{Branching random walk (BRW)}\label{Branching random walk (BRW)}  
In the literature, a branching random walk on the real line is described as follows (see, for example, \cite{Biggins_fst_lst} \cite{Biggins_Mtgl} \cite{Biggins_Chernoff}). In generation zero, an initial particle is located at the origin on the real line $\mathbf{R}$. It splits into a random number of child particles who form generation one. The children's displacements, relative to their parent, correspond to a point process on $\mathbf{R}$. The children in turn split too to form the second generation, and so on. If the average split number (branching factor) is greater than one, then with positive probability the number of the descendants grows exponentially through generations. Classical BRW studies typically address models where offspring and their displacements are independent across generations (e.g. \cite{Biggins_Mtgl}, \cite{Biggins_Chernoff} \cite{BIGGINS_et1997}, \cite{Hu&Shi2009}, \cite{Gao_BRW2014}). The law of large numbers and central limit theorem type results about the distribution of position are established, under some conditions of independence (for instance, i.i.d. branching and walking).

Considering i.i.d. offspring (and hence independent of the parent's position), and i.i.d. displacement, Harris \cite{Harris63} conjectured that the distribution of the descendants' position in the $n$th generation approaches a Gaussian distribution. This conjecture was proved by \cite{Stam66}, \cite{Kaplan_II}. Its extended generation-dependent versions, where offspring and displacement distribution depend on generation $n$, were proved by \cite{Biggins_CLT}, \cite{Klebaner82}, to mention a few. Problems where offspring and displacement are dependent on their parents' positions are studied through model-specific approaches though (e.g. \cite{Yoshida}). Concerning the deviation from the expectation, \cite{Biggins_Chernoff} showed that for any fixed $\delta > 0$, the proportion of particles whose scaled position is less than $\mu - \delta$ (or greater than $\mu + \delta$) converges almost surely to zero. In another direction of quantifying the rarity of the large deviation, extremum has been well studied (e.g. \cite{DEKKING1991} \cite{MCDIARMID_95}, \cite{BACHMANN2000}, \cite{BRAMSON&Zeitouni 2009} \cite{Hu&Shi2009}, \cite{Addario2009}  \cite{Aidekon 2013} ). \cite{DEKKING1991} showed tightness for $M_n - \mathbb{E} M_n$, where $M_n$ is the minimal position in $n$th generation. \cite{MCDIARMID_95} gave probability bound for the deviation of $M_n$, i.e. $\Pr(M_n - \mathrm{Med}_{n} > x) < e^{-\delta x}$, where $\mathrm{Med}_n$ is the median of $M_n$. A Chernoff bound for deviations of the descendants' positions in the $n$th generation has only recently been established for the aforementioned i.i.d.\ case \cite{C. Liu Chernoff Bound BRW}. To our knowledge, this is the only existing work on Chernoff type concentration inequalities for BRW.

\subsection {Motivation and notation}\label{Motivation and notation}
Concentration inequalities estimate tail probabilities:
\begin{equation} \label{RW Pr}
    \int_{|S_n - E(S_n)| > \lambda}  p_{1}(x_1 | u_0)p_{2}(x_2 | u_1) \cdots p_n(x_n| u_{n-1})dx_1dx_2 \cdots dx_n
\end{equation}
where $S_n = X_1 + X_2 + \cdots + X_n$, $p_{1}(x_1 | u_0)p_{2}(x_2 | u_1) \cdots p_n(x_n| u_{n-1})$ is the probability density of $(x_1, \ldots, x_n)$ (here and throughout we use $p_i(x_i)$ for $p_i(x_i| u_{i-1})$ for notational simplicity).

In the algorithm analysis and the study of random graphs, which are modelled as stochastic processes of graph evolution (see e.g. \cite{Karp_Sipser 1981}, \cite{Wormald1}, \cite{Wormald2}, \cite{Wormald Ann. App Prob}, \cite{Wormald in Warsaw} and references therein), the differential equation method (DEM) originated in the 1980s and generalized by in \cite{Wormald Ann. App Prob}. Recently DEM has obtained some important results and received increasing attention. The procedure of DEM (see \cite{Wormald1} for a general framework) consists of (a) verifying that the random parameters concerned concentrate around their means, and (b) finding solutions to the differential equations satisfied by the means.
 
However, existing concentration inequalities, including those proved in DEM, are formulated in the setting of random-walk processes (RW), whereas a large class of problems, such as the sequential generation of ensembles of $K$-SAT and $q$-COL instances, are modelled as branching random walks (BRWs). 

Let's first fix some notation. It is suggested \cite{C. Liu Chernoff Bound BRW} that a BRW be defined by a sequence of pairs
\[
            \Big( a_i(u_{i-1}), p(x_i \mid u_{i-1}) \Big)_{i=1, 2, \hdots } 
\]
where the branching factor $a_i$ (also called birth-rate) is the expectation of offsprings which depends on the birthplace $u_{i-1}$, and
 \[
      p(x_i \mid u_{i-1})dx : = \Pr(X_i \in dx \mid \sigma(u_{i-1}))
 \]
is the probability density function (pdf) of children's displacements which depends on the birthplace as well. Here $a_i$ can be either greater than 1 (so called supercritical) or less than 1 (subcritical).
The probability law of the spine (a single lineage of descendants) is (\cite{C. Liu Chernoff Bound BRW})
\begin{equation}\label{law of spine} 
             \frac { \displaystyle 
                     \prod_{i=1}^{n} a_i(u_{i-1}) p_i(x_i \mid u_{i-1}) \,\renewcommand{\thefootnote}{$\star$} \footnotemark 
                   }
                   { \displaystyle 
                     \int \prod_{i=1}^{n} a_i(u_{i-1}) p_i(x_i \mid u_{i-1}) \, dx_1 dx_2 \cdots dx_n
                   } 
                   \, dx_1 dx_2 \cdots dx_n 
\end{equation}
which can be interpreted as the proportion of spines $(x_1, x_2, ..., x_n)$ in the whole forest. In the context of $K$-SAT, it is this tail probability that we are concerned with
\begingroup
\renewcommand{\thefootnote}{$\star$}
  \footnotetext{$\displaystyle u_{i-1} = x_0 + x_1 + \cdots + x_{i-1}$. 
               }
  \addtocounter{footnote}{-1}%
\endgroup%
\begin{equation} \label{BRW Pr}
    \int_{\left|\, S_n - E(S_n) \, \right| > \lambda} 
             \frac { \displaystyle 
                     \prod_{i=1}^{n} a_i(u_{i-1}) p_i(x_i \mid u_{i-1}) 
                   }
                   { \displaystyle 
                     \int \prod_{i=1}^{n} a_i(u_{i-1}) p_i(x_i \mid u_{i-1}) \, dx_1 dx_2 \cdots dx_n
                   } 
                   \, dx_1 dx_2 \cdots dx_n 
\end{equation}
Except in special cases where the factor $a_i$ can be cancelled out, e.g., $a_i$ is independent of $u_{i-1}$, (\ref{BRW Pr}) cannot be reduced to the corresponding random walks' tail probability (\ref{RW Pr}). In this paper, we develop concentration inequalities directly for BRWs; this constitutes our second major contribution. Although originally motivated by the problems of $K$-SAT/$q$-COL, we believe the results are of independent interest. For the applications, the readers are referred to section \ref{Applications to CSP Thresholds}. 
   
If $a_i$ is constant, the BRW is equivalent to random walk process where $a_i~=~1$. In other words, random walk and branching random walk is mathematically equivalent if $a_i$ is constant. Probability density function (pdf) of BRW can be found in  (\ref{BRW Pr}). In $K$-SAT, for instance, $a_{n+1} = 1-u_{n}^{\!_K}$. That is, if a particle's position is $u_n$, the size of its offsprings is $1-u_n^{\!_K}$.
 
Intuitively, if the branching is ``squeezing''---i.e. if the branching factor is smaller in the regions farther from the mean trajectory ---then atypical paths receive less population weight than in the corresponding random walk. This ''squeezing'' effect makes the BRW distribution more concentration around it mean.  In other words, reducing the birth rate in distant regions pulls the population towards its mean trajectory, resulting in a more concentrated distribution. In context of $K$-SAT, the instances with fewer constraint have more descendants than those with more constraints; in particular, the reverse cannot hold.

Formally, let $Z(u_i)$ be the total progeny of the particle at position $u_i$. Given fixed $u_{i-1}$, this is a function of $X_i$, which we write as $Z(X_i)$. We make the following assumption.

\subsection{Concentration inequality for BRW}\label{Concentration inequality for BRW}
~ \\
\textbf{Assumption (negative association)}.  
If $f(\cdot)$ is a monotonically increasing function, then
\begin{equation}\label{neg_ass_ineq}
   \int f(|X_i|)\,Z(X_i)\,P(dX_i)
   \;\leq\;
   \left( \int f(|X_i|)\,P(dX_i) \right)
   \left( \int Z(X_i)\,P(dX_i) \right).
\end{equation}
If $Z(X_i)$ is decreasing in $X_i$, then the negative association inequality (\ref{neg_ass_ineq}) holds. $K$-SAT satisfies this assumption.

In a random walk (RW), a particle moves endlessly --- it never dies, never mates, and therefore never reproduces. The expectation of $(\cdot)$ is
\[
     \mathbb{E} [\, \cdot\, ]  =  \int (\, \cdot \, )P\big(dX_1 \cdots dX_n\big)
\]
where $P(dX_1 \cdots dX_n) = p_1(X_1) \cdots p_n(X_n) \ dX_1 \cdots dX_n$, with a slight abuse of notation.
In BRW, each particle produces descendants generation by generation. The population average of a quantity $(\cdot)$ is defined as
\begin{equation}\label{definition of E^M}
    \mathbb{E}^{\!^{(M)}} [\, \cdot \, ] =
    \frac { \displaystyle 
             \int (\, \cdot \,) \prod_{j=1}^M a_j \ P\big(dX_{1} \cdots dX_{\!_M}\big)
          }
          { \displaystyle
             \int \prod_{j=1}^M a_j   \ P\big(dX_{1} \cdots dX_{\!_M}\big)
          } 
\end{equation}
where $a_i = a_i(u_{i-1})$.  This is dependent on future generations, whereas in a random walk (RW) such statistics are independent of the future.

Define conditional expectation, with $P(\, \cdot \, \mid u_i) =   P\big(\, \cdot \, \mid \sigma(u_i)\big) $,
\begin{equation}\label{E_i_M}
   \mathbb{E}_{i}^{\!^{(M)}}[\, \cdot \,]    :=   \mathbb{E} {\!^{(M)}} [\, \cdot \mid u_i \,]  =
        \frac { \displaystyle 
                 \int (\, \cdot \,)   a_{i+1}a_{i+2}\cdots a_M   \ P\big(dX_{i+1} \cdots dX_{\!_M} \mid u_i \big)
              }
              { \displaystyle 
                 \int                 a_{i+1}a_{i+2}\cdots a_M   \ P\big(dX_{i+1} \cdots dX_{\!_M} \mid u_i \big)
              }
\end{equation} 
For example, 
\begin{align} \label{E^M}
\begin{split}
    &\mathbb{E}^{\,\!^{(M)}}_i [\, X_\ell  \,]  : =\mathbb{E} {\!^{(M)}} [\, X_\ell \mid u_i \,]   
    =  \frac { \displaystyle 
             \int X_\ell  \prod_{j=i+1}^M  a_j   \ P\big(dX_{i+1} \cdots dX_{\!_M} \mid u_i\big)
          }
          { \displaystyle 
             \int         \prod_{j=i+1}^M  a_j   \ P\big(dX_{i+1} \cdots dX_{\!_M} \mid u_i\big)
          }
\end{split}          
\end{align}
where $u_i = u_0 + X_1/N + \cdots + X_i/N$.  

Define
\begin{equation*}
           P^{\,\!^{(M)}}_i \Big(dX_{i+1} \cdots dX_M\Big)  =  
             \frac {   
                       a_{i+1}(u_{i}) \cdots  a_{\!_{M}}(u_{\!_{M-1}})  \ P\big(dX_{i+1} \cdots dX_M \mid  u_i \big)
                   }
                   { \displaystyle 
                  \int a_{i+1}(u_{i}) \cdots  a_{\!_{M}}(u_{\!_{M-1}})  \ P\big(dX_{i+1} \cdots dX_M \mid  u_i\big)
                   }
\end{equation*}
So the equality (\ref{E^M}) can be written as 
\begin{equation}\label{EX_l}
          \mathbb{E} {\!^{(M)}} [\, X_\ell \mid u_i \,]  
          =   \int X_\ell \  P^{\!^{(M)}}_i \Big(dX_{i+1} \cdots dX_\ell \cdots dX_M\Big) 
\end{equation}
When the branching factors $a_j$ are constant (and hence cancel out), the BRW conditional measure reduces to that of the underlying RW:  
\begin{equation*}
           P^{\!^{(M)}}_i \Big(dX_{i+1} \cdots dX_M \Big)  =  
               {   
                  \ P\big(dX_{i+1} \cdots dX_M  \mid  u_i\big)
               }              
\end{equation*}
and equality (\ref{EX_l}) becomes 
\begin{align*} 
          \mathbb{E} {\!^{(M)}} [\, X_\ell \mid u_i \,]  
         & =   \int X_\ell \  P\big(dX_{i+1} \cdots dX_M \mid  u_i\big) 
         \\
         & =   \int X_\ell \  P\big(dX_{i+1} \cdots dX_{\ell} \mid  u_i\big) 
\end{align*}
The second equality holds because, when the branching factors are constant, the future process does not affect the earlier behaviour.

By the law of iterated expectations
\[
     \mathbb{E} ^{\,\!^{(M)}} [\, \cdot \,] = \mathbb{E}^{\,\!^{(M)}} \big(\, \mathbb{E}_{i}^{\,\!^{(M)}} [\, \cdot \,] \, \big)
\]
that is, the BRW expectation satisfies the usual law of total expectation, despite being defined through weighted (branching) measures. We distinguish between two types of expectation. One is the standard (unweighted) expectation
\[ 
    \mathbb{E} \left[ \, \cdot  \mid u_i \right] := \int (\, \cdot \,) \, P(dX_{i+1} \mid u_i) 
                                                      = \int (\,\cdot\,) \, p_{i+1}(X_{i+1} \mid u_i)dX_{i+1}  
\] 
and the other is the BRW-weighted expectation $\mathbb{E} ^{\,\!^{(M)}}$ or $\mathbb{E}^{\,\!^{(M)}}[\, \cdot \mid u_i \,]$ 

Define
\[
     a_i(u_{i-1}) := \mathbb{E} \big[ a_{i} (u)  \mid u_{i-1}  \big] 
\]
where $u_{i} = u_{i-1} + X_{i}/N $. 

The following lemma is the BRW counterpart of Lemma \ref {con_extension}. Its proof is identical and therefore omitted.

\begin{lemma}\label{con_inequality_BRW}  
 Let $A$ denote \  $1 + L \displaystyle \frac{M}{N}$,  where $M=O(N)$. If  
\begin{equation}\label{Lipschitz BRW}  
   \left| \,  
                 \mathbb{E}^{\,\!^{(M)}}_{i-1}(X_\ell \mid X_i) - \mathbb{E}^{\,\!^{(M)}}_{i-1}(X_\ell \mid X_i')  \, 
   \right | 
              \le L \displaystyle \frac{|X_i - X_i'|}{N}  
              ,\quad \text{for $ i < \ell $}                                     
\end{equation} 
and 
\begin{equation}\label{Exp moment existence}
   \mathbb{E}^{\,\!^{(M)}}_{i-1}(e^{\delta A|X_i|}) \le K \quad \textnormal{(finite MGF)}
\end{equation}  
for a constant $\delta > 0$ , $i=1, 2, \ldots  $ then for $n \le M$
\begin{equation}\label{Extension for BRW}
         \Pr[\ |S_n - \mathbb{E}^{ } (S_n) | \ge \lambda ] \le 2\exp 
               \left(
                     - \min  \left \{
                                       \frac{\delta^2}{4K^2} \frac{\lambda^2}{n}, \frac{\delta \lambda}{2} 
                            \right \}
              \right) 
\end{equation}
where $\mathbb{E}^{ } (S_n) $ is given by (\ref{definition of E^M}). 
\end{lemma}
\/*****************************************************************************************
Note, here without superscript,

\[ 
          \mathbb{E} [\, \cdot \,] := \int (\cdot)P(dX_{i+1}) = \int (\cdot)p_{i+1}(X_{i+1})dX_{i+1}  
\] 
which we call ``local'' expectation. 
We assume branching factor is a smooth function of birth-place $u$.
******************************************************************************************/
\begin{cor}\label{coro con_extension BRW} Suppose, in addition, that $|S_n| < Cn$ for some constant $C > 0$.  
%
Then, for every $\lambda > 0$,  
\begin{equation}\label{con ineq sub Gaussian BRW}
       \Pr(|S_n - \mathbb{E} S_n| \ge \lambda) \le 2\exp 
                                  \left(
                                           -  c\frac{\lambda^2}{n}  
                                  \right)                                   
\end{equation}  
where 
\[   
       c = \min \left\{\frac{\delta^2}{4K^2}, \frac{\delta}{4C}  \right \}       
\]     
\end{cor}

\textbf{Remark}.  
\begin{itemize}
\item Condition (\ref{Lipschitz BRW}) in Lemma \ref{con_inequality_BRW} is equivalent to the following Lipschitz continuity
\begin{equation*}\label{Lipschitz continuity BRW} 
       \left| \mathbb{E}^{\,\!^{(M)}} (X_\ell \mid u_i) - \mathbb{E}^{\,\!^{(M)}}(X_\ell \mid u_i') \right| \le L\left| u_i - u_i' \right| 
                         \quad \mbox{for $i < l$} 
\end{equation*}
where 
\[ 				u_i  = \displaystyle \frac{1}{N}  (X_1 + \cdots + X_{i-1} + X_i), \quad
          u_i' = \displaystyle \frac{1}{N}  (X_1 + \cdots + X_{i-1} + X_i').
\]      
 
\end{itemize}

The following theorem is the counterpart of Theorem \ref{con_ineq}, which gives easily verifiable conditions for BRW concentration inequality (\ref{con ineq sub Gaussian BRW}).

\begin{boldthm}[Concentration inequality in BRW]\label{con_ineq_BRW}   
If $\mathbb{E}_{i-1}(e^{\delta A|X_i|}  )\le K $, $S_n \le Cn$, and the negative association (\ref{neg_ass_ineq}) holds, 
then, in the domain where $a_i(u)$, $\mu_{i}(u)$ and $\nu_i(u)$ are smooth, the concentration inequality holds. 
\end{boldthm}

\begin{proof}  
    See Appendix~\ref{Proof of con_ineq_BRW}. 
\end{proof}

Thus we have shown that the negative association assumption (\ref{neg_ass_ineq}) is sufficient to establish concentration inequalities for BRW, provided the underlying RW is not too scattering. Is it also necessary? Now consider the classical random walk of coin tossing, $S_n = X_1 + \cdots + X_n$ with $X_i \in [1, -1]$ (so $\mathbb{E}(S_n)$~=~0), which is also a BRW with branching factor $a_i = 1$. It is easy to check that if the branching factor is $a= 1 + \delta |u_i|$ for $\delta > 0$, then the BRW is dispersive. This suggests that, to some extent, the negative association assumption is also necessary.

In the situation of bounded $X_i$, a particle always stays around the mean trajectory as long as its birthplace $u_{i-1}$ is near that trajectory. In the unbounded case, however, a particle may enter regions near singularities (e.g., critical points in $K$--SAT) where $u_i$ is not necessarily smooth and the exponential moment condition $\mathbb{E}(e^ {\delta Xi}) < K $ may fail. Nevertheless, the good news is, since concentration inequality holds in the neighbourhood area of the mean trajectory---away from the the singularities---the outliers account for a negligible proportion of the population. Crucially, pruning these rare extremal paths does not weaken the concentration. The following theorem formalizes this observation. 

\begin{thm}\label{thm2} Let $\epsilon$ be small compared with $\delta$, $N^{1/2} \lesssim \lambda < \epsilon N$, and as before, 
\[
 S_i :=\sum_{t=1}^i X_t, \quad   \Delta S_i  :=  |S_i - \mathbb{E} S_i|, \quad  u_i := \frac{S_i}{N} 
\] 
Define the ``good neighborhood'' around the mean path
\[
        D_{i-1} := \Big( \bar u_{i-1} - \frac\lambda {N} ,  \bar u_{i-1} + \frac\lambda {N} \Big)
        , \quad \bar u_{i-1} : = \frac {\mathbb{E}S_{i-1}}{N}
\]
Assume as before
\[
        \mathbb{E} \left[ e^{\delta A|X_i|} \, \big| \, u_{i-1} \in D_{i-1} \right] \le K 
        \quad \textnormal{\textnormal{(bounded MGF)}}
\]
Then
\begin{equation}\label{around_Ex_A} 
     \begin{aligned}
      \Pr \Big( \displaystyle \max_{1 \leq i \leq n} \Big | \,S_i - \mathbb{E}[S_i] \, \Big |  \leq  \lambda    \Big)
          \geq  \prod_{i=1}^{n}\Big(1 - 2e^{-{c\lambda^2}/{i}} \Big) 
     \end{aligned} 
\end{equation}    
\end{thm} 
In our context, $N^{1/2} \ll \lambda < \epsilon N$ and $n < N$. Thus
\[
    \Pr \Big( \displaystyle \max_{1 \leq i \leq n} \Big | \,S_i - \mathbb{E}[S_i] \, \Big |  \ge  \lambda \Big) 
    \leq  2n \exp \left( -\frac{c\lambda^2}{n} \right)
\]
which is extremely (exponentially) small. 
\begin{proof}[\textnormal{\textsc{Proof} of} (\ref{around_Ex_A})]
The proof proceeds by induction. For each $i \ge 1$, set 
\[
         A_i := {\Delta S_i \le \lambda}
\]
Then, the following equivalence holds:
\[
    \displaystyle \max_{1 \leq i \leq n} \big| \,S_i - \mathbb{E}[S_i] \, \big |  \leq  \lambda 
     \Longleftrightarrow A_1  \wedge A_2 \wedge \cdots \wedge A_n 
\]
Note the implications  
\begin{align*}
    A_{i-1} &\iff |S_{i-1} - \mathbb{E}S_{i-1}| \leq \lambda \\
            &\iff \left| u_{i-1} - \bar{u}_{i-1} \right| \leq \frac{\lambda}{N} \\
            &\iff u_{i-1} \in D_{i-1}
\end{align*}

For the base case, we have $u_0 \in D_0$ where 
\[
       D_0 := \Big( u_0 - \frac\lambda {N} ,  u_0 + \frac\lambda {N} \Big);
\]
bounded MGF is satisfied for $i=1$. Thus the concentration inequality holds as 
\[
      \Pr \Big( |\, S_1 - \mathbb{E}[S_1] \,|  \leq  \lambda    \Big) 
          \geq   \Big(1 - 2e^{-{c\lambda^2}/1} \Big) 
\]
Assume the claim (\ref{around_Ex_A}) holds for all $i$ through to $n-1$, i.e,   
\[
    \Pr \Big(A_1  \wedge A_2 \wedge \cdots \wedge A_{n-1} \Big) \geq  \prod_{i=1}^{n-1}\Big(1 - 2e^{-{c\lambda^2}/{i}} \Big) 
\]
Then, given 
\[ 
      u_1 \in D_1, u_2 \in D_2, \ldots,  u_{n-1} \in D_{n-1}   
\]
the standard concentration inequality can be established for $S_n$ by the same argument as in Lemma \ref{con_extension} and Theorem \ref{con_ineq} :  
\[
    \Pr \Big( \big|\, S_n - \mathbb{E}S_n \, \big | \ge \lambda   \, \Big | \,
                                                 u_1 \in D_1, \ldots ,  u_{n-1} \in D_{n-1}  
        \Big)    
        \le 2\exp \Big( -c\frac{\lambda^2}{n} \Big)
\]
so
\[
    \Pr \Big( \big|\, S_n - \mathbb{E}S_n \, \big | \le \lambda   \, \Big | \,
                                                 u_1 \in D_1, \ldots ,  u_{n-1} \in D_{n-1}  
        \Big)     \ge    1 - 2\exp \Big( -c\frac{\lambda^2}{n} \Big)
\]
Thus 
\begin{align*}
    \Pr \Big(A_1 \wedge \cdots \wedge A_{n-1} \wedge A_n \Big) 
       &=  \Pr \Big(A_n \, \big| \, A_1 \wedge \cdots \wedge A_{n-1} \Big) \cdot   
           \Pr \Big(A_1  \wedge   \cdots \wedge A_{n-1}  \Big) 
      \\&
       \ge \Big( 1 - 2 e^ {-c\frac{\lambda^2}{n} }     \Big) \prod_{i=1}^{n-1}\Big(1 - 2e^{-{c\lambda^2}/{i}} \Big) 
\end{align*}
The claim follows. 
\end{proof}

\section {Applications to Phase Transitions in CSPs}\label{Applications to CSP Thresholds}
\subsection {\textbf{$\bm{K}$-\textup{SAT}: Introduction}}
A constraint satisfaction problem (CSP) consists of $N$ variables, $x_1, ..., x_{\!_N}$, and a set of $C_1, ..., C_M$. $K$-SAT is a most extensively studied CSP. Computer scientists observed that under-constrained K-SAT formulae are almost surely satisfiable and those over-constrained are almost surely unsatisfiable (\cite{Chao and Franco 1986} \cite{Simon 1986} \cite{Chvatal 1992}). Further, experimental evidence strongly suggests that there exists a critical value $r_k$ of the ratio $r = M/N$ (clause density) such that almost all $K$-SAT formulas are satisfiable with $r < r_k$ and almost all unsatisfiable with $r > r_k$. For 2-SAT, it has been proved that $r_2 = 1$. (\cite{Chvatal 1992}  \cite{Goerdt 1992} \cite{Fernandez 1992} \cite{scaling window of the 2-SAT 2001}). For $K = 3, 4, ... $, both the location and existence of threshold still remain unproven conjectures. A big step was made by Friedgut \cite{Friedgut} who, using discrete Fourier analysis, proved that $K$-SAT has sharp threshold; i.e. for clause density $r < r_k(n)$ almost all instances are satisfiable and for $r > r_k(n)$ almost all are unsatisfiable. For $k \geq 3$, researchers focus on proving upper bound and lower bound and asymptotic threshold (\cite{Kirousis1998}, \cite{Dubois 2000}, \cite{Kaporis2002} \cite{Frieze&Suen1996}, \cite{Achlioptas_Nature}, \cite{Coja2014}). Basically first moment is used to get upper bounds. Analysis of certain algorithms can give some lower bounds. Second moment method \cite{Achlioptas JAMS 2004} improves the lower bounds significantly. For large $K$, employing the second moment method, \cite{Ding NAE-SAT} \cite{Ding Sly & Sun large k} rigorously proved the existence of satisfiability threshold for both $K$-SAT and its important variant NAE-SAT. Since late 90s, insightful though mathematically highly non-rigorous methods from statistical mechanics provide precise conjectures for $K$-CNF thresholds (\cite{Monasson_Nature}, \cite{Franz et al 4.3962}, \cite{Mezard Parisi Zecchina2002}, \cite{paper alpha_d}).

\subsection {\textbf{Galton-Watson branching process and $\bm{\mathbb{E}\big(e^{\delta A|X_i|} \mid u\big)\le K}$}}
We now verify 
\[ 
    \mathbb{E}\big(e^{\delta A|X_i|} \mid u\big)\le K
\] 
which Theorem \ref{con_ineq_BRW} requires.  

In an instance of $K$-SAT, the implication exploration process can be modelled by a Galton-Watson branching tree (see, e.g. \cite{Feller} \cite{Janson2012}). At the root stage (generation zero), we randomly select an unfrozen variable $v_1$ as the ancestor in the formula and fix its truth value. In the next generation, the ancestor implies several variables---its offspring $\xi$. 

By symmetry and independence, the offspring of each variable follow the same distribution. Since the number of occurrences of each variable in a $K$-SAT (about the number of the clauses containing it) is approximately Poisson distributed, the offspring variable $\xi$ has finite exponential moments. Each clause containing $v_1$ may imply other variables appearing in the same clause with $v_1$, whose offspring in turn goes on to imply the second generation of unfrozen variables, and so on. 

In the domain where $\mu(u)$ is smooth, this implication chain stops after finitely many steps---it cannot cause infinitely many variables to become frozen. In the terminology of Galton–Watson trees, this process is referred as subcritical \cite{Janson2012}. For a subcritical Galton–Watson process \cite{Nakayama et al.2004} proves that $\xi$  has a finite exponential moment if and only if the total progeny $Z$ has a finite exponential moment (a simpler proof, obtained with the help of ChatGPT 5.2, is given in Appendix \ref{finite Ee^ᶿZ}). Therefore, random $K$-SAT satisfies the conditions of Theorem \ref{con_ineq_BRW} and hence \textbf{$\bm{u}$ concentrates around its mean $\bm{\bar u}$} with sub-Gaussian concentration tails. 

\subsection{\textbf{\textup{Derivation of}  $\bm{K}$\textup{-SAT PDE}}}  \label{Derivation of K-SAT}
Define
\[
        \mathcal{S}_i = \{x \in \{0, 1\}^N: x_1 = \cdots = x_i = 1 \}
\]
i.e, the $(N-i)$-dimensional subcube of $\{0,1\}^N$ obtained by fixing the first $i$ variables. Let
\[
              \text{SAT}_i := \{F:F \text{ has a satisfying assignment in } \mathcal{S}_i \}
\]
Then, for fixed $x_1, x_2, \cdots , x_i$, the following two events are equivalent with respect to satisfiability:
\begin{enumerate}
\item[(a)] $F \text{ is satisfiable in } \mathcal{S}_i$\,; namely $F \in \text{\,SAT}_i$. 
\item[(b)] $F \, x_1 x_2 \cdots x_i \text{ is satisfiable}$; 

\end{enumerate}
Therefore, for random $F$
\[
                  \Pr(F \in \text{\,SAT}_i)   =  \Pr(F x_1 x_2 \cdots x_i \text{ SAT} ) 
\]   
By symmetry, the satisfiability probability is independent of both the choice of the $i$ variables and their assigned values; only $i$ matters.  Let $F_m$ be a random formula of length $m$ (i.e., $m$ constraints). For $0 \le i \le N$,  define
\[
   P(i,m) :=  \Pr(F_m \in \text{\,SAT}_i) = \Pr( F_m \, x_1 x_2 \cdots x_i \text{ SAT}) ;
\] 
which, for brevity, we may write as $\Pr(F_m\, x_1   \cdots x_i )$. Thus, the satisfiability probability $\Pr(F_m)$ of the classical random $K$-SAT is the special case $P(0,m)$ in our generalized $K$-SAT framework.

Let $u$ be the fraction of backbones of $F$ (the backbone, for short). Note 
\[
\Pr(Fx \notin \textnormal{SAT}_i \mid F \in \textnormal{SAT}_i) = \mathbb{E} [u], 
\] 
The proof of the following proposition is straight from Appendix \ref{sub-Gaussian<=> moment ≤ C_p/N^{p/2}}.  
%
\begin{proposition}\label{prop8}                          
If $u$ is \emph{concentrated around $\bar u$} with a sub-Gaussian tail bound, then, for $k \ge 2$ 
\[
          \mathbb{E} [ u^k ] = {\bar u}\,^k + O\left( N^{-1} \right)
\]
\end{proposition} 
\begin{proof} Since 
\[
     u^k = (\bar u + \Delta)^k,\quad   \Delta = u - \bar u,  
\] 
we have
\[
     u^k - \bar u^k = k\bar u^{k-1}\Delta + \sum_{j=2}^{k} \binom{k}{j} \bar{u}^{k-j} \Delta^j 
\]
Sub-Gaussian tail bound entails $\mathbb{E} | \Delta | ^j   \le \; O\left( N^{-j/2} \right)$ (Appendix \ref{sub-Gaussian<=> moment ≤ C_p/N^{p/2}}). Noting that $\mathbb{E}\Delta = 0$, we have
\[                 
        \mathbb{E} [ u^k ] -  \mathbb{E} [ \bar u^k ] =  O\left( N^{-1} \right) +  O\left( N^{-3/2} \right) 
                           = O\left( N^{-1} \right)
\]           
\end{proof}
Generally, we have 
\begin{proposition}\label{generally prop8} If $u$ is \emph{concentrated around $\bar u$} with a sub-Gaussian tail bound and  $f(u)$ is a smooth function, then 
$\mathbb{E} \left[f(u)\right] = f(\bar u) + O\left( N^{-1} \right) $                          
\end{proposition}
\begin{proof}  By Taylor series,
\begin{align*}
    \mathbb{E} \left[ f(u) \right] 
             & =  \mathbb{E} \left [ f(\bar u) + f'(\bar u) \Delta  + \frac{1}{2} f''(\xi)\Delta ^2 \right ] 
             \\[3pt]
             & =  f(\bar u) + \frac{1}{2}f''(\xi^*) \mathbb{E} \Delta ^2   \qquad (\mathbb{E}\Delta = 0)
             \\[6pt]
             & =  f(\bar u) +  O\left( N^{-1} \right)      \qquad (\text{by Appendix \ref{sub-Gaussian<=> moment ≤ C_p/N^{p/2}}})
\end{align*}
\end{proof}

We now obtain a recurrence for $P(i,m)$ of $K$-SAT, sometimes omitting ``$\in~\!\!\!\!\text{SAT}_i$'' for notational conciseness. Recall that $u(F_m)$ denotes the backbone of a $K$-SAT formula $F_m$. Then
\begin{align*} 
        u(F_m) &= \mathbb{E} \left(  \mathbf{1}_{\{ F_m x \notin \text{SAT}_i \}} \mid F_m   \right)
        ,\qquad F_m \in \mathrm{SAT}_i 
\end{align*}
That is, conditional on $F_m\in\mathrm{SAT}_i$, $F_m \, x \notin \mathrm{SAT}_i$ if and only if $x$ is the negation of a backbone literal.
Taking expectation of $1 - u$,   
\[
\begin{aligned}
      1-\bar u
      &=
      \mathbb E\!\left[
          \mathbf 1_{\{F_mx\in\mathrm{SAT}_i\}}  \mid   F_m\in\mathrm{SAT}_i
      \right] 
%
\end{aligned}
\] 
which averages over both $x$ and $F_m$. 
To evaluate $1 - \bar u$, we apply the law of total expectation with respect to the partition
\[
              A_1 := x \in \{x_1, \ldots ,x_i\} 
      , \quad A_2 := x \in \{\neg x_1, \ldots , \neg x_i\} 
      , \quad A_3 := x \in \{x_{i+1}, \neg x_{i+1}, \ldots, x_N, \neg x_N\} 
\]
Then, with the notation ``$\in \text{SAT}_i$'' omitted
\[
\begin{aligned}
     1 - \bar u = 
      \mathbb E\!\left[
          \mathbf 1_{\{F_mx\}}  \mid   F_m 
      \right]  
       & =  \mathbb{E} \Big[ \mathbf 1_{\{F_mx\} }   \mid F_m, A_1 \Big] \Pr(A_1 \mid F_m)  
      \\[4pt] 
       &   \,+\, \mathbb{E} \Big[ \mathbf 1_{\{F_mx\} }   \mid F_m, A_2 \Big] \Pr(A_2 \mid F_m)
      \\[4pt] 
       &   \,+\, \mathbb{E} \Big[ \mathbf 1_{\{F_mx\} }   \mid F_m, A_3 \Big] \Pr(A_3 \mid F_m)
\end{aligned}
\]
We now evaluate the three summands on the right-hand side in turn.

\vspace{0.5\baselineskip} 

\textbf {The first summand}. 
$\mathbb{E} \Big[ \mathbf 1_{\{F_mx\} }   \mid F_m, A_1 \Big] \Pr(A_1 \mid F_m)$ is
\[
\begin{aligned}  
        \mathbb{E} \Big[ \mathbf 1_{\{F_mx \in~\text{SAT}_i \} } \mid F_m \in~\text{SAT}_i, A_1 \Big] 
                                                                                     \Pr(A_1 \mid F_m \in \mathrm{SAT}_i) 
\end{aligned}
\]
where
\[
    \mathbb{E} \Big[ \mathbf 1_{\{F_mx \in~\text{SAT}_i \} }   \mid F_m \in~\text{SAT}_i, A_1 \Big]
      = \Pr (F_mx  \in \mathrm{SAT}_i  \mid F_m \in \mathrm{SAT}_i,  A_1 ) 
\] 
From the definition of $\mathcal{S}_i$, the right-hand side
\[
     \Pr\Big(F_m \, x \in \mathrm{SAT}_i\} \,\Big|\, F_m \in \mathrm{SAT}_i, x \in \{x_1, \ldots ,x_i\} \Big) = 1
\] 
Moreover, since $x$ is chosen independently of $F_m$,
\[
      \Pr(A_1 \mid F_m \in \mathrm{SAT}_i) = \Pr(A_1) = \frac{i}{2N}
\]
Consequently,  the first summand equals  $\displaystyle \frac{i}{2N}$

\vspace{0.5\baselineskip}  

\textbf {The second summand}.
$\mathbb{E} \Big[ \mathbf 1_{\{F_mx\} }   \mid F_m, A_2 \Big] \Pr(A_2 \mid F_m)$ is
\[
\begin{aligned}  
        \mathbb{E} \Big[ \mathbf 1_{\{F_mx \in~\text{SAT}_i \} } \mid F_m \in~\text{SAT}_i, A_2 \Big] 
                                                                                     \Pr(A_2 \mid F_m \in \mathrm{SAT}_i) 
        = 0
\end{aligned}
\]
because $x = \neg x_r = 0$, $x_r \in \mathcal{S}_i $. 

\vspace{0.5\baselineskip}  
\textbf{The third summand}.
$\mathbb{E} \Big[ \mathbf 1_{\{F_mx\} }   \mid F_m, A_3 \Big] \Pr(A_3 \mid F_m)$ is 
\[        
\begin{aligned}
    \mathbb{E} \Big[ \mathbf 1_{\{F_mx \in~\text{SAT}_i \} } 
                    &  \mid F_m \in~\text{SAT}_i, A_3  
               \Big] 
                     \Pr(A_3 \mid F_m \in \mathrm{SAT}_i) 
         \\[4pt]
                   & =  \mathbb{E} \Big[ \mathbf 1_{\{F_m x_{i+1} \in~\text{SAT}_i \} } 
                                          \mid F_m \in~\text{SAT}_i 
                                   \Big]  
                                 \left( \frac{2N-2i}{2N} \right)  
         \\[4pt]
                   & =  \Pr \Big( F_m\, x_{i+1} \in~\text{SAT}_i  
                                          \mid F_m \in~\text{SAT}_i 
                           \Big) 
                                 \left( 1- \frac{i}{N} \right)  
\end{aligned}                        
\]
The first equality holds because 
\begin{enumerate}
\item[(a)] the choice of $x$ is independent of $F_m$  
\item[(b)] $x$ is randomly chosen from $2N$ literals 
\item[(c)] for each literal of $\{x_{i+1}, \neg x_{i+1}, \ldots, x_N, \neg x_N \}$ the expectation is the same due to symmetry.
\end{enumerate}
Further
\[ 
\begin{aligned}   
 \Pr \Big( 
                     F_m\, x_{i+1}   \in \mathrm{SAT}_i      \mid F_m  \in \text{SAT}_i  
      \Big )  
          & =  \frac{ 
                      \Pr \Big( x_{i+1}F_m \in \mathrm{SAT}_i \, \land \, F_m \in \mathrm{SAT}_i \Big)
                    }
                    {
                      \Pr(F_m \in \mathrm{SAT}_i)
                    }
\\[4pt]   
           & =  \frac{ 
                      \Pr \Big( x_{i+1}F_m \in \mathrm{SAT}_i  \Big)
                    }
                    {
                      \Pr(F_m \in \mathrm{SAT}_i)
                    }  
\\[4pt]     &
           =  \frac{ 
                      \Pr \Big(F_m \in \mathrm{SAT}_{i+1}  \Big)
                    }
                    {
                      \Pr(F_m \in \mathrm{SAT}_i)
                    }  \qquad (\text{definition of $\mathrm{SAT}_{i+1}$})
\\[4pt]                   
           &      =     \frac {
                                P(i+1, m)
                              }
                              {
                                P(i, m)
                              }                              
\end{aligned} 
\]
Hence, the third summand is
\[ 
        \Big( 1- \frac{i}{N} \Big)  
                                     \frac {
                                              P(i+1, m)
                                            }
                                            {
                                              P(i, m)
                                            }
\]
Finally, we obtain the first recursion for $P(i, m)$
\begin{align}\label{b(i,.)}
\begin{split}
       1 - \bar u &=\frac{i}{2N} +  \Big(1 - \frac{i}{N} \Big)\frac { P(i+1, m) }
                                                                   {
                                                                     P(i, m) 
                                                                   }
\end{split}
\end{align}  
For the second recursion, 
\begin{align*}
       P(i, m) &= \Pr \Big( F_{m-1} (y_1 \lor \cdots \lor y_{\!_{K}}) \in \text{\,SAT}_i \Big)  
       \\[6pt]
         & =   \Pr \Big(F_{m-1} \in \mathrm{SAT}_i \Big) \Pr 
                    \Big( 
                             F_{m-1} (y_1 \lor \cdots \lor y_{\!_{K}}) \in \text{\,SAT}_i \mid  F_{m-1} \in \text{\,SAT}_i
                    \Big)  
       \\[6pt]       
        & =   \Pr \Big(F_{m-1} \in \mathrm{SAT}_i \Big)   
            \mathbb{E} \Big[ \underbrace{
                \Pr \Big( 
                      F_{m-1}y_1 \lor \cdots \lor F_{m-1}y_{\!_{K}}  \in \mathrm{SAT}_i \;\Big|\, F_{m-1} }
                                     _
                                       { 
                                           \hspace{5.em}1 - u^K, \;\; u:=u(F_{m-1}) 
                                       }                           
                    \Big) 
                                    \;\Big | \, F_{m-1}  \in \mathrm{SAT}_i                     
                       \Big]  
                     \\ & \hspace{28em} \text{(tower property)}
       \\[0pt]
               &= \Pr \Big( F_{m-1}  \in \mathrm{SAT}_i \Big)   \Big( 1 - \mathbb{E} \left[ u^K \right] \Big) 
       \\[6pt]
               &= P(i, m-1) \Big( 1 -  {\bar u}\,^K  - \frac{1}{2}K(K-1) \zeta^{K-2} \, \mathbb{E}(u-\bar u)^2    \Big) 
               \, \quad \text{(proposition \ref{generally prop8} )}
\end{align*}   
It follows, 
\begin{align}\label{a(.,m)}
    \frac{P(i,m)}{P(i,m-1)} 
       =   1 -  {\bar u}\,^K   \, + \, \sim \mathrm{Var}(u)
\end{align}

\/********************************************************************************************************************* 
\begin{proposition}\label{prop6}                        
Let $\{y_j\}_{j=1}^k$ be $k$ independent random literals, and Then 
\[
        \Pr \Big ( \bigwedge_{j=1}^k F\,y_j \in \textnormal{SAT}_i \,\Big|\,  F \in \textnormal{SAT}_i \Big) 
        = \mathbb{E}(1-u)^k
\]           
\end{proposition} 

\begin{proposition}\label{prop9}              
For independent literals $y_1, y_2, \ldots, y_k$, $k \ge 2$, we have
\begin{align*} 
        & \Pr \Big ( Fy_1 \in \textnormal{SAT}_i \land    \cdots \land Fy_k \in \textnormal {SAT}_i  \Big)  
        \\[3pt]
        & =  \Pr \Big (\bigwedge_{j=1}^k F\,y_j \in \textnormal{SAT}_i \; \Big|\,   F \in \textnormal {SAT}_i \Big)
                                                                                      \Pr( F \in \textnormal {SAT}_i)
        \\[3pt] 
        & =   \left(  (1 - \bar u)^k + O(N^{-1 + \varepsilon}) \right)   \Pr( F \in \textnormal {SAT}_i) 
             \quad  \textnormal{(by Propositions \ref{prop6} and \ref{prop8})}
\end{align*}

\end{proposition} 
*********************************************************************************************************************/

For large $N$, we treat $P(i,m)$ as a smooth function $P(x,z)$ where $x = i/N$ and $z = m/N$ (here $z$ is used instead of  $\alpha$ to align with PDE notational convention). Small shifts in $i$ or $m$ correspond to $\Delta x = 1/N$ and $\Delta z = 1/N$. So the equations  (\ref{b(i,.)}) and (\ref{a(.,m)}) can be written as
\begin{equation}\label{recursion of P}
            \frac { P(i+1, m) }{ P(i, m) } = \frac{1 - \bar u - x/2}{1 - x}
            ,\qquad    \frac{P(i,m)}{P(i,m-1)} =  1 -  {\bar u}\,^K   \, + \, \sim \mathrm{Var}(u)
\end{equation}  
Take logarithms of both sides: 
\[
     \frac 1 N \, \frac{\ln P(i+1, m) - \ln P(i, m)}{1/N} = \ln  \frac{1 - \bar u- x/2}{1-x}
\]
\[
     \frac 1 N \, \frac{\ln P(i, m) - \ln P(i, m-1)}{1/N} = \ln \Big( 1 -  {\bar u}\,^K   \, + \, \sim \mathrm{Var}(u) \Big)
\]
namely, 
\[
     \frac 1 N  \partial_x \ln P(x,z)  =  \ln  \frac{1 - \bar u- x/2}{1-x}, \qquad     
     \frac 1 N  \partial_z \ln P(x,z)  = \ln \Big(1 - \bar u^K \, + \, O( \mathrm{Var}) \Big) 
\]
Applying the symmetry of mixed partial derivatives (Clairaut's theorem), 
\[
           \partial_{zx} \ln P(x,z) =  \partial_{xz} \ln P(x,z) 
\]
we obtain 
\begin{equation}\label{KSAT PDE}
      \left(1 - \frac{x}{2} - u\right) \frac{K u^{K-1}}{\rule{0pt}{10pt} 1 - u^K}\,  u_x \;-\; u_z = 0
\end{equation} 
where the $O(\mathrm{Var})$  and  $O(\mathrm{Var}')$ terms, each of order $ O(N^{-1})$ (see \eqref{Var'} in the proof of Theorem \ref{con_ineq}), are neglected. 
In the equation above and hereafter, we use ``$u$'' for $\bar u$ in view of the sub-Gaussian concentration of $u$ around $\bar u$; there is no need to distinguish between $\bar u$ and $u$ for our purpose, as is common in physics. 

The PDE \eqref{KSAT PDE}, first obtained in \cite{My dissertation} as a conjecture, is referred to as the $K$-SAT PDE. For convenience, we use ``the PDE'' to refer to both the equation itself and its solution.  With the initial condition $u=x/2$ when $z=0$, the solution of (\ref{KSAT PDE}) is
\begin{equation}\label{K-SAT}
            z = \frac{2(1-u^K)}{\rule{0pt}{10pt} Ku^{K-1}} \ln  \frac {1 - u - \frac{x}{2}}  {1-2u}
\end{equation}
 
\vspace{0.5\baselineskip} 
From  (\ref{recursion of P}), we know that adding a $K$-clause reduces the proportion of satisfiable formulae by a factor of $a(x,z) :=1-u^K(x,z)$, and increasing $i$ by 1 decreases $P^{(i)}_m $ by a factor of $ \frac{1-x/2-u}{1-x}$. Therefore if we write  
\begin{equation}\label{int_C}
    P(i,m) = \exp \left(N \cdot \int_C dE(x,z) \right)
\end{equation}
where $C$ is a curve joining two points in the $x$-$z$ plane, then some calculation gives
\begin{equation}\label{dE}
     dE(x,z) =  \ln \big(1-u^K(x,z)\big)dz + \ln \frac{1-x/2-u}{1-x} dx :=  Q\,dz + P\,dx 
\end{equation}
The curve integral in (\ref{int_C}) is path independent because $P^{(i)}_m $ depends only on the endpoint $(x,z)$; independent of the way the $K$-CNF is generated. Indeed, one checks that the vector field $(P,Q)$ is curl-free: 
\begin{equation}\label{curl_zero}
     \frac{\partial P}{\partial z}  =  \frac{\partial Q}{\partial x}  \;
\end{equation}
%
%
%
To verify this, note that
\[
          Q = \ln (1 - u^K), \quad P = \ln \frac {1 - u - \frac{x}{2}}{1 - x}. 
\]
Hence, 
\[
    \frac {\partial Q}{\partial x} = - \frac{Ku^{K-1}}{1-u^K}\cdot \frac{\partial u}{\partial x}, \quad
    \frac {\partial P}{\partial z} = \frac {-1}{1 - u - \frac{x}{2}} \cdot \frac {\partial u}{\partial z}  
\]
and the equality \eqref{curl_zero} follows from the $K$-SAT PDE (\ref{KSAT PDE}). One can easily check that $\mathbb{E}(x,0)=0$, and $\mathbb{E}(0,z) = 0$ when $u=0$.

\subsection{\textbf{\textup{2-SAT}}}\label{subsection 2-SAT} 
%
%
%
%

So far, the best result for 2-SAT establishes the scaling window of the SAT/UNSAT transition as $\Theta(N^{-1/3})$ \cite{scaling window of the 2-SAT 2001}. In the following theorem, we provide a more precise relationship between the clause density and satisfiability, including the the prefactor function within the critical window. We use $y$ for the 2-SAT clause density, i.e. $m/N$, instead of $\alpha$. 


\begin{thm}\label{2-sat threshold} Let $\Pr(y)$ denote the satisfiability probability at $y$. In the large $N$ limit,
\[
\begin{cases}
             \Pr(y) = 1          & y < 1    \\[8pt]                   
      \displaystyle        y =          1 + \frac{3}{\sqrt[3]{4}}\sqrt[3]{ \ln \frac{\Pr(1)}{\Pr(y)}} \cdot N^{-1/3} 
                                 &          \textnormal{$y $ in the critical  window}                \\[16pt]                    
              \displaystyle     \Pr(y) = \Pr(1)\exp\!\left(-N\dfrac{4}{27}(y-1)^3\right) 
                                  &   y > 1  \textnormal{ and } |y-1| \ll 1                         
\end{cases}
\] 
\end{thm}
Namely\[
\begin{cases}
             \Pr(\alpha) = 1          & \alpha < 1    \\[8pt]                   
      \displaystyle  
       \alpha -1 =  \frac{3}{\sqrt[3]{4}}\sqrt[3]{ \ln \frac{  \Pr(1) }{\Pr(\alpha)}}
      \cdot N^{-1/3} 
                                 &          \textnormal{$ \alpha $ in the critical  window}                \\[16pt]                    
              \displaystyle     \Pr(\alpha) = \Pr(1)\exp\!\left(-N\dfrac{4}{27}(\alpha-1)^3\right) 
                                  &   \alpha > 1  \text{ and } |\alpha-1| \ll 1                         
\end{cases}
\] 
\begin{proof} For $K=2$, equation \eqref{K-SAT} gives 
\[
     y = \frac{(1-u^2)}{u} \ln  \frac {1-u - \frac{x}{2}} {1-2u}      
\]       
For small $u$ and $x$, employing Taylor expansion and eliminating negligible terms, we have (see Appendix \ref{u=0 y<1 2-SAT})
\begin{equation}\label{(y = 1 + 3u/2}
   2(1-y)u + 3u^2 = x + O(xu, x^2, u^3)
\end{equation} 
This implies $u=0$ along the line $x=0$ for $y < 1$. Therefore
\[
       \prod^{yN}~(1-u^2) = 1, \text{ for $y < 1$}
\]
Thus the first line of the theorem is true.

Now for $1 - y_{\varepsilon} = \Theta(N^{-1/3})$, let 
\[ 
           \displaystyle \Pr(y_{\varepsilon})=\prod^{y_{\varepsilon}N}~(1-u^2) = 1 -\varepsilon 
\]  
Also from (\ref{(y = 1 + 3u/2}), when $x=0$ for small $u$ such that $u^2 \ll u$, it holds
\[
          y =  1 + \frac {3} 2u\; \Rightarrow  dy = \frac 3 2 u   
\]
We have 
\[
    \Pr(1)  = \Pr(y_\varepsilon) \exp \left(  N\int_{y_\varepsilon}^1 \ln \left(1 - u(t)^2 \right)dt  \right)  
\]
where \vspace{-1em}
{\allowdisplaybreaks
\begin{align*}
        &  \exp \left(  N\int_{y_\varepsilon}^1   \ln \left(1 - u(t)^2 \right)dt  
             \right) 
         =   \exp \left( -  N\int_{y_\varepsilon}^1 \left(u(t)^2 + O(u^4) \right) dt
                     \right) \\[6pt]
        &  =    \exp \left( -  N\int_{y_\varepsilon}^1  u(t)^2   dt  + o(1) \right) 
        \quad \text{ 
                     \big($u^4 \sim O(N^{-4/3})$\big) \renewcommand{\thefootnote}{$\ddagger$}\footnotemark   
                   }                                        
                     \\[6pt]  
        & =   \exp \left( -  N \cdot    \frac{4}{27}(y-1)^3  \,   \Bigg|_{y_{\varepsilon}}^1
                                        \right)  
                \\[6pt]  
        & =   \exp \left( - N\frac{4}{27}(1 - y_{\varepsilon})^3     \right)  
\end{align*}%
}%
Thus
\[
    \Pr(1)  = (1-\varepsilon)  \exp \left(-N\frac{4}{27}(1 - y_{\varepsilon})^3  \right) 
\]
Simple algebra gives
\begin{equation}\label{y < 1 in thm}
       y_\varepsilon = 1 + \frac{3}{\sqrt[3]{4}}\sqrt[3]{\displaystyle \ln \frac{\Pr(1)}{\Pr(y_\varepsilon)}} \cdot N^{-1/3}
\end{equation} 
%
%
%
%
which gives the second line of the theorem for $y < 1$.
For $y > 1$,  
\begin{equation}\label{y>1} 
                \Pr(y)  = \Pr(1)\exp \left( N\frac{4}{27} 
                                                                                                (1-y)^3   
                                    \right)          
\end{equation}
which gives
\begin{equation}\label{y > 1 in thm}
           y = 1 + \frac{3}{\sqrt[3]{4}}\sqrt[3]{\ln\frac{\Pr(1)}{\Pr(y)}} \cdot N^{-1/3}
\end{equation} 
The combination of \eqref{y < 1 in thm} and \eqref{y > 1 in thm} gives the second line of the theorem, and \eqref{y>1} the third. 
\end{proof}
\begingroup
    \renewcommand{\thefootnote}{$\ddagger$}
    \footnotetext{Recall we use $u$ and $\bar u$ interchangeably.}    
    \addtocounter{footnote}{-1}%
\endgroup%
Assume $\Pr(y_\varepsilon)\approx 1$ and $\Pr(1) = 50\%$.  Then,  from  \eqref{y < 1 in thm}
\begin{equation}\label{y < 1 y50}
              1 \approx y_\varepsilon + 1.67 N^{-1/3}
\end{equation}  
For $y>1$, if $\Pr(1) = 1$ and $\Pr(y) = 50\%$, \eqref{y > 1 in thm}  yields
\[
    y_{50} \approx 1 + 1.67 N^{-1/3}
\]      
   
The tables below list two series of calculated $y_{50}$, truncated to a few decimal places for comparison with the experimental results. We have found only two sets of published data for 2-SAT.
The last column lists fitted formulae of the form $y_{50} = a + b \cdot N^{-1/3}$, obtained by linear regression.

\begin{table}[ht]
\centering 
\begin{tabular}{l c c  c c c c l } 

    N : & 50 & 100   & 200  & 300   & 400 & 500  &   Regression formula   \\ [0.5ex] 
\hline 
\rule{0pt}{10pt} 
   \textbf {$y_{{50}}$}
               & \textbf{1.45} &\textbf{1.36} &\textbf{1.29}
                                 & \textbf {1.25}  & \textbf {1.23}  &\textbf {1.21}
                                                                                 &   $1.01 + 1.64N^{-1/3}$  
                                                       \\ 
Simon et al \cite{Simon 1986}
               & 1.40  & 1.40  & 1.23 &1.22  & 1.22 &1.18
                                                                                & $0.98 + 1.65N^{-1/3}\, (R^2: 87\%) $
                                 \\  
\hline 
\end{tabular}
\label{table:z50 vs Simon} 
\end{table}
 
\begin{table}[ht] 
\centering
\begin{tabular}{l c c c c c    c c l} 
N : & 25 & 50 & 100 & 250 & 500 & 1000 & 1500 & Regression formula \\ [0.5ex] 
\hline 
\rule{0pt}{12pt} 
\textbf{$y_{50}$}
 & \textbf{1.60} & \textbf{1.45} & \textbf{1.36} & \textbf{1.265} & \textbf{1.21} & \textbf{1.167} &  
                                                                  \textbf{1.146} & $0.99 + 1.75N^{-1/3}$ \\
Singer  & 1.64 & 1.48 & 1.36 & 1.252 & 1.20 & 1.147 & 1.127          & $0.94 + 2.0N^{-1/3}$ $(R^2: 99.7\%)$\\  
et al \cite{Singer}*  \\
\hline 
\multicolumn{9}{l}{
  \small $^*$ \begin{tabular}[t]{@{}l l@{}}
    (1)  Their algorithms employed are based on local search. \\
    (2)  The first 7 points of theirs are used here, with the same number of instances each.
  \end{tabular}
} \\ 
\end{tabular}
\end{table}

\begin{figure}[!ht] 
  \begin{minipage}{0.55 \textwidth} 
    \centering 
    \vspace{1em}
    \includegraphics[height=20em, width=25em, trim=0 0 5 10, clip]{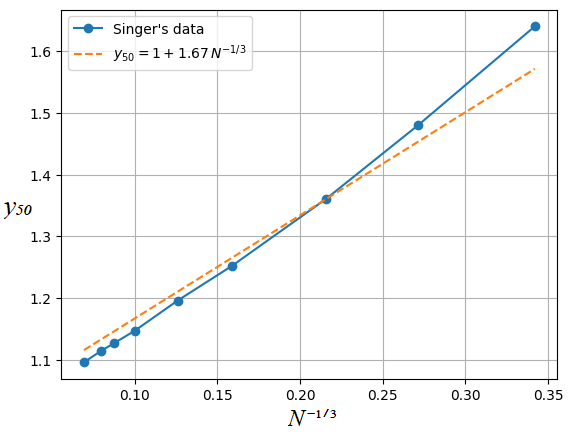}  
    \vspace{-1em}      \caption{\normalfont{Singer et al.'s data} \cite{Singer} vs. $y_{50}$}
    \label{fig: Singer's data}  
  \end{minipage} 
\end{figure}

\FloatBarrier 

\subsection{\textbf{$\bm{ (2+p) }$-\textup{SAT}}}
The classification of phase transition orders originated in statistical mechanics. 2-SAT has second order of phase transition, in the sense that the backbone fraction (i.e. ``backbone'' for short) is continuous when phase transition occurs. It is observed empirically that 3-SAT's phase transition is of the first order as its backbone is discontinuous at the point of phase transition. 

Intuitively, the onset of computation hardness around the satisfiability threshold is related to the emergence of a backbone, because it costs many iterations to identify assignments consistent with the backbone. To understand the relationship between the backbone and computational hardness, mixed $2+3$-SAT model, namely $(2+p)$-SAT, is studied \cite{2+p-SAT Achlioptas} \cite{BasseOConnor2023} \cite{Biroli&Monasson&Weigt2000} \cite{MonassonZecchina97} \cite{Monasson_Nature} \cite{MonassonZecchina1998} (see \cite{2+p-SAT Achlioptas} for brief survey there). Various other mixed models of SAT/COL are studied in \cite{Walsh2002}.

A $(2+p)$-SAT formula is a Boolean CNF formula mixed with $(1-p)m$ 2-SAT clauses and $pm$ 3-SAT clauses. The parameter $p$ thus provides a smooth interpolation between the 2-SAT and 3-SAT models. Based on experiments, it was conjectured that there exists a critical value $p_c$ such that for $p < p_c$, for any proportion of 3-SAT clauses, the $(2+p)$-CNF exhibits 2-SAT like phase transition, i.e. the backbone varies continuously at the phase transition point, and after $p_c$, the $(2+p)$-CNF undergoes a first-order phase transition, i.e., discontinuous backbone. 
This conjecture attracted significant interest from researchers. It was conjectured, based on experiments on  spin glass model, that $p_c$ is around 0.41 \cite{pc=0.41}. \cite{2+p-SAT Achlioptas} rigorously proved that $0.4 \le p_c < 0.695$. A slightly improved upper bound, 0.6846, is given in \cite{Zhou_Gao2013}. 

We prove that the critical point $p_c=0.5$. Moreover, the following theorem presents a characterization of the critical window of the phase transition, analogous to Theorem~\ref{2-sat threshold} of 2-SAT. Let $y$ be scaled length of 2-CNF and $z$ of 3-CNF. We have the following theorem



\begin{thm}\label{Thm 2+p-SAT}
Let $z < 1$ and $\Pr(y)$ denote the satisfiability probability at $y$. In the large $N$ limit,
\[
\begin{cases}
       \Pr(y) = 1                           & y < 1    \\[8pt] 
 \displaystyle  y = 1 + \frac{3}{\sqrt[3]{4}} \sqrt[3]{  (1-z)^2   \ln \frac{\Pr(1)}{\Pr(y)}} \cdot N^{-1/3} 
                                 &          \textnormal{$y$ in the critical  window}                \\[16pt]                                                                   
 \displaystyle      
       \Pr(y) = \Pr(1)\exp\!\left(-N\dfrac{4}{27(1-z)^2}(y-1)^3\right)       
                                            &   y > 1  \text{ and } |y-1| \ll 1
\end{cases}
\] 
\end{thm} 

\begin{proof} 
The solution of the $(2+p)$-SAT PDE can be obtained either by solving the 3-SAT PDE with  $u(x,y)$ from 2-SAT as the initial condition, or vice versa. It is
\begin{equation}\label{2+p-SAT PDE}
%
%
%
%
%
        z \frac {3u^2}{2(1-u^3)} +  y\frac{u}{1-u^2} = \ln \frac {1-u-x/2}{1-2u}
\end{equation}
When $x=0$, for small $u$, we have (see Appendix \ref{u=0 y<1 (2+p)-SAT})
\[ 
          0=(1-y)u+\frac32(1-z)u^2    
\]
This implies $u=0$ along the line $x=0$ for $y < 1$. Thus, the first line of Theorem \ref{Thm 2+p-SAT} holds. 

Letting $x=0$ and using Taylor expansion for small $u$ gives
\begin{equation}\label{y=1+(1-z)*...}
                    y = 1 +  (1 - z) \frac {3u}{2} \quad \Rightarrow dy = (1-z)\frac 3 2 du
\end{equation}
Then, from 
\[
    \Pr(y)  = \Pr(y_\varepsilon) \exp \left(  N\int_{y_\varepsilon}^y \ln \left(1 - u(t)^2 \right)dt  \right)  
\]
we have, 
\begin{equation}
                \Pr(y)  = \Pr(y_\varepsilon)\exp \left( -N\frac{4}{27(1-z)^2} 
                                                                          \Big[  
                                                                                  (1-y_\varepsilon)^3 - (1-y)^3 
                                                                          \Big]
                                                     \right )
\end{equation}
Similarly to 2-SAT, the rest of Theorem~\ref{Thm 2+p-SAT} follow by setting $y=1$ and $y_\varepsilon = 1$, respectively.  
\end{proof} 
\vspace{-.5em} 
From the above analysis, given $z < 1$, as $y$ approaches 1 from below, $u$ begins to grow linearly with $y$. The discrepancy from 2-SAT lies only in the prefactor in the critical window of the phase transition. 
\begin{figure}[ht] 
  \begin{minipage}{0.3 \textwidth} 
    \centering         
    \includegraphics[height=10em, width=15em, trim=0 0 0 0, clip]{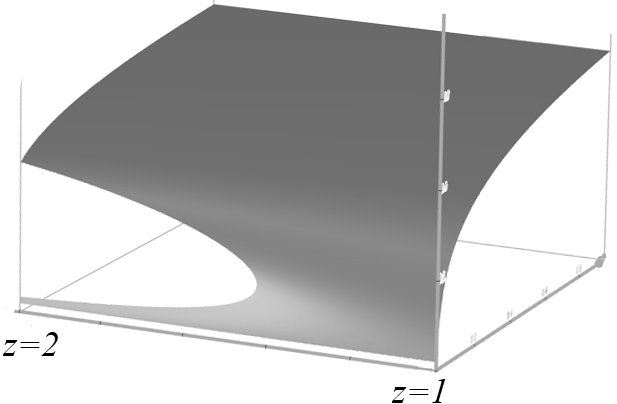}    
    \caption{\centering $u(x,z)$, $y=0.95$}
    \label{y=0.95}  
  \end{minipage}
  \hfill 
  \begin{minipage}{0.3\textwidth}     
    \centering   
    \includegraphics[height=10em, width=15em, trim=0 10 0 0, clip]{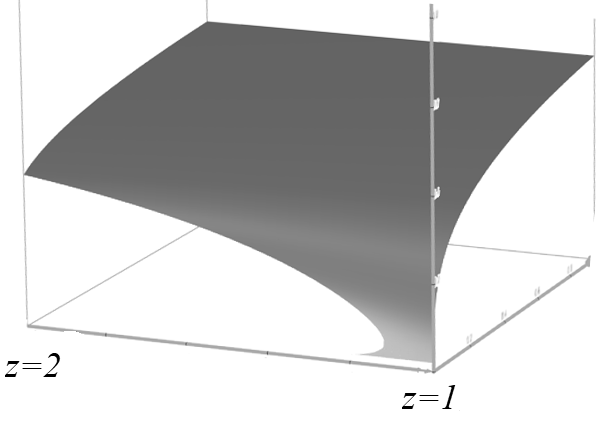} 
    \caption{ $u(x,z)$, $y=0.995$ }
    \label{y=0.995}  
  \end{minipage}
  \hfill 
  \begin{minipage}{0.3\textwidth}     
    \centering 

    \includegraphics[height=11.0em, width=15em, trim=0 0 0 0, clip]{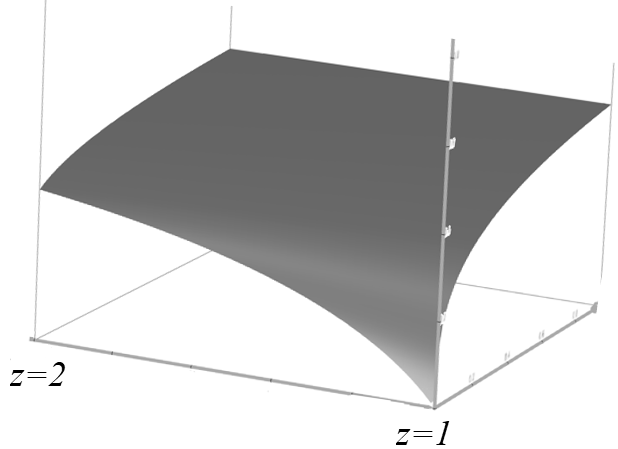} 
    \caption{$u(x,z)$, $y=1$  }
    \label{y=1}  
  \end{minipage}  
\end{figure}
\FloatBarrier

This 2-SAT-like behaviour vanishes when $z=1$. Specifically, for $z \ge 1$, when $y=0$, the $(2+p)$-SAT PDE \eqref{2+p-SAT PDE} becomes the 3-SAT PDE, in which the phase transition occurs at $r_3$ (believed to be approximately $4.267$). Then, as 2-SAT clauses are added, the position of phase transition decreases in $z$, as does the location of the singularity in $(x,z)$), until $y$ reaches 1 when the singularity and the phase transition are merged into a singular point at $(y,z)=(1,1)$ (Figs.~\ref{y=0.95},  ~\ref{y=0.995} and   
~\ref{y=1}). 
In the following, we derive the behaviour of $u$ around this critical point. 

Expanding each term in \eqref{2+p-SAT PDE}  for $x=0$, we obtain (see Appendix \ref{z=1+8u/9}) 
\[
      y - 1 = \frac 3 2 (1 - z)u + \frac 4 3 u^2 + \frac {15} 4 u^3 \quad \text{\big(omitting $O(u^4)$ \big)}
\]
      %
      %
When $y=1$,  
\[ 
      z = 1+\frac 8 9 u + O(u^2)  
\]
This determines the prefactor in the critical window for the phase transition in $z$. 

\subsection{\textbf{\textup{$K$-SAT}}}  
%
%
%
One can check that the frozen fraction of assignments for a SAT instance satisfies the same PDE\,\footnote{Obviously, all the preassigned variables are backbones as well.}.  
  
\begin{figure}[ht] 
  \begin{minipage}{0.3 \textwidth} 
    \centering         
    \includegraphics[height=10em, width=15em, trim=0 0 0 0, clip]{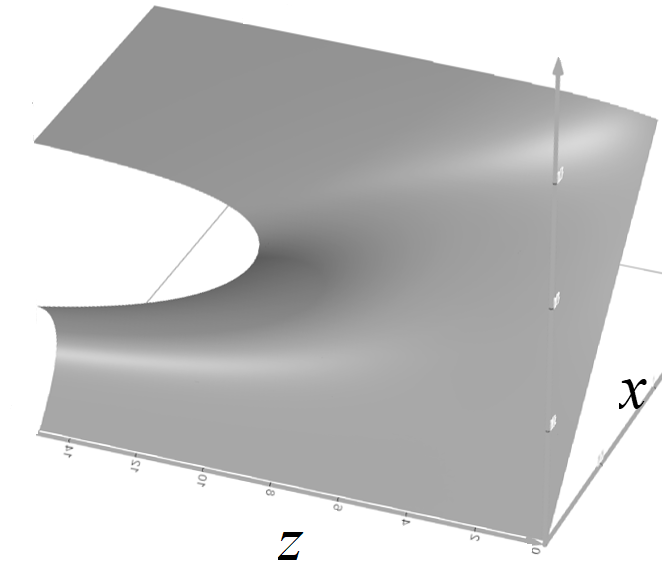}    
    \caption{\centering  $u(x,z)$, 4-SAT}     
    \label{4-SAT}  
  \end{minipage}
  \hfill 
  \begin{minipage}{0.3\textwidth}     
    \centering   
    \includegraphics[height=10em, width=15em, trim=0 0 0 0, clip]{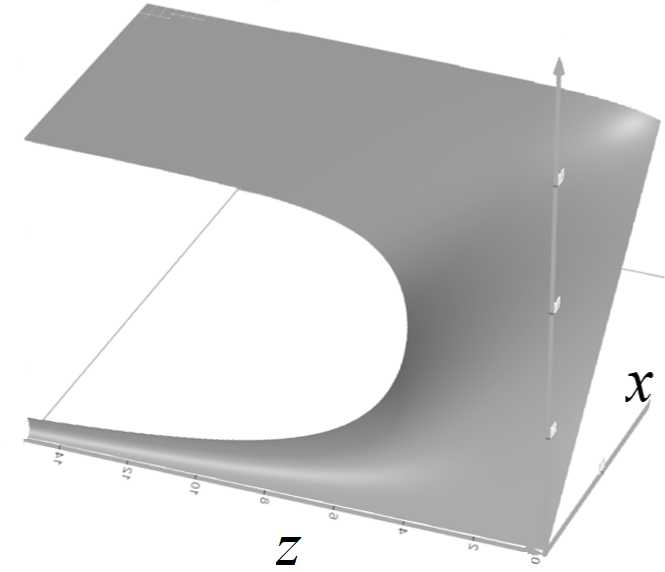} 
    \caption{$u(x,z)$, 3-SAT}
    \label{3-SAT}  
  \end{minipage}
  \hfill 
  \begin{minipage}{0.3\textwidth}     
    \centering 

    \includegraphics[height=9.0em, width=15em, trim=0 0 0 0, clip]  {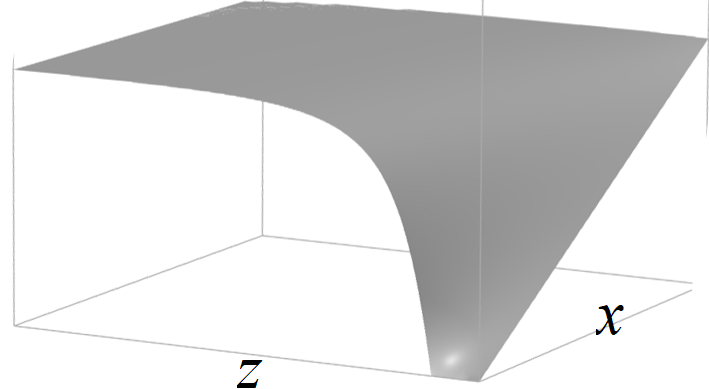}     
    
    \caption{$u(x,z)$, 2-SAT}
    \label{2-SAT}  
  \end{minipage}  
\end{figure}
\FloatBarrier 

\textbf{$\bm {\alpha_d}$ of $K$-SAT}. The $K$-SAT PDE shows that there is a singularity at which solutions undergo an abrupt transition. Let $z=\alpha_d$ denote the corresponding critical value at which this transition occurs. In 3-SAT, the singularity is at  $\{ x_0,z_0 \}=\{ 0.145, 3.183 \}$, where $u_z$ and $u_x$ become infinitely large. Indeed, a cusp catastrophe forms at the singularity; in 3-SAT, it is described by 
\[ 160\Delta u^3 + (78\Delta x) \Delta u - (\Delta z + 9.809 \Delta x) = 0,
\]
   
Given $x$, from the start $ z = \alpha_d(x)$, two additional branches emerge: one on the upper surface $S_2$ and the other the lower surface $S_1$ (denoted by $u_\text{U} (x,z)$ and $u_\text{L} (x,z)$ respectively). Thus, $\alpha_d$ can be found by solving $\partial_u z = 0$ and  $\partial_u x = 0$. The following table lists $\alpha_d(x=0)$, together with the empirical results of Mertens et al. \cite{paper alpha_d} for comparison.  

\begin{table}[ht]
\centering 
\begin{tabular}{c c c  c c c c c c} 

    ~~~~~~~~~ $K$ : &3 &4  &5  &6  &7  &8  &9 &10   \\ [0.5ex] 
\hline 
~~~~~~~   \textbf {$\alpha_d$}
       & \textbf{4.0029} &\textbf{8.360} &\textbf{16.16} &\textbf{30.51}
             & \textbf {57.21}  & \textbf {107.21}  &\textbf {201.29} &\textbf {379.01}
                                                       \\ 
Mertens et al.\cite{paper alpha_d}
           &3.927    &8.297    &16.12  &30.50  &57.22  &107.24   &201.35 &379.10
                                 \\
\hline 
\end{tabular}
\label{table:z50 vs Simon} 
\end{table}

Below $\alpha_d$, $u(0,z)$ is strictly zero (see Fig. \ref{fig: u(0,z), 3-SAT}). Thus, we establish 4.0029 as a new lower bound for the 3-SAT threshold and 8.360 for 4-SAT. 
The best rigorous lower bounds are 3.52 \cite{HAJIAGHAYI2003}\cite{KAPORIS2006} and 7.91 \cite{Achlioptas JAMS 2004} respectively. Beyond $\alpha_d$, the backbone of 3-CNF formulas and that of the solution clusters diverge. 


Asymptotically, some calculation on the $K$-SAT PDE (\ref{K-SAT}) gives $\alpha_d = \frac {2^K}{K}(\ln K + d^{*})$ where $d^{*}$ satisfies
$          d^{*} = \ln (\frac {1}{2}\ln K + \frac {1}{2}d^{*})   
$.
In the glass spin formulation analysed via the cavity method \cite{paper alpha_d}, 
\[
      \alpha_d =  \frac {2^K}{K}\big(\ln K + d^{*}) \exp \left( \frac{e^{-d^{*}}}{2}  \right)
\]  
That is, the exponential factor $\exp \left( \frac{e^{-d^{*}}}{2}  \right)$ accounts for the discrepancy between the $K$-SAT PDE and the cavity method analysis.

\textbf{$\bm {\alpha_e}$ of $K$-SAT}. Also beginning from the onset of the cusp-catastrophe, there is another unique line, referred to as the ``$\alpha_e$ line'', on $x$-$z$ plane corresponding to two curves, one on each surface, along which the ``energies'' $E(x,z)$ on the two surfaces are equal. As an aside, calculations give that the $\alpha_e$ line starts with slope $dz/dx \approx -9.81$ at the onset of the cusp catastrophe. Along the $\alpha_e$ line, the proportions of solution clusters with the larger backbone (more frozens; surface $S_2$) and those with a smaller backbone (fewer frozens; surface $S_1$) are equal (sometimes, we also use ``backbone'' to refer to the frozen variables in an assignment, for convenience).  Furthermore, as an analysis indicates (see Appendix \ref{dE_U vs dE_L}), in the regime below the $\alpha_e$ line (i.e., in the less constrained regime), clusters of solutions with larger backbone dominate, and vice versa. Beyond the $\alpha_e$, the clusters of solutions with less frozen variables dominate. In other wards, except at the point $\alpha_e$, either the small-backbone or large-backbone is dominant in the number of solution clusters. 

Let $dE_\text{U}$ denote $dE$ on the upper surface,  $dE_{\text{L}}$ the lower surface. From the formula of $dE$, (\ref{dE}), $dE_{\text{L}}=dE_\text{U}$ yields
\[
       \frac {dz}{dx} = \frac{\ln (1-\frac x 2 - u_{\text{U}}) - \ln (1-\frac x 2-u_{\text{L}})}
                           {\ln (1-u_\text{U}^3) -  \ln (1-u_{\text{L}}^3)}
\]
This determines the $\alpha_e$ curve, whose endpoint in 3-SAT is at $(x=0, z=4.3962)$. The following table lists some of the results (bold numbers).
\begin{table}[ht]
\centering 
\begin{tabular}{c c c  c c c c } 
 & K : & 3 & 4 & 5 & 6 & 7 \\ [0.5ex] 
\hline 
&best upper bound
     & 4.51   & 10.23    & 21.33  & 43.51   & 87.88 \\ 
&   $\boldsymbol{\alpha_e}$ 
     & \textbf {4.3962} & \textbf {10.077} & \textbf {21.234} & \textbf {43.45}   & \textbf {87.84}    \\
&  \qquad   $\alpha_c$ (Mertens et al.\cite{paper alpha_d}) & 4.267 & 9.931 & 21.117 & 43.37 & 87.79  \\
& best lower bound       & 3.52            & 7.91             & 18.79 & 40.62  & 84.82   \\
\hline 
\end{tabular}
\label{table:K-SAT results} 
\end{table}
It is interesting to note that Silvio Franz et al. \cite{Franz et al 4.3962} obtained 4.3962, exactly the same value as our $\alpha_e$, as SAT-UNSAT critical point within the 1-step replica symmetry breaking (1RSB).

\/******************************************************************************************************************************
\subsection {Reasons for the difference between 4.267 and 4.3962 }
In the $K$-SAT study by statistical physicists, it is observed that in the low clause density area, there is one giant cluster of solutions. At $\alpha_d$, the cluster shatters into exponentially many clusters. Afterward, the number of these clusters decreases exponentially (due to its frozen variables, i.e., those taking the same value in all solutions in a cluster) until 4.267 where clusters vanish. There are set of assumptions in the cavity method used by statistical physicists. The most critical one underlying the difference between 4.267 and 4.3962, we argue, is the assumption that at 4.267 no solutions survive at all; as explicitly claimed $\alpha_c$ = $\alpha_s$. The theory presented in this paper, however, suggests that after $\alpha_d$ there remain a very small number of clusters with zero fraction of frozen variables and, although exponentially fewer than other clusters, they survive after 4.267 up to 4.3962.

In computer science, experimental thresholds range from 4.2 to 4.3. First, in their SAT models used, clauses with duplicate variables are forbidden. Therefore, these models are supposed to have smaller thresholds than the uniform model, because the correlation between clauses is stronger; the fewer the number of variables, the stronger this kind of correlation is. Due to this factor, the experimental threshold is expected moving toward to the right as $N$ increasing. Second, they use 50\% point as the test point and assume its moving trend is monotonic with $N$ increasing. This assumption, we argue, is questionable. It could be that when $N$ increase further, the fifty-percent point moves back toward the right. In fact in \cite{Crawford 1996}, their results reveal that seventy-percent point, for example, is moving toward the left for $N \le 100$ and then the right after $N > 100$.
*******************************************************************************************************************************/

\subsection{\textbf{\textup{Other SAT problems}}} 
In a general case, one can replace $1 - u^K$ with $a(u)$ in (\ref{recursion of P}) and obtain the PDE as
\begin{align*}
                  (1 - \frac x 2  -  u) \frac{a'(u)}{a(u)} u_x + u_z = 0
\end{align*}
If the initial condition is $u(x,0)=\frac x 2 $, the solution of the PDE is given by
\[   
        z = -  \frac{2a(u)}{a'(u)} \ln \frac{1 - \frac{x}{2} - u}{1 - 2u}
\]
For examples,  
\begin{enumerate}[label=${\bullet}$]
  \item  3-XORSAT: \,$ a =  1 - 4u^3$
  \[
      z = \frac{1 - 4u^3}{6u^2} \ln \frac{1-u-\frac{x}{2}}{1 - 2u}
  \]
  \item  1-in-3 SAT: $ a =  1 - 3u^2 - u^3 $%
  \[
      z =\frac{2(1 - 3u^2 - u^3)}{6u + 3u^2} \ln \frac{1-u-\frac{x}{2}}{1 - 2u}
  \]  
  \item NAE-3-SAT:   $ a =  1 - 2u^3$%
  \[
      z =\frac{1 - 2u^3}{3u^2} \ln \frac{1-u-\frac{x}{2}}{1 - 2u}
  \]    
  \item 2-COL/1-in-2-SAT   :   $ a =  1 - 2u^2$%
 \begingroup
 \renewcommand{\thefootnote}{$\dagger$}
 \footnote{
 More precisely, $ a =  1 - 2u^2 - \gamma$, where $\gamma = O\left(N^{-1}\right)$ and $\gamma'=O\left(N^{-1}\right)$. The proof of these bounds is omitted, as it is analogous to the 3-COL argument; see (\ref{3-COL a}) in Appendix~\ref{3-COL PDE} \textbf{(3-COL PDE system)} and Appendix~\ref{gamma} as well. The same remark applies to the other three SAT problems. Intuitively, the larger the $K$, the smaller the error term $\gamma$.
} 
      \addtocounter{footnote}{-1}     
 \endgroup  
  \[
      z =\frac{1 - 2u^2}{2u} \ln \frac{1-u-\frac{x}{2}}{1 - 2u}  
  \] 
\end{enumerate}  
\begin{figure}[ht] 
  \begin{minipage}{0.22 \textwidth} 
    \centering         
    \includegraphics[height=9em, width=10em, trim=0 0 0 0, clip]{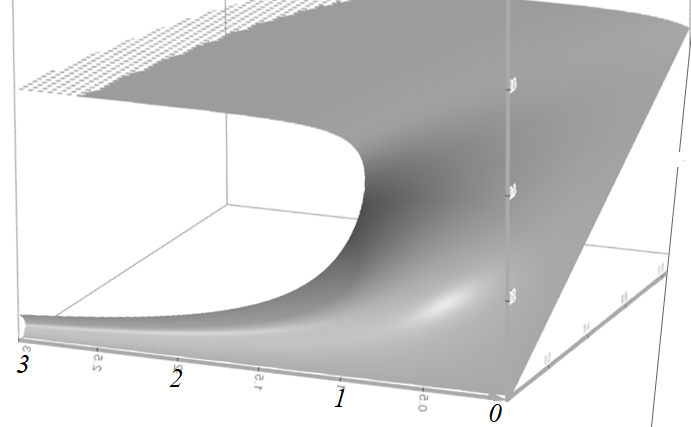}    
    \caption*{\centering  3-XORSAT}     
    \label{3-XORSAT}  
  \end{minipage}
  \hfill 
  \begin{minipage}{0.22\textwidth}     
    \centering   
    \includegraphics[height=9em, width=10em, trim=0 0 0 0, clip]{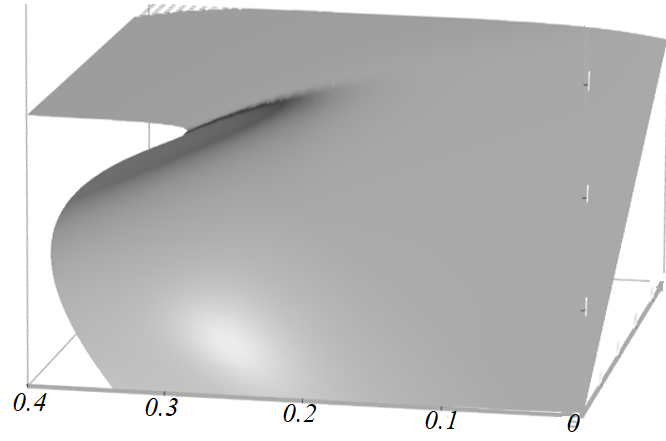} 
    \caption*{1-in-3 SAT}
    \label{1-in-3 SAT}  
  \end{minipage}
  \hfill 
  \begin{minipage}{0.22\textwidth}     
    \centering   
    \includegraphics[height=9.0em, width=11em, trim=0 0 0 0, clip]{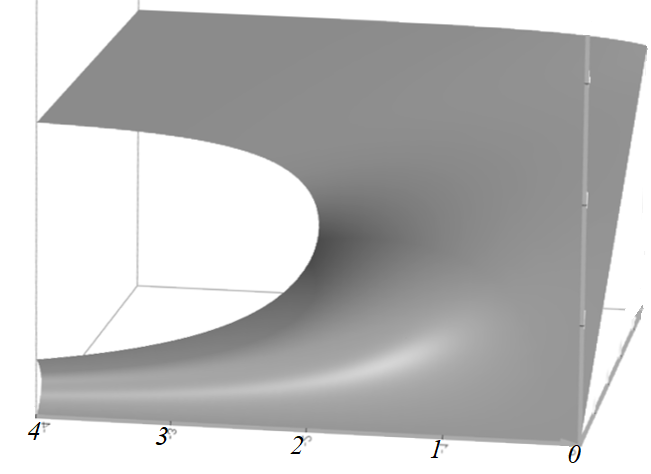}         
    \caption*{NAE-3-SAT}
    \label{NAE-3-SAT}  
  \end{minipage} 
  \hfill 
  \begin{minipage}{0.3\textwidth}     
    \centering 

    \includegraphics[height=9.0em, width=12em, trim=50 0 0 0, clip]{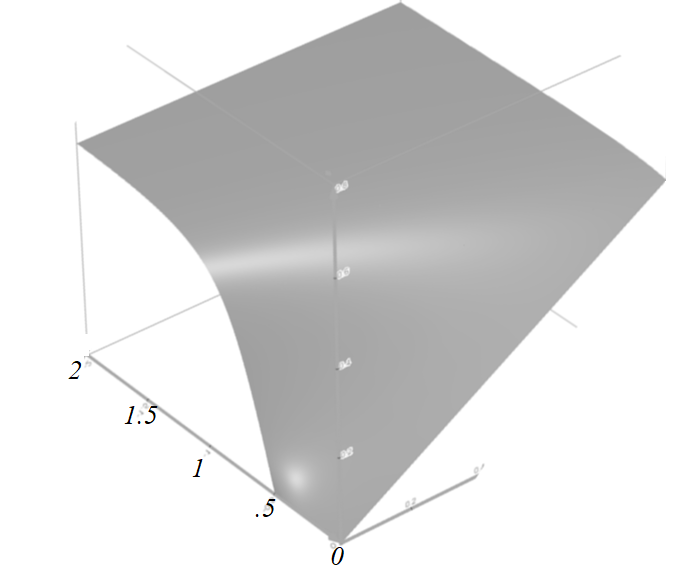}\\       
    \caption*{2-COL} 
  \end{minipage}     
\end{figure}
\FloatBarrier 

\subsection{\textbf{$\bm{ q }$-\textup{COL PDE}}}   
We use 3-COL to illustrate the method; the argument extends naturally to $q$-COL. A 3-colouring problem can be formulated as the following binary CSP: 
\begin{equation}\label{3-col}
   \bigwedge_{i=1}^{m} (x_i \oplus y_i) = \text{True}
\end{equation}
where $x_i$ and $y_i$ take truth values from three truths 0/1/2 (or red/green/blue) and ``$\oplus$'' here a ternary logic operator defined by:  $x_i \oplus y_i$ = True iff $x_i \not = y_i$. 

We extend the notion of backbone to the case of three-value. Given a satisfiable formula 
\[ F_m = \bigwedge_{i=1}^{m} x_i \oplus y_i, 
\]
analogous to $K$-SAT, we use $F_m \in \mathcal{S}_{i,j}$ to mean that $F_m$ is satisfiable in a partial solution space, where
\begin{enumerate}[label={${\bullet}$}, leftmargin=3em, style=nextline]
\item $i$ variables are preassigned to \{0\}, \{1\} and \{2\} (corresponding to red, greeg, blue), respectively, 
\item $j$ variables are preassigned to \{0,1\}, \{0,2\}, \{1,2\}, respectively; e.g., if variable $v$ is preassigned \{0,1\}, then the value 2 is unavailable for $v$; i.e., $v$ is forbidden from taking the value \{2\}. 
\end{enumerate}
We define $x := \frac i N$, and $y := \frac j N$. Let $u$ denote the proportion of vertices that are frozen by $F_m$ to a specified color (namely $u$ is backbone), and let $u_2$ denote the proportion of variables restricted to a specified pair of colors---pair color backbone. 

We obtain the following PDE system for 3-COL (see Appendix \ref{3-COL PDE} for the derivation)  
\[ 
\begin{cases}
   \displaystyle  
      \frac{\partial (2u + u_2) }{\partial z}  - \frac {2u(1 - x - 2y - 2 u - u_2)}{1 - 3u^2} \frac{\partial u }{\partial x}= 0
       \\[16pt]
   \displaystyle    
      \frac{\partial (3u + 2u_2)}{\partial z} - \frac{2u(1 - y - 3u - 2u_2)}{1-3u^2} \frac{\partial u}{\partial y} = 0
\end{cases}     
\] 
with the following initial condition at $z=0$
\[ \hspace{5em}
\begin{cases}
         u(x,y,0) = x,  &    \\
         u_2(x,y,0) = y
\end{cases}
\]
Finding a closed-form solution to this system remains an open problem.


\/******************************************************************************  
\section { }
See \cite{Liu K-SAT/q-COL} 

                            Foc16 II.pdf  This introduces E(x,z): Pₘ⁽ⁱ⁾ = exp(N·∫_C dE(x,z)  
                             Extended Focs16 27page(II)&(I).pdf
                             ChatGPT: Solving dx/du = 0 at = 3.183 gives two fold solutions in (0,1/2):
                                                   uₛ₁ ≈ 0.2154608098       
                                                   uₛ₂ ≈ 0.2199421645 
                            For both, the corresponding x-value is essentially the same: 
                            
                                      xₛ(uₛ₁; 3.183) ≈ 0.1453232623,  
                                      xₛ(uₛ₂; 3.183) ≈ 0.1453225287
                                                   
                           Focs16 II.pdf & Full STOC 2016.pdf are two versions of K-SAT parts.
                           ********************************************************************************/
\/*********************************************************************************************************
\clearpage
\appendix 
\section{\centering} \label{finite Ee^ᶿZ}  
 
***********************************************************************************************************/

\/**********************************************************************************************************          
\section{\centering} \label{appendix:a}      
\begin{proof}[\textnormal{Proof of law of iterated expectations}]
$ \mathbb{E}^{(M)}[\, F\,] = \mathbb{E}^{(M)} \Big( \mathbb{E}^{(M)}_i[F] \Big) $

BRW joint probability measure:
\begin{equation*}
           P^{\,^{(M)}} (dX_1 \cdots dX_M)  =  
             \frac {   \prod_{k=1}^{M} a_k   \ P( dX_1 \cdots dX_{\!_M} )
                   }
                   { Z  }
\end{equation*}
\begin{equation*}
               Z : = \int \prod_{k=1}^M a_k P(dX_1 \cdots dX_M)
\end{equation*}
\end{proof}
\clearpage
***********************************************************************************************************/

\appendix 
\section{\centering}\label{sub-Gaussian<=> moment ≤ C_p/N^{p/2}}  
\begin{center}  
\boldmath
              \textbf{sub-Gaussian} $\Rightarrow$ $\mathbb{E}  |x|^p  \le \; \sim   N^{-p/2}$
\end{center} 
\begin{proof}  
By the tail-integral formula,  
\[
        \mathbb{E}  |X|^p    = \int_0^\infty  \Pr(|X|^p \ge x)   dx = \int_0^\infty  \Pr(|X| \ge x^{1/p})  dx
\]
Making the substitution of $x^{1/p} = t$, 
\[
        \mathbb{E}   |X|^p  = p \int_0^\infty  \Pr(|X| \ge t)  t^{p-1} dt
\]
By sub-Gaussian tail, 
\[
        \mathbb{E}  |X|^p \le C \cdot p \int_0^\infty  e^{-ct^2/n}   t^{p-1} dt
\]
The substitution $ t = \sqrt{n}s$ gives
\[
        \mathbb{E}   |X|^p   \le C\cdot p \int_0^\infty  e^{-cs^2}  n^{(p-1)/2} s^{p-1}  \sqrt{n}s\, ds
                                     = C \cdot p n^{p/2} \int_0^\infty  e^{-cs^2}   s^{p-1}   s\, ds
\]
Since the last integral above is finite for $p \ge 1$,  
\[
        \mathbb{E}   |X|^p     \le   C'_p   n^{p/2}  
\]
By scaling, it follows that
\[
                \mathbb{E}  \left|
                                  \frac{X}{N} 
                            \right| ^p
                           \le    {C_p}{N^{-p/2}}
\]

\end{proof}

\clearpage
\section{\centering}\label{Proof of con_ineq_BRW}  

\begin{thm}\label{con_ineq_BRW in Appendix}   
If $\mathbb{E}(e^{\delta A|X_i|} \mid u_{i-1})\le K $, $S_n \le Cn$, and the negative association (\ref{neg_ass_ineq}) holds, 
then, in the domain where $a_i(u)$, $\mu_{i}(u)$ and $\nu_i(u)$ are smooth, the concentration inequality holds. 
\end{thm}

\begin{proof} Let $Z_i^{(m)}(u_i)$ denote the population descended from the birthplace $u_i = u_{i-1} + \frac{X_i}N$ after $m$ generations:
\begin{align}\label{Z_i^M}
    Z_i^{(m)}(u_i) & = \int a_{i+1} \cdots a_{\!_{i+m}}  P(dX_{i+1}\cdots dX_{\!_{i+m}} \mid u_i)   ,\quad a_{i+1} = a(u_i) 
    \\ \nonumber
                   & := \int  dZ_i^{(m)}(u_i)
\end{align}
Define
\[
    \mathbb{E}_{i}^{(m)}[\, \cdot \,]    :=   \mathbb{E} {^{(m)}} [\, \cdot \mid u_i \,]   
          = \frac{\displaystyle 
                            \int (\, \cdot \,)  dZ_i^{(m)}(u_i)}{Z_i^{(m)}(u_i)}
\]
we have
\begingroup
    \renewcommand{\thefootnote}{$\ddagger$}
    \footnotetext{ $ P(dX_i \mid u_{i-1}) = p_i(X_i \mid u_{i-1})dX_i. $
                 }
    \addtocounter{footnote}{-1}%
\endgroup 
\begin{align*} 
    \mathbb{E}_{i-1}^{ (m)}[\,  e^{\delta A|X_i|}   \,] 
    &= \frac { \displaystyle 
                 \int e^{\delta A|X_i|} \,    Z_i^{(m)}(u_i) \ P(dX_i|u_{i-1})  
                                            \;  \renewcommand{\thefootnote}{$\ddagger$}   \footnotemark  
              }
              { \displaystyle 
                 \int     Z_i^{(m)}(u_i) \ P(dX_i|u_{i-1})  
              }   
\end{align*}
If $Z_i^{(m)}(u_{i-1}+\frac{X_i}N)$ is decreasing in $|X_i|$ for fixed $u_{i-1}$ (e.g., in the domain where backbone is not decreasing and so $X_i \ge 0$, in the $K$-SAT case; for $X_i < 0$, see the remark at the end of theorem proof), the negative association (\ref{neg_ass_ineq}) implies, 
\begin{align} \label{E^M < K}
    \mathbb{E}_{i-1}^{(m)}[\,  e^{\delta A|X_i|}   \,] 
    & \le \frac { \displaystyle 
                 \int e^{\delta A|X_i|}  \ P(dX_i|u_{i-1})  
                 \int   Z_i^{(m)}(u_i) \ P(dX_i|u_{i-1})  
                                            \;  
              }
              { \displaystyle 
                 \int   Z_i^{(m)}(u_i)  \ P(dX_i|u_{i-1})  
              }   
               \le K                
\end{align}
Thus, the exponential-moment condition in Lemma \ref{con_inequality_BRW}, inequality (\ref{Exp moment existence}), holds.

Now, similar to Theorem~\ref{con_ineq}, we only need to prove  
\[
       \Big |\, \mathbb{E}^{(m)}(X_\ell \mid u_i) - \mathbb{E}^{(m)}(X_\ell \mid u_i') 
                  \, \Big| \,  \le \, L|u_i - u_i'|  ~~~~ for ~ i < l
\]
i.e. $\mathbb{E}^{(m)}_{i}(X_\ell)$ is Lipschitz continuous in $u_i$.

Recall that in the proof of Theorem \ref{con_ineq}, smoothness of $\mu_{i} (u_i)$ and $\nu_i (u_i)$ was employed to establish the Lipschitz continuity of the trajectory with respect to the history. In an RW, ``local'' smoothness of $u_i$ implies the required Lipschitz continuity with respect to the history because the descendant's population is $1$. In a BRW, it remains to show that the future descendant population affects the corresponding Lipschitz constant by at most a bounded factor. In the present branching random walk (BRW) setting, the corresponding conditions are the smoothness of 
\[\mu_{i}^{(m)} (u_i) :=  \mathbb{E}_i^{ (m) }(X_{i+1}|u_i)
    , \quad  \nu_i^{ (m) } (u_i):=  \mathbb{E}_i^{ (m) }(X_{i+1}^2|u_i),
\]    
together with
\[ 
    \beta_i^{ (m) } (u_i)  := \mathbb{E}_{i}^{(m)}\! \big( a(u_{i+m}) \mid u_i \big).
\]
         
To show $\mu_{i}^{ (m) } (u_i)$, $\nu_i^{ (m) } (u_i)$ and $\beta_i^{ (m) } (u_i)$ are smooth functions of $u_i$,  we proceed by induction on the generations, recalling that the hypotheses in the theorem---the assumptions on the smoothness of $\beta_{i} = a_i(u)$, $\mu_{i}$ and $\nu_{i}$ (without the superscript ``$^{{(m)}}$'')---ensure that the base case (generation 0) holds. The following concerns generation $m$, where we use $n$ instead of $i$ for ease of comparison with the proof of Theorem~\ref{con_ineq}.

By \eqref{Z_i^M}, given birthplace $u_n$, its descendant population is
{\allowdisplaybreaks
\begin{align}\label{prod_m} 
\begin{split} 
  Z_n^{(m)}(u_{n}) &= Z_n^{(m)}(u_{n-1} + \frac{X_n}{N})  
   \\
   &= \int \Big(\prod_{i=1}^m a_{n+i} \Big) \  P(dX_{n+1} \cdots dX_{n+m} \mid u_{n}) 
              , \quad a_{n+i} = a(u_{n+i-1})   
   \\
  &=\int \Big(\prod_{i=1}^{m-1} a_{n+i}\Big) \ P(dX_{n+1} \cdots dX_{n+m-1} \mid u_{n}) \int a_{n+m} P(dX_{n+m} \mid u_{n-1+m})
   \\
   &=  \int  a_{n+m} dZ_n^{(m-1)}  
   \\    
   &=  Z_n^{(m-1)}(u_{n})    \cdot  
    \frac { \displaystyle
            \int  a_{n+m} dZ_n^{(m-1)}   
          }
          { \displaystyle
              Z_n^{(m-1)}(u_{n})   
          }  
          \\ 
    &  \hspace{4em}     \text{(one fewer branching factor in the denominator than its numerator.)}
     \\[5pt] 
   &= Z_n^{(m-1)}(u_{n})   \cdot  \mathbb{E}_n^{ (m-1) }  \left[\, a_{n+m}  \, \right]  
     \\[5pt]     
   &= Z_n^{(m-2)}(u_{n})   \cdot  \mathbb{E}_n^{ (m-2) }  \left[\, a_{n+m-1} \, \right]   
                                  \mathbb{E}_n^{ (m-1) }  \left[\, a_{n+m}   \, \right] 
     \\     
   & \vdots  
   \\   
   &= \prod_{k={0}}^{m-1} \mathbb{E}_{n}^{(k) }  \left[\, a_{n+k+1} \, \right]     ,
                                   \quad  \,\mathbb{E}_n^{(0)}[\, a_{n+1}\,] := a_{n+1} = a(u_n)  
\end{split}
\end{align}
}%
where
\[
    \mathbb{E}_{n}^{{(k)}}\left[\, a_{n+k+1} \, \right] =
    \frac {\displaystyle 
             \int  a_{n+k+1} \, dZ_n^{(k)}(u_n)
          }
          { \displaystyle 
              Z_n^{(k)}(u_n)
          }
\] 
\/*******************************************************************************************
{\allowdisplaybreaks 
\begin{align}\label{Z_{n+1}u_n for checking} 
\begin{split} 
  &Z_{n+1}(u_{n} + 0)  
  \\
   &=  \int \prod_{n+2}^m a_i \ P(dX_{n+2} \cdots dX_{\!_m} | u_{n}\!+\!0) 
   \\
   &=  \int \prod_{n+2}^{m-1} a_i \ P(dX_{n+2} \cdots dX_{\!_{m-1}} | u_{n}\!+\!0) \int a_{\!_m} P(dX_{\!_m} | u_{\!_{m-1}})
   \\
   &=  \int \prod_{n+2}^{m-1} a_i \ P(dX_{n+2} \cdots dX_{\!_{m-1}} | u_{n}\!+\!0) \, \beta(u_{\!_{m-1}})
   \\    
   &= \int \prod_{n+2}^{m-1} a_i \ P(dX_{n+2} \cdots dX_{m-1} | u_{n}\!+\!0)   \cdot  
    \frac { \displaystyle 
             \int \prod_{n+2}^{m-1} a_i \  P(dX_{n+2}  \cdots dX_{\!_{m-1}} | u_{n}\!+\!0) \beta(u_{\!_{m-1}}) 
          }
          { \displaystyle
               \int \prod_{n+2}^{m-1} a_i \ P(dX_{n+2}  \cdots  dX_{\!_{m-1}} | u_{n}\!+\!0) 
          }
     \\
   &= \int \prod_{n+2}^{m-1} a_i \ P(dX_{n+2} \cdots dX_{m-1} | u_{n}\!+\!0)  \cdot  \mathbb{E}_{n+1}^{\,\!^{(m-1)}} 
                                                          \left[\, \beta(u_{\!_{m-1}}) \mid u_{n}\!+\!0 \, \right]
     \\
   & \vdots  
   \\ 
   &= \int a_{n+2}(u_{n+2}) P(dX_{n+2} | u_{n}+0) \ \cdot \prod_{k={n+2}}^{m-1} \mathbb{E}_{n+1}^{(k)} 
                                                       \left[\, \beta(u_{k}) \mid u_{n}\!+\!0 \, \right]
   \\          
   &= \prod_{k={n+1}}^{m-1} \mathbb{E}_{n+1}^{(k) }  \left[\, \beta(u_{k} \mid u_{n}\!+\!0) \, \right]     
                                   ,
                                      \quad  \,\mathbb{E}_{n+1}^{(n+1)}[\, \beta(u_n+0)\,] = \beta(u_n + 0) =   \beta(u_n) 
\end{split}
\end{align}
}%
where
\[
    \mathbb{E}_{n+1}^{{(k)}}\left[\, \beta(u_{k}) \mid u_{n}\!+\!0)  \, \right] =
    \frac {\displaystyle 
             \int \beta(u_{k}|u_{n}\!+\!0) \, a_{n+2} \cdots a_{{k}} \ P(dX_{n+2} \cdots dX_{k} \mid u_{n}\!+\!0) 
          }
          { \displaystyle 
             \int a_{n+2} \cdots a_{{k}} \ P(dX_{n+2} \cdots dX_{k}  \mid u_{n}\!+\!0) 
          }
\] 
*****************************************************************************************/
Similarly,  
\[
    Z_{n+1}^{(m-1)}(u_{n + 1}) = \int \Big(\prod_{i=n+2}^{n+m} a_{i} \Big) \  P(dX_{n+2} \cdots dX_{n+m} \mid u_{n+1}) 
                               = \prod_{k={0}}^{m-2} \mathbb{E}_{n+1}^{(k) }  \left[\, a_{n+1+k+1} \, \right] 
\]    
Consequently, letting $X_{n+1} = 0$ yields
\begin{align}\label{Z_{n+1}u_n} 
     Z_{n+1}^{(m-1)}(u_{n}) = \prod_{k={0}}^{m-2} \mathbb{E}_{n+1}^{(k) }  \Big[ a_{n+1+k+1} \mid u_{n}\!+\!0  \Big]  
\end{align} 
where
\[
    \mathbb{E}_{n+1}^{{(k)}}\left[\, a_{n+1+k+1} \mid u_{n}\!+\!0 \, \right] =
    \frac {\displaystyle 
             \int  a_{n+1+k+1}\, dZ_{n+1}^{(k)}(u_n + 0)
          }
          { \displaystyle 
                          Z_{n+1}^{(k)}(u_n + 0)
          }
\] 
Given birthplace $u_{n+1} \left(=u_n + \frac {X_{n+1}}{N} \right)$, $\mathbb{E}_{n+1}^{(k) } \left[ a_{n+1+k+1} \right]$ is a function of $u_{n+1}$. 
Define
\begin{align} \label{mu_n^k}
\begin{split}
        \mu^{(m)}_n(u_n)  :&=  \mathbb{E}_n^{(m)} \big[\, X_{n+1}|u_n \, \big]   
            = \frac{ \displaystyle 
                   \int X_{n+1} \, dZ_n^{(m)}(u_n) 
                  }
                  { \displaystyle 
                          \int dZ_n^{(m)}(u_n) 
                  }
         \\[8pt]
         & =           
         \frac{ \displaystyle 
                \int X_{n+1} \cdot a( u_n) \, P(dX_{n+1} \mid u_n) Z_{n+1}^{(m-1)}(u_{n+1})
              }
              {    Z_{n}^{(m)}(u_{n})
              }
                , \qquad a_{n+1} = a(u_n)   
         \\
         & =  
      \frac{ \displaystyle 
              \int X_{n+1} a( u_n )  \, P(dX_{n+1} \mid u_n)    \frac { Z_{n+1}^{(m-1)}(u_{n+1}) } { Z_{n+1}^{(m-1)}(u_n + 0) }
           }
           {   Z_n^{(m)}(u_n)/Z_{n+1}^{(m-1)}(u_n + 0)
           }
\end{split}
\end{align}
\big(In random walks, $a_{n+1}a_{n+2} \cdots = 1$ and $  \mu^{(m)}_n(u_n) = \mu_{n} (u_n)$\big). 

To estimate $\mathbb{E}_n^{(m)} \big[\, X_{n+1}|u_n \, \big]$, we proceed by induction. Let 
{\allowdisplaybreaks
\begin{align*} 
        \overline X_{n+1}^{(1)}  & := \mathbb{E}_n^{(1)} \left[ X_{n+1}|u_n \right ]  
        = \frac{ \displaystyle 
                   \int X_{{n+1}}  a(u_{{n}})  \, P(dX_{{n+1}} \mid u_n)
                  }
                  { \displaystyle 
                    \int           a(u_{{n}})  \, P(dX_{{n+1}} \mid u_n) 
                  } 
        =   \int X_{{n+1}}  \, P(dX_{{n+1}} \mid u_n)   
\end{align*}
}%
which is the smooth function $\mu_n(u_n)$ (see Theorem \ref{con_ineq}). 
To estimate the population ratios in the numerator and denominator of \eqref{mu_n^k}, we look at
\[        
    \frac 
         {
            \mathbb{E}_{n}^{(k+1) }[\, \cdot \,]
         }          
         {\mathbb{E}_{n+1}^{(k) }\Big[\, \cdot \mid X_{n+1}=0 \, \Big]
         }
\] 
Recall that the average birth rate of generation $k$ after the $n\text{th}$, is
{\allowdisplaybreaks
\begin{align*}
      \mathbb{E}_{n}^{(k)} \left[\, a_{n+k+1} \, \right] & =  \frac { \displaystyle 
             \int   a_{n+k+1} dZ_n^{(k)}(u_n)
          }
          { \displaystyle
               Z_n^{(k)}(u_n)
          } 
     \\[6pt] 
     &=  \frac { \displaystyle 
      \int a_{n+1}P(dX_{n+1} | u_n) \, \int a_{n+k+1} \, dZ_{n+1}^{(k-1)}(u_{n+1})\,   
          }
          { \displaystyle  
                Z_n^{(k)}(u_n)
          }   
    \\[6pt] 
      &= \frac{  \displaystyle      \int a_{n+1}P(dX_{n+1} |u_n)  Z_{n+1}^{(k-1)}(u_{n+1})
                                                  \,  \mathbb{E}_{n+1}^{(k-1)} \left[\, a_{n+k+1} \, \right]
           }
           { 
              Z_n^{(k)}(u_n)
           } 
    \\[6pt] 
      &= \frac{  \displaystyle      \int     \,  \mathbb{E}_{n+1}^{(k-1)} \left[ a_{n+k+1} \right]  dZ_n^{(k)}(u_n)
           }
           { 
              Z_n^{(k)}(u_n)
           }
    \\
     &= \mathbb{E}_{n}^{(k) }   \left[
                                    \mathbb{E}_{n+1}^{(k-1) } \left[\, a_{n+k+1} \, \right]
                                \right] 
\end{align*} 
}%
That is, for all $n$ and $k \ge 1$
\begin{equation}\label{E_n^{(k)}}
\mathbb{E}_{n}^{(k)} \left[\, a(u_{n+k}) \, \right] =
         \mathbb{E}_{n}^{(k) }   \left[
                                    \mathbb{E}_{n+1}^{(k-1) } \left[\, a(u_{n+k}) \, \right]
                                \right] \quad 
\end{equation}
We next prove $\mathbb{E}_{n}^{(1)} \left[\, a(u_{n+1}) \, \right]$ is continuous in $u_n$, using the continuity of  $\overline X_{n+1}^{(1)}$, and then show the continuity of $\overline X_{n+1}^{(2)}$ using  $\mathbb{E}_{n}^{(1)} \left[\, a(u_{n+1}) \, \right]$, and so on: 
\[
\overline X_{n+1}^{(1)} \;\Rightarrow\; \mathbb{E}_n^{(1)} \;\Rightarrow\; \overline X_{n+1}^{(2)} \;\Rightarrow\; \mathbb{E}_n^{(2)} \;\Rightarrow\; \cdots
\] 
leading to smoothness of $\mu^{(m)}_n(u_n)$ and $\mathbb{E}_{n}^{(m)}\left[\, a(u_{n+m}) \, \right]$ for all $m \lesssim N$. The smoothness of $\nu_n^{{(k)}} (u_n)$ can be proved with smooth $\mu^{(k)}_n(u_n)$ and $\mathbb{E}_{n}^{(k-1)}\left[\, a(u_{n+k-1}) \, \right]$.

When $k=1$, for any $i$
\begin{align*}
      \mathbb{E}_i^{(1)} \left[\, a(u_i + X/N) \, \right] 
       & = \frac { \displaystyle 
                         \int a(u_i)  \, P(dX  | u_i ) \, a\Big( u_i +  \frac{X}{N} \Big) 
             }
          { \displaystyle
               \int a(u_i) \, P(dX | u_i)
          } 
       =  \int  a\Big( u_i +  \frac{X}{N} \Big)   P(dX | u_i)  
\end{align*}%
is a smooth function of $u_i$ (meaning it has continuous derivatives up to some required order) since $a(u)$ is uniformly smooth in $u$ and the conditional distribution $P(dX_{i+1} | u_i)$ varies smoothly with $u_i$. In $K$-SAT, $a(u) = 1 - u^K$. For instance, 
\begin{equation}\label{E_n^{(1)}}
    \mathbb{E}_n^{(1)} \left[\, a \Big( u_{n} +  \frac {X_{n+1}}N \Big) \, \right]    
     = a(u_n) + \frac{a'(u_n)}{N} \overline X_{n+1}^{(1)} + O(N^{-2})
\end{equation} 

Write
\begin{align*}  
      \mathbb{E}_{n+1}^{(k) } \left[\, a(u_{n+1+k})  \, \right]  := f_{n+1}^{(k)} \left (u_n + \frac {X_{n+1}}N \right)  
\end{align*}
Then,
\begin{align*}  
      \mathbb{E}_{n+1}^{(k) } \left[\, a(u_{n+1+k}) \mid X_{n+1}=0\, \right] 
                               = f_{n+1}^{(k)} \left (u_n\right)  
\end{align*}
Let $f_i^{(0)}(u_i)$ denote  $a(u_i)$ \big(namely, $\mathbb{E}_n^{(0)}[\, a(u_n)\,] = a(u_n)$\big). Write 
\[ \overline X_{n+1}^{(k)} :=  \mathbb{E}_n^{(k)} (X_{n+1}) 
\] 
thus (\ref{E_n^{(1)}}) can be written as
\begin{equation}\label{f^{(1)}}
          f_n^{(1)}(u_n) = f_{n+1}^{(0)}(u_n) + \frac{\overline X_{n+1}^{(1)}}{N} {f_{n+1}^{(0)}}'(u_n)  + O(N^{-2})
\end{equation}
and
(\ref{E_n^{(k)}}) becomes 
\[
          f_n^{(k)}(u_{n}) = \mathbb{E}_{n}^{(k)}  
                                \left[
                                       f_{n+1}^{(k-1)}(u_{n+1}) 
                                \right] 
\]

Now assume for $k \ge 2$, $f_{n+1}^{(k-1)}(u_{n+1})$ is smooth in $u_{n+1}$ for any $n$.  
Employing a Taylor expansion, we have
\begin{align} \label{E_n^k}
\begin{split}
  f_n^{(k)}(u_{n})   
  & = \mathbb{E}_n^{(k)}\!\left[ f_{n+1}^{(k-1)} \Big( u_{n} +  \frac {X_{n+1}}N \Big) \right]
  \\
  & = f_{n+1}^{(k-1)}\!\left(u_n\right)  +  \frac{\overline X_{n+1}^{(k)}}{N}{f_{n+1}^{(k-1)}}' (u_n)    +  O\!\left(N^{-2} \right)
\end{split}  
\end{align}%
which is smooth in $u_n$ for all $n$, provided $\overline{X}_{n+1}^{(k)}$ is smooth. Observe that the term $\displaystyle \frac{\overline X_{n+1}^{(k)}}{N}{f_{n+1}^{(k-1)}}' (u_n) $ involves no second order derivatives of $f_{n+2}^{(k-2)}$ which is absorbed into the $O(N^{-2})$ term.  
Thus, we have proved that for any $k$, $\mathbb{E}_{n}^{(k)} \left[\, a(u_{n+k}) \, \right]$ is smooth in $u_n$.  Hence, by (\ref{E_n^k}),
{\allowdisplaybreaks
\begin{align}\label{E/E} 
\begin{split}
   \frac {\mathbb{E}_{n}^{(k+1)} \left[\, a(u_{n+1+k}) \, \right]  }   
         {\mathbb{E}_{n+1}^{(k)}\big[\,  a(u_{n+1+k}) | X_{n+1}=0 \, \big] } 
  &=\frac{ \displaystyle 
           f_{n+1}^{(k)}\!\left(u_n \right) +  \frac{\overline X_{n+1}^{(k+1)}(u_n)}{N} {f_{n+1}^{(k)}}'\!\left(u_n\right)  
                                                                                        +  O\!\left(N^{-2} \right)
         }
         { 
           f_{n+1}^{(k)}  \left( u_n + 0 \right) 
         }    
  \\[10pt]
  & = 1 
      + \frac{\overline X_{n+1}^{(k+1)}(u_n) }{N} \cdot \frac{{f_{n+1}^{(k)}}'\!\left(u_n\right)} 
                                                          {f_{n+1}^{(k)}  \left( u_n \right) } 
      +  O\!\left(N^{-2} \right)
\end{split}
\end{align}
}%
Notice that $Z_{n+1}(u_n)$ is $Z_{n+1}\big(u_n +  {X_{n+1}}/{N}\big) \big | _{X_{n+1}=0} $\,. \,From (\ref{prod_m}) and (\ref{Z_{n+1}u_n}), we have 

(a) the population ratio in the numerator of (\ref{mu_n^k}): 
{\allowdisplaybreaks
\begin{align*}
  \frac{Z_{n+1}^{(m-1)}(u_{n+1})} {Z_{n+1}^{(m-1)}(u_{n})}  
       & = \frac{  
                  \displaystyle \prod_{k=0}^{m-2} \mathbb{E}_{n+1}^{(k)}\big[\,  a(u_{n+1+k}) \, \big]
                }
                { \displaystyle \prod_{k=0}^{m-2} \mathbb{E}_{n+1}^{(k)}\big[\,  a(u_{n+1+k}) | X_{n+1}=0 \, \big]
                } 
  \\
  & = \prod_{k=0}^{m-2}  
       \frac{ \displaystyle  
          f_{n+1}^{(k)}\!\left(u_n\right) +  {f_{n+1}^{(k)}}'\!\left(u_n\right) \frac{X_{n+1}}{N} +  O\!\left(N^{-2} \right)
            }
            { \displaystyle  f_{n+1}^{(k)}  \left( u_n  \right) 
            }
            \quad  \big(\text{smooth $f_{n+1}^{(k)}$}\big)
  \\
  & = \prod_{k=0}^{m-2}  \left(1  + \frac{X_{n+1}}{N} \frac{{f_{n+1}^{(k)}}'(u_n)}{f_{n+1}^{(k)}  \left( u_n \right) }     
                                  +   O\!\left(N^{-2} \right)
                           \right)
\end{align*}
}%
 
(b) the population ratio in the denominator of (\ref{mu_n^k}): 
{\allowdisplaybreaks
\begin{align*}
   \frac  {Z_n^{(m)}(u_n)} {Z_{n+1}^{(m-1)}(u_{n})}                 
  & = \frac{ \displaystyle \prod_{k=0}^{m-1}   \mathbb{E}_{n}^{(k)} \big[\, a(u_{n+k}) \, \big] 
               }
               { \displaystyle \prod_{k=0}^{m-2} \mathbb{E}_{n+1}^{(k)}\big[\,  a(u_{n+1+k}) | X_{n+1}=0 \, \big]
               }
   \\            
  &=\frac{ \displaystyle a(u_n) \prod_{k=0}^{m-2} \; \mathbb{E}_{n}^{(k+1)} \big[\, a(u_{n+1+k}) \, \big] 
         }
         { \displaystyle \prod_{k=0}^{m-2} \mathbb{E}_{n+1}^{(k)}\big[\,  a(u_{n+1+k}) | X_{n+1}=0 \, \big]
         }
       \\[10pt]
   &= a(u_n) \, \prod_{k=0}^{m-2}  
      \left(  1   
              +  \frac{\overline X_{n+1}^{(k+1)}(u_n) }{N} \cdot \frac{{f_{n+1}^{(k)}}'\!\left(u_n\right)} 
                                                                      {f_{n+1}^{(k)}  \left( u_n \right) } 
              +  O\!\left(N^{-2} \right)
     \right) \quad  \big(\text{by \eqref{E/E}}\big)
\end{align*}
}%
where  $  \overline X_{n+1}^{(k+1)} := \mathbb{E}_{n}^{(k+1) }  \left[X_{n+1}  \mid u_n \right]$. 

Write
\[
           A_k(u_n) =   \frac{{f_{n+1}^{(k)}}'(u_n)} {f_{n+1}^{(k)}(u_n)} \Big( X_{n+1} - \overline X_{n+1}^{(k+1)}(u_n) \Big)
\] 
Then, (\ref{mu_n^k})---that is, $\mu^{(m)}_n(u_n)$---becomes, with terms of order $O\!\left(N^{-2} \right)$ omitted,  
\begin{align*}
  \overline  X_{n+1}^{(m)}(u_n)  
      &=  \int \!  X_{n+1} \prod_{k=0}^{m-2} \! \left(
       1   
          +  
        \frac{1}{N}  A_k(u_n)
                                                 \right)   \! P(dX_{n+1}|u_n) 
\end{align*}  
where $a(u)$ cancels out.  This is a smooth function of $u_n$, since $f_{n+1}^{(k)}$, ${f_{n+1}^{{(k)}}}'$ and $\overline X_{n+1}^{(k+1)}(u_n)$ are smooth by induction for all $k+1 < m-1$,  and $f_{n+1}^{(k)}$ are bounded away from zero (recall $a(u_i) = 1 - u_i^K$ in $K$-SAT); as $m = O(N)$, expressions of the form $\exp\!\left(N^{-1}\sum  A_k(u_n) \right)$ are uniformly bounded,  
In the same way, we can prove the smoothness of $\nu_n^{{(m)}} (u_n)$ and $\mathbb{E}_{n}^{(m)} \left[\, a(u_{n+m}) \, \right]$. 

\textbf{Remark}. 
\begin{itemize} 
\item   In the $K$-SAT case, in the domain of 1-clause decreasing (i.e., $X_i < 0$) and $m = o(N)$ 
\begin{align*}
\mathbb{E}_{i-1}^{(m)}[\,  e^{\delta A|X_i|}   \,]  
              &= \int \!  e^{\delta A|X_i|}  \prod_{k=0}^{m-2} \! \left(
                   1   
                      +  
                    \frac{1}{N}  A_k(u_{i-1})
                                                             \right)   \! P(dX_{i}|u_{i-1})  
              \\
              &
                \le K + o(1)
\end{align*}

Here the $o(1)$ term is uniform. Thus on a vertical segment of the process where only 1-clause decreases occur, with $m = o(N)$, the exponential moment is bounded uniformly along the segment, and the concentration inequality holds. Immediately after passing this segment, the increase in the number of $K$-clauses preserves the negative association condition. 

\hspace{1em} On the other hand, the effect of decreasing a 1-clause can be offset (the case where $X_i < 0$), through negative association, by adding several 3-clauses\,\footnote{Roughly, by $k$ 3-clauses such that 
  $(1 + u)(1 - u^3)^k \le 1$
}, and afterward, the negative association condition is preserved and strengthened through the addition of 3-clauses. Over the entire 2D area of $N \times N$ points, such 1-clauses account for only $O(N)$, since the total number 1-clauses is $O(N)$. Therefore, the finite  exponential moment condition holds on a proportion $1 - O(N^{-1})$ of the whole domain. 
 
\end{itemize}%

\noindent The rest of the proof is the same as in Theorem \ref{con_ineq}. It is worthy noting that to complete the proof we need to use
\[
   \mathbb{E}^{ (m) } |u_n - \bar u_n|^p \le \frac {C_p} {N^{p/2}} ~~~~ \mbox { for  $p \ge 1 $
                                           }
\]
This in turn requires concentration inequality for $x_1 + x_2 + \cdots  x_n$ which holds by induction.
\end{proof}

\clearpage 
\section{\centering} \label{finite Ee^ᶿZ}  

\begin{thm}\label{Ee^ᶿZ ≤ ∞}  Let $Z$ denote the  total number of descendents, $\xi$ the number of offspring of the ancestor.
For a subcritical Galton–Watson process, 
\[
          \mathbb{E} [ e^{\theta \xi} ] < \infty, \textnormal{ for some } \theta > 0 \Longleftrightarrow 
          \mathbb{E} [ e^{\delta   Z} ] < \infty, \textnormal{ for some } \delta > 0           
\]
\end{thm}

\begin{proof} $\Rightarrow$: 
We prove the following equivalent proposition:
\[                           
          \mathbb{E} [ t^\xi ] < \infty \; \textnormal{ for some $t > 1$}    \Rightarrow
          \mathbb{E} [ s^Z ]   < \infty \; \textnormal{ for some $s > 1$} 
\]   
Define increasing functions $f(x) := \mathbb{E} [ x^\xi ] $.  Let $h(x) = x/f(x)$. Thus $h(1)=1$ and
\[
            h(x)' = \frac{f(x) - xf'(x) }{{f(x)}^2} 
           \Rightarrow h(1)' = 1 - f'(1) = 1 - \mathbb{E}(\xi) > 0\,, 
\] 
due to that $ E(\xi) = m < 1$ for subcritical processes. Therefore, there is $x_0 > 1$ sufficiently close to 1 such that 
$x_0/f(x_0) > 1$; moreover $f(x_0) > 1$. Define 
\begin{equation}\label{sf(x)=x}
      s := \frac{x_0}{f(x_0)} > 1
\end{equation}
Let $Z_n$ denote the population in generation $n$, and let $a_n := \mathbb{E}[s^{Z_n}]$. Then
\begin{align*}
    a_{n+1} &\equiv \mathbb{E}[s^{Z_{n+1}}]
    \\[4pt]
            &=  \mathbb{E} \left[ s^{ 1 + \sum_{i=1}^\xi Z_{n}^{(i)}} 
                  \right]
              \quad (\text{$\xi$: number of offspring of the root)}
    \\[4pt]              
            &=   \mathbb{E} \left[s \prod_{i=1}^\xi s^{Z_{n}^{(i)}}
                  \right]              
    \\[4pt]              
            &= s\cdot \mathbb{E} \left[ \prod_{i=1}^\xi \mathbb{E} \left[ s^{Z_{n}} \right]
                  \right]              
              \quad \text{(i.i.d. reproduction of individuals)} 
    \\[4pt]              
            &= s\cdot  \mathbb{E} \left[ \,  {a_n}^\xi
                  \right]  
    \\[4pt]              
            &= s  f( {a_n})                   
\end{align*}
Thus, since $Z_0=1$ so that $a_0 = s$, and $Z_n$ is increasing, we have an increasing sequence $a_n$. Moreover, $a_n$ is upper bounded by $x_0$. Indeed, note 
\[
    a_0 = s = \frac{x_0}{f(x_0)}< x_0 \quad \text{\big(by (\ref{sf(x)=x}) and  $f(x_0) > 1$\big)}
\]      
If $a_n < x_0$, then $a_{n+1} =  sf(a_n) \le  sf(x_0) = x_0 $ from (\ref{sf(x)=x}). Hence $a_n \le x_0 $ for all $n$. It follows from the monotone convergence theorem (MCT), 
$
      a_n \uparrow L
$. 
Likewise, for subcritical Galton–Watson process, 
$
                             Z_n \uparrow Z.
$
Therefore, 
\[
       s^{Z_n} \uparrow  s^Z. 
\]       
Then, 
\[
     \lim_{n\to\infty} \mathbb{E}[s^{Z_n}] = \mathbb{E}[s^Z] = L < \infty
\] 
$\Leftarrow$:  
Since $\xi \le Z$, for  $0 < \theta \le \delta  $, $ e^{\theta \xi} \le e^{ \delta Z}$.  
Taking expectation yields 
\[               
              \mathbb{E} [\, e^{\theta \xi} \,] \le \mathbb{E} [\, e^{\delta Z} \,] < \infty
\]

\end{proof}

\clearpage 
\section{\centering} \label{u=0 y<1 2-SAT}  

We start from
\begin{equation}\label{B.1}
    y = \frac{1-u^2}{u} \ln \frac{1-u-\frac{x}{2}}{1-2u}. 
\end{equation} 
Employing the Taylor series 
\[
         \ln \frac{1+x}{1-x} = 2\Big( x + \frac { x^3} 3 + \frac { x^5} 5   + \cdots \Big)
\] 
we expand the logarithmic term:
\begin{align*}
   \ln \frac{1-u-\frac{x}{2}}{1-2u}   & =
              \ln \frac{1 + \frac{u- {x}/{2}}{2-3u-{x}/{2}}}
                       {1 - \frac{u- {x}/{2}}{2-3u-{x}/{2}}}
                                      =    \frac{u-x/2}{1-3u/2-x/4} + O(u^3)
      \\[6pt]
           &=           \left(u - \frac x 2 \right)\left(1 + \frac32 u  + \frac x 4 + O(u^2) \right) + O(u^3) 
      \\[6pt]     
           &=\left(u-\frac x 2\right) + \frac 32 u  \left(u - \frac x 2 \right) + \frac x 4 \left(u - \frac x 2 \right) 
                       +  O(u^3)
      \\[6pt]     
           &= u + \frac 3 2 u^2 - \frac x 2  +  O(x^2, x u, u^3) 
\end{align*}
Plugging in (\ref{B.1}),  
\begin{align*}
     yu  & = (1-u^2)\left(u + \frac 3 2 u^2 - \frac x 2  +  O(x^2, x u, u^3) \right)
     \\[6pt]
         & =  u + \frac 3 2 u^2 - \frac x 2  +  O(x^2, x u, u^3)  
\end{align*}
Rearranging,
\[ 2(1-y)u + 3u^2 = x + O(xu, x^2, u^3).
\] 
In particular, when $x=0$,
\[ 
        2(1-y)u + 3u^2 = 0,
\]
which implies that $u=0$ for $y<1$. Incidentally, $ \displaystyle \frac{du}{dx} =  \frac{1}{2(1-y)}$\,, which diverges when $y$ approaches to 1. 

\clearpage
\section{\centering} \label{u=0 y<1 (2+p)-SAT} 
We start with the $(2+p)$-SAT PDE 
\[
        z \frac {3u^2}{2(1-u^3)} +  y\frac{u}{1-u^2} = \ln \frac {1-u-x/2}{1-2u}
\]
Expand each term for small $u$:
\[
        \frac {3u^2}{2(1-u^3)} = \frac{3}{2}u^2 +  O(u^5), \quad \frac{u}{1 - u^2} = u + u^3 + O(u^5)
\]
and     
\begin{align*}
   \ln \frac{1-u-\frac{x}{2}}{1-2u}   = u + \frac 3 2 u^2 - \frac x 2  +  O(x^2, x u, u^3) 
\end{align*}
Thus, the $(2+p)$-SAT PDE yields 
\[        z \frac {3u^2}{2}    +  yu + yu^3        = u + \frac 3 2 u^2 - \frac x 2  +  O(x^2, x u, u^3) 
\]
Combining like terms: 
\begin{align*}
                   (z - 1)\frac 3 2 u^2  +  (y-1)u  =  - \frac x 2  +  O(x^2, xu, u^3) 
\end{align*}
In particular, when $x=0$, 
\[ 
          0=(1-y)u+\frac32(1-z)u^2   \quad \text{(neglecting $O(u^3)$)}
\]
which implies that $u=0$ for $y<1$. Incidentally, $\displaystyle \frac{du}{dx} = \frac{1}{2(1-y)}$ for small $x$, the same as the case of 2-SAT.  
%
%
%
%

\clearpage 
\section{\centering} \label{z=1+8u/9}  
More accurate expansion with $x=0$:
\[
        z \frac {3u^2}{2(1-u^3)} = \frac{3}{2}zu^2 +  O(u^5),  \quad y\frac{u}{1 - u^2} = yu + yu^3 + O(u^5)
\]
For the logarithm, 
\begin{align*}
   \ln \frac{1-u}{1-2u}   & =
              \ln \frac{1 + \frac{u}{2-3u}}
                       {1 - \frac{u}{2-3u}}
                                      =  2\left(\frac{u}{2-3u} + \frac 1 3 \frac{u^3}{(2-3u)^3}   \right) + O(u^5)
      \\[6pt]
           &=   u + \frac 3 2 u^2 + \frac 7 3 u^3 + \frac {15}4 u^4 + O(u^5)  
\end{align*}
Plugging into the PDE: 
\[
        \frac{3}{2}zu^2  + yu + yu^3  = u + \frac 3 2 u^2 + \frac 7 3 u^3 + O(u^4)  
\]
Combining like terms and rearranging, 
\[
          0 = (1 - y)u + \frac{3}{2}(1 - z)u^2 + \left(\frac 7 3 - y \right)u^3 + \frac {15} 4 u^4 + O(u^5) 
\]   
Dividing by $u$, 
\[
          0 = (1 - y) + \frac{3}{2}(1 - z)u + \left(\frac 4 3 + (1-y) \right)u^2  + \frac {15} 4 u^3  + O(u^4) 
\]    
Thus, when $y=1$, we have  
\[ 
      z =  1  + \frac 8 9u + \frac {5} 2 u^2 + O(u^3) 
\] 
When $z=1$,    
\begin{align*}
      (y - 1) &=   \frac 4 3  \frac{u^2}{(1+u^2)} + \frac {15} 4 u^3 \frac{1}{(1+u^2)}  + O(u^4)
         \\[6pt]
              & =   \frac 4 3 {u^2}  + \frac {15} 4 u^3 + O(u^4)  
\end{align*} 
\[   \hspace{-13em}  \Rightarrow \quad   y - 1 = \frac 4 3 u^2  +  \frac {15} 4 u^3  
\]

\clearpage 
\section{\centering} \label{dE_U vs dE_L}   
\begin{center}  
                  \textbf{$dE_\text{U}/dE_\text{L}$ along paths deviating from $\alpha_e$}
\end{center}
   
From (\ref{recursion of P}), namely 
\[
       \frac{ \Pr(F_{m+1} ) }{ 
                                          \Pr(F_{m} ) 
                                        }
                    =   (1 - u^K), \qquad     
                   \frac{ \Pr ( x_{i+1}F_m ) } { \Pr(F_m) } =  \frac{1 - u - x/2}{1 - x}
\]
we have,  
\[
         \Pr(F_{m + \Delta m} ) = \Pr(F_m)(1- u^K)^{\Delta m }
         ,\quad \Pr(x_{i+\Delta i} F_m) = \Pr(x_i F_m) \left( \frac{1 - u - x/2}{1 - x} \right)^{\Delta i}
\]
Thus, between the points of $(m, i)$ and   $(m+\Delta m, i + \Delta i)$
\[
      \Delta \Pr(\text{SAT}) = \left(1-u^K\right)^{\Delta m} \left(  
                                               \frac{1  - u - x/2} {1 - x}
                                       \right)^{\Delta i}
\]
Hence 
\[    R :=   \frac 
           {
             \Delta \Pr(\text{SAT})_\text{U}
           }
           {
              \Delta \Pr(\text{SAT})_\text{L}
           } 
           = \left(
                   \frac{
                           1  - {u_\text{U}}^K
                        }{
                           1  - {u_\text{L}}^K
                        }
              \right)^{\Delta m}  
             \left(
                   \frac{
                          1  - {u_\text{U}} - x/2
                        }{
                          1  - {u_\text{L}} - x/2
                        }
              \right)^{\Delta i}   
\]
On the $\alpha_e$ curve, $R = 1$. Since $u_\text{U} > u_\text{L}$, both factors in the parentheses are less than 1. It follows that decreasing $\Delta i$ with $\Delta m$ unchanged, or decreasing $\Delta m$ with $\Delta i$ unchanged, both increase $R$, so that $R>1$. Conversely, increasing $\Delta i$ or $\Delta m$ decreases $R$, so that $R < 1$. 
  
\/******************************************************************************************************************************
I already derived this: 

                                    Pr(i+1, m)      1  - u - x/2
          Pr(xᵢ₊₁Fₘ) | Fₘ SAT) =  ------------ = ----------------  
                                     P(i,m)           1 - x        
                                     
If i is changed to i + 1, Pr(SAT) decreased by (1 - u - x/2)/(1 - x). So i -> i - 1 Pr -> Pr/[(1 - u - x/2)/(1 - x)]

If Δm and Δi  

              ΔPr(SAT) = (1 - u³)^Δm * {(1 - u - x/2)/(1 - x)}^Δi

     ΔPr_U(SAT) = (1 - u_U³)^Δm * {(1 - u_U - x/2)/(1 - x)}^Δi 

     ΔPr_L(SAT) = (1 - u_L³)^Δm * {(1 - u_L - x/2)/(1 - x)}^Δi

If  ΔPr_U(SAT)/ΔPr_L(SAT) = 1 given Δm/Δi = dz/dx = z'_e (e.g.  dz/(-dx) = -9.8091   )
              
                (1 - u_U³)^z'_e * {(1 - u_U - x/2)/(1 - x)} 
            [ ----------------------------------------------- ]^Δi = 1
                (1 - u_L³)^z'_e * {(1 - u_L - x/2)/(1 - x)}

Now, if z'_e increase, e.g., |Δi| increasing while Δm unchanged. 

                     (1 - u_U - x/2)/(1 - x)             1 - u_L - x/2  
                [ --------------------------- ]^(-1) = ------------------- > 1  
                     (1 - u_L - x/2)/(1 - x)             1 - u_U - x/2)

Therefore, when z'_e increases entails that ΔPr_U(SAT)/ΔPr_L(SAT) increases. 

The same for the situation when Δm decreases while Δi is unchanged. In this case, (1 - u_U³)^Δm/(1 - u_L³)^Δm (which < 1) 
increases.

variation of paths from from $\alpha_e$
                                      
*******************************************************************************************************************************/

\clearpage 
\section{\centering} \label{3-COL PDE}
\begin{center}  
                  \textbf{3-COL PDE system}
\end{center}

Define:    

\begin{enumerate}[label={}, leftmargin=3em, style=nextline]
  \item  $i_{\{0\}}$: variables preassigned to \{0\}
  \item  $i_{\{1\}}$: variables preassigned to \{1\}   
  \item  $i_{\{2\}}$: variables preassigned to \{2\};  \\
   
  \item  $j_{\{01\}}$: variables preassigned to pair \{0, 1\}  
  \item  $j_{\{02\}}$: variables preassigned to pair \{0, 2\}   
  \item  $j_{\{12\}}$: variables preassigned to pair \{1, 2\}; \\   
     \hspace*{4em} e.g.,  $v \in j_{\{01\}}$ iff $F_m$ is satisfiable by both $v = 0$ and $v = 1$ only. \\
    
    
    $r := \{x_i\}_{i=1}^N   \big\backslash \Big( i_{\{0\}} \cup i_{\{1\}}  \cup i_{\{2\}} 
                                                 \cup j_{\{01\}} \cup j_{\{02\}}  \cup j_{\{12\}} 
                                     \Big)
                                $; i.e. variables not preassigned. 
      
\end{enumerate}  
Assume symmetry among $i_{\{0\}}$, $i_{\{1\}}$, $i_{\{2\}}$, and symmetry among $j_{\{01\}}$, $j_{\{02\}}$, $j_{\{12\}}$. Set 
\[
      i : =  |i_{\{0\}}|  = |i_{\{1\}}|  = |i_{\{2\}}|     
\]      
\[
      j : =  |j_{\{0,1\}}|   =  |j_{\{0,2\}}|   =  |j_{\{1,2\}}|     
\]      
Let $u_{\{0\}}$ denote the backbone of 0, i.e. the set of variables frozen to 0 in $F_m$. By a slight abuse of notation, we also use the same symbol for both the set and its cardinality when no confusion can arise.  By symmetry among the three colours, let
\[
         u:= u_{\{0\}} = u_{\{1\}}  =  u_{\{2\}}
\]
\[
         u_2:= u_{\{0,1\}} = u_{\{0,2\}} =  u_{\{1,2\}}
\]

Consider the probability of $F_m v \in \text{SAT}_{i,j}$ when $v$ restricted to colour 0, namely
$$
            \Pr(F_m (v=0) \in \text{SAT}_{i,j}   \mid F_m \in   \text{SAT}_{i,j}) 
$$
Given $v = 0$, let the probability that $F_m$ is UNSAT (by $v=0$) be denoted by
$$
            \Pr\Big( \neg F_m (v=0)    \;\big|\, F_m \in   \text{SAT}_{i,j} \Big) 
$$
which equals $2u + u_2$, since the chance that $u_{\{1\}} \cup u_{\{2\}} \cup u_{\{1,2\}}$ run into $v$ (butt runs into arrow), thus making $F_m$ UNSAT, is $2u + u_2$.  Therefore, under colour symmetry, 
\[
           \Pr(F_m (v=0) \in \text{SAT}_{i,j}   \mid F_m \in   \text{SAT}_{i,j}) = 1 - 2\bar u - \bar u_2
\]   
Hereafter we write $u$ for $\bar u$ and  $u_2$ for $\bar u_2$, for the sake of notational conciseness. 
Similarly to the $K$-SAT case \big(see, \eqref{b(i,.)}\big), by partitioning the population of variables that $v$ scans over, we obtain  
{\allowdisplaybreaks
\begin{align*}
          1 - 2 u - u_2 = & ~\Pr\big(v \in i_{\{0\}}\big) 
                        \\[4pt]
                         & + \Pr\big(v \in j_{\{0,1\}} \cup j_{\{0,2\}} \big)
                        \\[4pt]  
                          & + \Pr(v \in r) \cdot \Pr\Big(F_m (v=0) \in \text{SAT}_{i,j} \;\big|\,  v \in r, F_m \in   \text{SAT}_{i,j}\Big)
                            \\[4pt]
                         =& ~ x + 2y + (1 - 3x - 3y)\frac{P(i+\frac 1 3,j,m)}{P(i,j,m)}
\end{align*}
}%
Thus, 
\[
        \frac{ 
               \Pr \big(i+1/3,j,m \big)
             }
             {
               \Pr \big(i,j,m \big)
             }             
             = \frac {1 - x - 2y - 2u - u_2}{1 - 3x - 3y}
\]     
So
\[
        \frac{1}{3N} \partial_x \ln P(x,y,z) = \ln \frac {1 - x - 2y - 2 u - u_2}{1 - 3x - 3y} 
\]

On the other hand,  assuming that the contribution from non-singleton forced-equality classes is $o(1)$, analogously to the $K$-SAT case we obtain ,
\begin{equation}\label{3-COL a}
    a(x,y,z) := \frac{\Pr(i,j, m)}{\Pr(i,j, m-1)} = 1 - 3u^2 - \gamma 
\end{equation}
where both $\gamma$ and $\gamma'$ equal $O(N^{-1})$ (as shown in Appendix~\ref{gamma}), and $3u^2$ is the probability that endpoints of an edge fall into singleton backbones and thereby falsify the formula. It follows

\[
     \frac{1}{N} \partial_z \ln P(x,y,z)  = \ln (1-3u^2 -\gamma)
\]
By Clairaut's theorem, 
\[
    \frac{-3}{1 - x - 2y - 2 u - u_2}\frac{\partial (2u + u_2)}{\partial z} 
    =   \frac{-6u}{1-3u^2} \frac{\partial u}{\partial x}
\]
i.e. 
\[
   \frac{\partial (2u + u_2) }{\partial z}  - \frac {2u(1 - x - 2y - 2 u - u_2)}{1 - 3u^2} \frac{\partial u }{\partial x}  = 0
\]

Now the case of pair color backbone:
\[
            \Pr\Big( F_m (v=0) \land F_m (v=1)   \; \big| \,  F_m \Big)  , \quad (\text{$ = u_2 + u_3 = 1 - 3u - 2u_2 $})
\]            
where ``$\in \text{SAT}_i$'' is omitted for notational conciseness. Partitioning variable set of $v$ gives
{\allowdisplaybreaks
\[                                            
\begin{aligned}
     1 - 3u - 2u_2 = &  \Pr\big(
                           v \in i_{\{0\}} \cup i_{\{1\}}  \cup i_{\{2\}}
                          \big) \cdot 0
                   \\[4pt]
                   & + \Pr\big(  v \in j_{\{0,1\}}  \big) \cdot 1 
                   \\[4pt]
                   & + \Pr\big(  v \in j_{\{0,2\}}  \cup j_{\{1,2\}}   \big) \cdot 0
                   \\[4pt]
                   & + \Pr(v \in r)\Pr\Big( 
                                           F_m(v=0) \land F_m(v=1) \; \big |\, v \in r, F_m
                                  \Big)              
                   \\[4pt]
                   &= y + (1 - 3x - 3y)    \frac{ 
                                                         \Pr \big(i,j+1/3,m \big)
                                                }
                                                {
                                                         \Pr \big(i,j,m \big)
                                                }            
\end{aligned}
\]
}%
Consequently, 
\[
     1 - 3u - 2u_2  =  y + (1 - 3x - 3y)    \frac{ 
                                                         \Pr \big(i,j+1/3,m \big)
                                                }
                                                {
                                                         \Pr \big(i,j,m \big)
                                                }          
\]
i.e. ,  
\[
        \frac{ 
               \Pr \big(i,j+1/3,m \big)
             }
             {
               \Pr \big(i,j,m \big)
             }             
             = \frac {1 - y - 3u - 2u_2}{1 - 3x - 3y}
\]     
Therefore
\[
           \frac{1}{3N} \partial_y \ln P(x,y,z) = \ln \frac {1 - y - 3u - 2u_2}{1 - 3x - 3y}
\]
By Clairaut's theorem,  
\[
      \frac{-3}{1 - y - 3u - 2u_2}\frac{\partial (3u + 2u_2)}{\partial z} = \frac{-6u}{1-3u^2} \frac{\partial u}{\partial y}
\] 
and hence
\[
    \frac{\partial (3u + 2u_2)}{\partial z} - \frac{2u(1 - y - 3u - 2u_2)}{1-3u^2} \frac{\partial u}{\partial y} = 0
\]

The PDE system is therefore 
\[ 
\begin{cases}
   \displaystyle  
      \frac{\partial (2u + u_2) }{\partial z}  - \frac {2u(1 - x - 2y - 2 u - u_2)}{1 - 3u^2} \frac{\partial u }{\partial x}= 0
       \\[16pt]
   \displaystyle    
      \frac{\partial (3u + 2u_2)}{\partial z} - \frac{2u(1 - y - 3u - 2u_2)}{1-3u^2} \frac{\partial u}{\partial y} = 0
\end{cases}  
\] 
With the initial conditions 
\[
          u(x,y,0) = x, \qquad u_2(x,y,0) = y. 
\]

\clearpage 
\section{\centering} \label{gamma}
Let $e = (v \oplus w)$ denote a new clause (edge). When both $v$ and $w$ belong to the singleton backbone, the contribution to
$\Pr\big(F_m \land e \text{ is UNSAT} \big)$ is $3u^2$. Consider now the case where $w$ does not belong to $u$. Define the $v$-induced frozen set:
\[
  s(v) = \{\text{variables forced to take the same value as } v  \text{ upon fixing } v \}.
\]    
There are $O(N)$ such sets, and $|s(v)| = O(1)$ on any domain bounded away from the singularity. 
Hence, the probability that the new clause $(v \oplus w)$ makes $F_m \land (v \oplus w)$ unsatisfiable is
\begin{align*}
   \gamma =   \frac{1}{N}  \sum_v \Bigg[
                 & \Pr\Big(  F_m \land e \text{ is UNSAT}  \; \big|\, w \in s(v) \Big)  \Pr\big(w \in s(v) \big)
           \\                                  
                 &  +   \Pr\Big( F_m \land e  \text{ is UNSAT}  \; \big|\, w \notin s(v) \Big)     \Pr\big(w \notin s(v) \big)  
                          \Bigg]
\end{align*}
When $w \notin s(v)$, $F_m \land (v \oplus w)$ is satisfiable since $w$ can take a feasible value different from that of $v$. Therefore, the second term vanishes. Thus, 
\begin{align*}
               \gamma  = \frac{1}{N} \sum_v \Pr\big( w \in s(v) \big) = \frac{1}{N^2} \sum_v |s(v)| = O(N^{-1}) 
\end{align*}
Moreover, varying $i$ or $j$ by $1$ changes the total number of induced frozen variables by only $O(1)$, i.e. 
\[
         \Delta \left( \sum_v |s(v)| \right) = O(1) 
\] 
and hence $ \Delta \gamma = O(N^{-2})$, leading to  
\[
      \gamma_x, \gamma_y = O\left(N^{-1} \right)
\]

\end{document}